%% file: main.tex
\documentclass{emulateapj}
\usepackage{graphicx}
\usepackage[space]{grffile}
\usepackage{latexsym}
\usepackage{textcomp}
\usepackage{longtable}
\usepackage{multirow,booktabs}
\usepackage{amsfonts,amsmath,amssymb}
\usepackage{natbib}
\usepackage{url}
\usepackage{CJKutf8}
\usepackage{hyperref}
\usepackage{enumerate}
\hypersetup{colorlinks=false,pdfborder={0 0 0}}
\newif\iflatexml\latexmlfalse
\usepackage[utf8]{inputenc}
\usepackage[english]{babel}
\usepackage{graphicx, color}

\usepackage{lineno}

\newcommand{\kms}{km~s$^{-1}$}

\newcommand{\be}{\begin{equation}}
\newcommand{\ee}{\end{equation}}
\newcommand*{\teff}{$T_{\rm eff}$}
\newcommand*{\logg}{$\log~g$}
\newcommand*{\feh}{[Fe/H]}
\newcommand*{\afe}{[$\alpha$/Fe]}

\newcommand{\halpha}{H$\alpha$}
\newcommand{\hbeta}{H$\beta$}

\newcommand{\Gaia}{{\it Gaia}~}
\newcommand{\gaia}{{\it Gaia}}

\newcommand{\vizier}{VizieR}

\definecolor{todo}{RGB}{200,0,0}

\newcommand{\revision}[1]{\textcolor{black}{#1}}

\shorttitle{SDSS Data Release 17}
\shortauthors{SDSS-IV Collaboration}

\begin{document}

\title{The Seventeenth Data Release of the Sloan Digital Sky Surveys: Complete Release of MaNGA, MaStar and APOGEE-2 Data}

\input{authors.tex}

\begin{abstract}
\input{abstract.tex}
\end{abstract}


\section{Introduction}
\input{intro.tex}\label{sec:intro}

\section{Scope of DR17}
\input{scope.tex}

\section{Data Access}
\input{dataaccess.tex}

\section{APOGEE-2 : Full Release} \label{sec:apogee}
\input{apogee.tex}

\section{MaNGA: Full Release of Final Sample} \label{sec:manga}
\input{manga.tex}
\section{MaStar: The MaNGA Stellar Library} \label{sec:mastar}
\input{mastar.tex}

\section{eBOSS Like Data} \label{sec:eboss}
While both the main eBOSS as well as the co-observed TDSS made their final, full catalog release in DR16, new eBOSS like data is released for both the SPIDERS sub-survey, and the eBOSS-RM program. A number of eBOSS related VACs are also released. 

\subsection{eBOSS VACs}
\input{eboss.tex}

\subsection{SPIDERS} 
\input{spiders.tex}
\subsection{eBOSS-RM}
\input{rm.tex}

\section{Conclusions and Future Plans}
\input{future.tex}

\section{Acknowledgements}

Funding for the Sloan Digital Sky Survey IV has been provided by
the Alfred P. Sloan Foundation, the U.S. Department of Energy Office of
Science, and the Participating Institutions. SDSS-IV acknowledges
support and resources from the Center for High-Performance Computing at
the University of Utah. The SDSS web site is www.sdss.org.

SDSS-IV is managed by the Astrophysical Research Consortium for the 
Participating Institutions of the SDSS Collaboration including the 
Brazilian Participation Group, the Carnegie Institution for Science, 
Carnegie Mellon University, the Chilean Participation Group, the French Participation Group, Center for Astrophysics | Harvard \& Smithsonian, 
Instituto de Astrof\'isica de Canarias, The Johns Hopkins University, 
Kavli Institute for the Physics and Mathematics of the Universe (IPMU) / 
University of Tokyo, Korean Participation Group, Lawrence Berkeley National Laboratory, 
Leibniz Institut f\"ur Astrophysik Potsdam (AIP),  
Max-Planck-Institut f\"ur Astronomie (MPIA Heidelberg), 
Max-Planck-Institut f\"ur Astrophysik (MPA Garching), 
Max-Planck-Institut f\"ur Extraterrestrische Physik (MPE), 
National Astronomical Observatories of China, New Mexico State University, 
New York University, University of Notre Dame, 
Observat\'ario Nacional / MCTI, The Ohio State University, 
Pennsylvania State University, Shanghai Astronomical Observatory, 
United Kingdom Participation Group,
Universidad Nacional Aut\'onoma de M\'exico, University of Arizona, 
University of Colorado Boulder, University of Oxford, University of Portsmouth, 
University of Utah, University of Virginia, University of Washington, University of Wisconsin, 
Vanderbilt University, and Yale University.

Co-authorship on SDSS-IV data papers is alphabetical by last name and offered to all collaboration members who have contributed at least 1 month FTE towards any of the surveys during the period up to the end of data collection; as well as to any formally approved external collaborator who has contributed at least 1 month FTE to work critical to the data release. 

The documentation workshop for DR17 was held over Zoom (``DocuZoom") in May 2021. This event was the main venue for documentation updates for DR17 (including the start of this paper) and was attended by Anne-Marie Weijmans, Joel Brownstein, Mike Blanton, Karen Masters, Renbin Yan, Sten Hasselquist, Michael Talbot, Hannah lewis, Christian Hayes, Ani Thaker, Gail Zasowski, David Law, Brian Cherinka, Kyle Westfall, Amy Jones, Lewis Hill, Jon Holtzman, Jos\'e S\'anchez-Gallego, Rachael Beaton, Scott Anderson, Jennifer Johnson, Caroline Swartz, Jordan Raddick, Julie Imig and a Llama. 

Figure \ref{fig:data} made by Joel Browstein; Figure \ref{fig:apogeedr17new} and \ref{fig:apogeenstars} made by Christian Hayes; Figure \ref{fig:alphafe} made by Jon Holtzman. Figure \ref{fig:manga_sample} by Jos\'e S\'anchez-Gallego; Figure \ref{fig:mastarhc2} made by Kyle Westfall; Figure \ref{fig:marvin_web} made by Brian Cherinka; Figure \ref{fig:mastar_hr} made by Renbin Yan; 

This publication uses data generated via the Zooniverse.org platform, development of which is funded by generous support, including a Global Impact Award from Google, and by a grant from the Alfred P. Sloan Foundation.

This publication makes use of data products from the Two Micron All Sky Survey, which is a joint project of the University of Massachusetts and the Infrared Processing and Analysis Center/California Institute of Technology, funded by the National Aeronautics and Space Administration and the National Science Foundation.

This work is based, in part, on observations made with the \emph{Spitzer} Space Telescope, which is operated by the Jet Propulsion Laboratory, California Institute of Technology under a contract with NASA.

This publication makes use of data products from the Wide-field Infrared Survey Explorer \citep{WISE}, which is a joint project of the University of California, Los Angeles, and the Jet Propulsion Laboratory/California Institute of Technology, funded by the National Aeronautics and Space Administration, and NEOWISE, which is a project of the Jet Propulsion Laboratory/California Institute of Technology. 
WISE and NEOWISE are funded by the National Aeronautics and Space Administration 

This work has made use of data from the European Space Agency (ESA) mission \gaia\ (\url{https://www.cosmos.esa.int/gaia}), processed by the \gaia\ Data Processing and Analysis Consortium (DPAC, \url{https://www.cosmos.esa.int/web/gaia/dpac/consortium}). Funding for the DPAC has been provided by national institutions, in particular the institutions participating in the \gaia\ Multilateral Agreement.

This research has made use of NASA's Astrophysics Data System.

This research has made use of the SIMBAD database, operated at CDS, Strasbourg, France. The original description of the SIMBAD service was published in \citet{simbad_2000}.

This research has made use of the \vizier\ catalogue access tool, CDS, Strasbourg, France (DOI: 10.26093/cds/vizier). The original description of the \vizier\ service was published \citet{vizier2000}.

This research makes use of data from the Green Bank Observatory. The Green Bank Observatory is a facility of the National Science Foundation operated under cooperative agreement by Associated Universities, Inc.

This research made use of \textsc{astropy}, a community-developed core \textsc{python} ({\tt http://www.python.org}) package for Astronomy \citep{2013A&A...558A..33A}; \textsc{ipython} \citep{PER-GRA:2007}; \textsc{matplotlib} \citep{Hunter:2007}; \textsc{numpy} \citep{:/content/aip/journal/cise/13/2/10.1109/MCSE.2011.37}; \textsc{scipy} \citep{citescipy}; and \textsc{topcat} \citep{2005ASPC..347...29T}.

\bibliography{references}{}
\bibliographystyle{aasjournal}

\end{document}

%% file: authors.tex
\email{spokesperson@sdss.org}
\author{Abdurro'uf\altaffilmark{1},
Katherine Accetta\altaffilmark{2},
Conny Aerts\altaffilmark{3},
V\'ictor Silva Aguirre\altaffilmark{4},
Romina Ahumada\altaffilmark{5},
Nikhil Ajgaonkar\altaffilmark{6},
N. Filiz Ak\altaffilmark{7},
Shadab Alam\altaffilmark{8},
Carlos Allende Prieto\altaffilmark{9,10},
Andr\'es Almeida\altaffilmark{11}
% ...
}
\altaffiltext{1}{Academia Sinica Institute of Astronomy and Astrophysics, 11F of AS/NTU, Astronomy-Mathematics Building, No.1, Sec. 4, Roosevelt Rd, Taipei, 10617, Taiwan}
\altaffiltext{2}{Department of Astrophysical Sciences, Princeton University, Princeton, NJ 08544, USA}
\altaffiltext{3}{Institute of Astronomy, KU Leuven, Celestijnenlaan 200D, B-3001 Leuven, Belgium}
\altaffiltext{4}{Stellar Astrophysics Centre, Department of Physics and Astronomy, Aarhus University, Ny Munkegade 120, DK-8000 Aarhus C, Denmark}
\altaffiltext{5}{Instituto de Astronom\'ia, Universidad Cat\'olica del Norte, Av. Angamos 0610, Antofagasta, Chile}
\altaffiltext{6}{Department of Physics and Astronomy, University of Kentucky, 505 Rose St., Lexington, KY, 40506-0055, USA}
\altaffiltext{7}{Department of Astronomy and Space Sciences, Erciyes University, 38039 Kayseri, Turkey}
\altaffiltext{8}{Institute for Astronomy, University of Edinburgh, Royal Observatory, Blackford Hill, Edinburgh EH9 3HJ, UK}
\altaffiltext{9}{Instituto de Astrof\'{\i}sica de Canarias (IAC), C/ Via L\'actea s/n, E-38205 La Laguna, Tenerife, Spain}
\altaffiltext{10}{Universidad de La Laguna (ULL), Departamento de Astrof\'{\i}sica, E-38206 La Laguna, Tenerife Spain}
\altaffiltext{11}{Department of Astronomy, University of Virginia, Charlottesville, VA 22904-4325, USA}
% ...

%% file: abstract.tex
This paper documents the seventeenth data release (DR17) from the
Sloan Digital Sky Surveys; the fifth and final release from the fourth phase
(SDSS-IV).  DR17 contains
the complete release of the Mapping Nearby Galaxies at Apache Point
Observatory (MaNGA) survey, which reached its goal of surveying over
10,000 nearby galaxies. The complete release of the MaNGA Stellar
Library (MaStar) accompanies this data, providing observations of
almost 30,000 stars through the MaNGA instrument during bright
time. DR17 also contains the complete release of the Apache Point
Observatory Galactic Evolution Experiment 2 (APOGEE-2) survey which
\revision{publicly releases} infra-red spectra of over 650,000 stars. The main
sample from the Extended Baryon Oscillation Spectroscopic Survey
(eBOSS), as well as the sub-survey Time Domain Spectroscopic Survey
(TDSS) data were fully released in DR16. New single-fiber optical
spectroscopy released in DR17 is from the SPectroscipic IDentification of
ERosita Survey (SPIDERS) sub-survey and the eBOSS-RM program.  Along
with the primary data sets, DR17 includes \revision{25} new or updated Value
Added Catalogs (VACs). This paper concludes the release of SDSS-IV survey
data. SDSS continues into its fifth phase with observations already
underway for the Milky Way Mapper (MWM), Local Volume Mapper (LVM) and
Black Hole Mapper (BHM) surveys.

%% file: intro.tex
The Sloan Digital Sky Surveys (SDSS)  have been almost continuously observing the skies for
over 20 years, since the project began with a first phase in 1998
(SDSS-I; \citealt{2000AJ....120.1579Y}). SDSS has now completed four phases of operations (with a fifth ongoing;
see \S \ref{sec:future}). Since 2017, SDSS has had a dual hemisphere view of the sky, observing from both Las
Campanas Observatory (LCO), using the du Pont Telescope and the Sloan Foundation 2.5-m
Telescope, \citep{2006AJ....131.2332G} at Apache Point Observatory
(APO). This paper describes data taken during the fourth phase of SDSS
(SDSS-IV; \citealt{2017AJ....154...28B}), which consisted of three main
surveys; the Extended Baryon Oscillation Spectroscopic Survey
(eBOSS; \citealt{Dawson16}), Mapping Nearby Galaxies at APO
(MaNGA; \citealt{2015ApJ...798....7B}), and the APO Galactic Evolution
Experiment 2 (APOGEE-2; \citealt{Majewski2017}). Within eBOSS, SDSS-IV
also conducted two smaller programs: the SPectroscopic
IDentification of ERosita Sources (SPIDERS; \citealt{Clerc2016,
Dwelly17}) and the Time Domain Spectroscopic Survey
(TDSS; \citealt{morganson15a}), and continued the SDSS
Reverberation Mapping (SDSS-RM) program to measure black hole masses
out to redshifts $z\sim 1$--2 using single fiber spectra. Finally, the use of dual observing modes with
the MaNGA and APOGEE instruments (\citealt{Drory2015, Wilson2019})
facilitated the development of the MaNGA Stellar Library (MaStar;
\citealt{Yan2019}), which observed stars using the MaNGA fiber bundles during
APOGEE-led bright time observing.  

This suite of SDSS-IV programs was developed to map the Universe on a range
of scales, from stars in the Milky Way and nearby satellites in APOGEE-2, to nearby galaxies in MaNGA, and out to cosmological scales
with eBOSS. SPIDERS provided follow-up observations of X-ray emitting
sources, especially from eROSITA (\citealt{Merloni12,
Predehl_2014_eROSITA}), and TDSS and SDSS-RM provided a spectroscopic
view of the variable sky.

The final year's schedule for SDSS-IV was substantially altered due to
the COVID-19 pandemic. Originally, the SDSS-IV observations were
scheduled to end at APO on the night of June 30, 2020 and at LCO on
the night of September 8, 2020. Closures in response to COVID-19
altered this plan. APO closed on the morning of March 24, 2020 and the
2.5-m Sloan Foundation Telescope reopened for science observations the
night of June 2, 2020.  The summer shutdown ordinarily scheduled in
July and August was delayed and instead SDSS-IV observations continued
through the morning of August 24, 2020. LCO closed on the morning of
March 17, 2020 and the du Pont Telescope reopened for science
observations the night of October 20, 2020. The du Pont Telescope was
used almost continuously for SDSS-IV through the morning of January
21, 2021. These changes led to different sky coverages than were
originally planned for SDSS-IV but still allowed it to achieve or exceed all
of its original goals.

This paper documents the seventeenth data release (DR17) from SDSS
overall, and is the fifth and final annual release from SDSS-IV
(following DR13: \citealt{2017ApJS..233...25A};
DR14: \citealt{2018ApJS..235...42A},
DR15: \citealt{2019ApJS..240...23A} and DR16: \citealt{DR16}). With this release SDSS-IV has completed the goals set out in \citet{2017AJ....154...28B}. 

A complete overview of the scope of DR17 is provided
in \S \ref{sec:scope}, and information on how to access the data can
be found in \S \ref{sec:access}. We have separate sections on
MaNGA (\S \ref{sec:manga}), MaStar (\S \ref{sec:mastar}) and
APOGEE-2 (\S \ref{sec:apogee}), and while there is no new main eBOSS
survey or TDSS data in this release, we document releases from SPIDERS
and the eBOSS-RM program as well as eBOSS related value added
cataloges (VACs) in \S \ref{sec:eboss}. We conclude with a summary of
the current status of SDSS-V now in active operations \revision{along with describing plans for future data releases}
(\S \ref{sec:future}).

%% file: scope.tex
\label{sec:scope}
SDSS data releases have always been cumulative, and DR17 follows that
tradition, meaning that the most up-to-date reduction of data in all
previous data releases are included in DR17. The exact data products
and catalogs of previous releases also remain accessible on our
servers. However, we emphatically advise users to access any SDSS data
from the most recent SDSS data release, because data may have been reprocessed using updated data reduction pipelines, and catalogs may have been updated with new entries and/or improved analysis methods. Changes between the processing methods used in DR17 compared to previous data releases are documented on the DR17 version of the SDSS website \url{https://www.sdss.org/dr17} as well as in this article.

\begin{figure}
\centering
\includegraphics[width=\columnwidth]{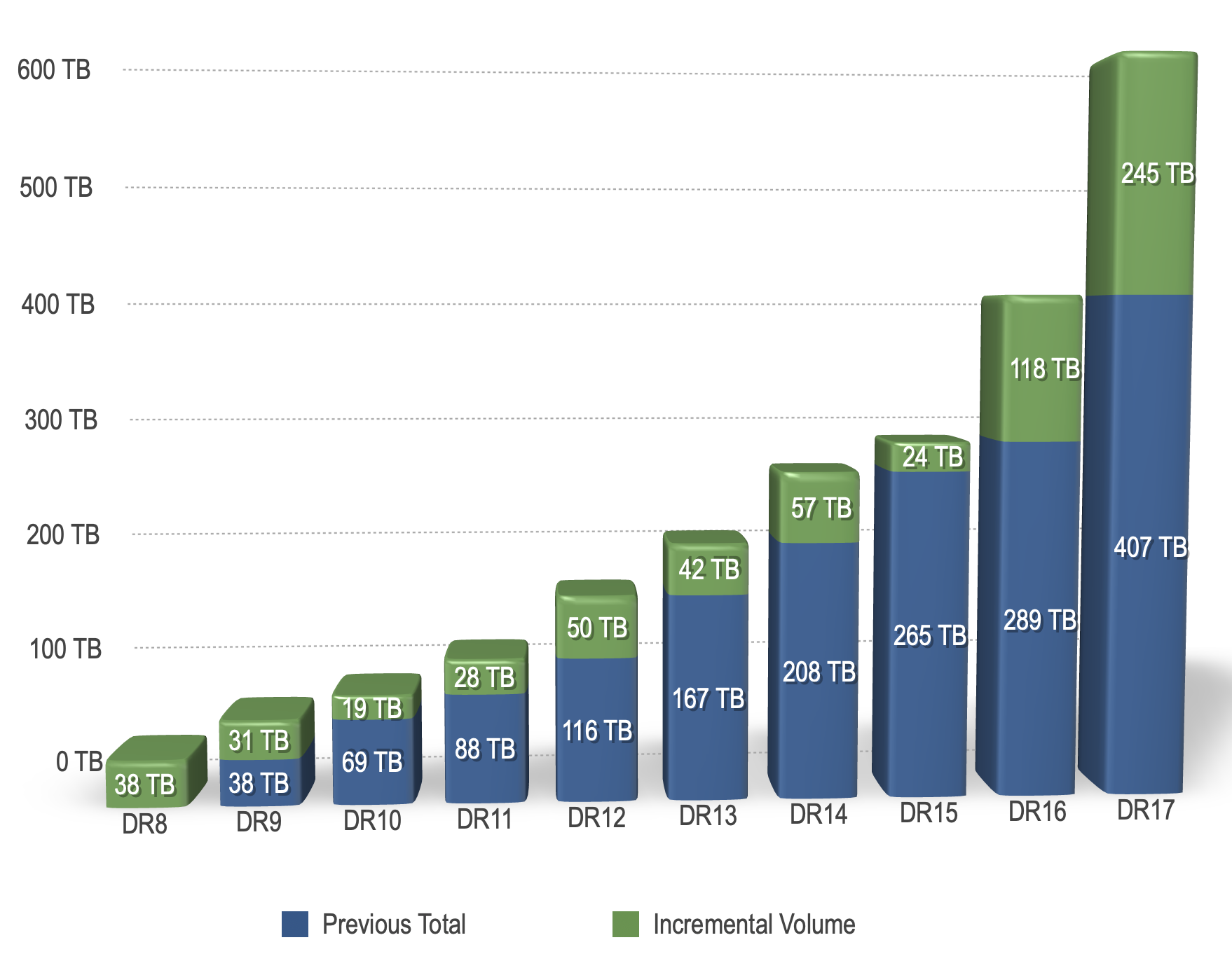}
\caption{The growth in data volume hosted by the SDSS Science Archive Server (SAS) since DR8. \revision{For a more detailed break down of data volume see \url{https://sdss.org/dr17/data\_access/volume}}}
\label{fig:data}
\end{figure}

This data release itself includes over 46 million new files totalling over 222 TB.  Although many of these files replace previous versions, the total volume of all SDSS files including all previous versions now exceeds 623 TB on the Science Archive Server (SAS).  The growth of the volume of data on the SAS since DR8 (which was the first data release of SDSS-III) is shown in Figure \ref{fig:data}. 

\begin{deluxetable*}{lrrrrr}
\tablewidth{5in}
\tablecaption{SDSS-IV spectroscopic data in all releases (DR13--DR17) \label{tab:scope}} 
\tablehead{ 
\colhead{Target Category} & \colhead{DR13} & \colhead{DR14}  & \colhead{DR15} & \colhead{DR16} & \colhead{DR17}}
\startdata
\multicolumn{4}{l}{APOGEE-2} \\ 
    \multicolumn{1}{r}{Main Red Star Sample}                  & 109376 & 184148 & 184148 & 281575 &  372458\\
    \multicolumn{1}{r}{AllStar Entries}                       & 164562 & 277371 & 277371 & 473307 &  733901\\
    \multicolumn{1}{r}{APOGEE-2S Main Red Star Sample}        & -      &      - &      - &  56480 &   96547\\
    \multicolumn{1}{r}{APOGEE-2S AllStar Entries}             & -      &      - &      - & 102200 &  204193\\
    \multicolumn{1}{r}{APOGEE-2S Contributed AllStar Entries} & -      &      - &      - &  37409 &   92152\\
    \multicolumn{1}{r}{NMSU 1-meter AllStar Entries}          & 894    &   1018 &   1018 &   1071 &    1175\\
    \multicolumn{1}{r}{Telluric AllStar Entries}              & 17293  &  27127 &  27127 &  34016 &   45803\\
\tableline 
\multicolumn{4}{l}{MaNGA} \\ 
\multicolumn{1}{r}{All Cubes} &  1390 &  2812 &  4824 & 4824 & 11273\\ 
\multicolumn{4}{l}{Main galaxy sample: } \\ 
\multicolumn{1}{r}{\tt PRIMARY\_v1\_2} &  600  & 1278  & 2126 & 2126 & 4621 \\ 
\multicolumn{1}{r}{\tt SECONDARY\_v1\_2} &  473  & 947  & 1665 & 1665 & 3724 \\ 
\multicolumn{1}{r}{\tt COLOR-ENHANCED\_v1\_2} & 216  & 447 & 710 &  710 & 1514 \\  
\multicolumn{1}{r}{Other targets\tablenotemark{3}} &  31 & 121 & 324 & 324 & 1420\\
\tableline 
\multicolumn{4}{l}{MaStar (MaNGA Stellar Library)} \\ 
\multicolumn{1}{r}{All Cubes} & 0 & 0 & 3321 & 3321 &  24130 \\
\tableline 
\multicolumn{4}{l}{eBOSS} \\ 
\multicolumn{1}{r}{LRG samples} & 32968 & 138777 & 138777 & 298762 & 298762\\
\multicolumn{1}{r}{ELG samples} & 14459 & 35094 & 35094 & 269889 & 269889\\
\multicolumn{1}{r}{Main QSO sample}  & 33928 & 188277 & 188277 & 434820 & 434820 \\	
\multicolumn{1}{r}{Variability selected QSOs} & 22756 & 87270 & 87270 & 185816 & 186625 \\
\multicolumn{1}{r}{Other QSO samples} & 24840 & 43502  & 43502 & 70785 & 73574 \\
\multicolumn{1}{r}{TDSS targets} & 17927 & 57675 & 57675  & 131552 & 131552 \\
\multicolumn{1}{r}{SPIDERS targets} &  3133 & 16394 & 16394  & 36300 & 41969 \\
\multicolumn{1}{r}{Reverberation Mapping} &   849\tablenotemark{1}  & 849\tablenotemark{1}  & 849\tablenotemark{1} & 849\tablenotemark{1}  & 849\tablenotemark{1} \\
\multicolumn{1}{r}{Standard Stars/White Dwarfs} &  53584 & 63880 & 63880 & 84605  & 85105 \\
\vspace{-0.1in}
\tablenotetext{1}{The number of RM targets remains the same, but the number of visits increases.} 
\tablenotetext{3}{Data cubes not in any of the 3 main galaxy samples, including both ancillary program targets and non-galaxy data cubes.} 
\enddata
\end{deluxetable*}

Table \ref{tab:scope} shows the growth of SDSS-IV data separated by survey and target types across our five annual data releases. \revision{These numbers are a mixture of counts of unique spectra and unique objects, and while correct to the best of our ability, can be subject to change based on which quality control flags are implemented.} We also summarize these information below: 

\begin{enumerate}
\item APOGEE-2 is including 879,437 new infrared spectra.\footnote{The number of spectra are tallied as the number of new entries in the AllVisit file. \autoref{tab:scope} conveys the numbers of unique targets that come from the AllStar file.} 
    These data come from observations taken from MJD 58302 to MJD 59160 (i.e., from July 2, 2018 to November 07, 2020) for APOGEE-2 North (APOGEE-2N) at APO and from MJD 58358 to MJD 59234 (August 29, 2018 to January 20, 2021) for APOGEE-2 South (APOGEE-2S) at LCO and the new spectra comprise both observations of 260,594 new targets and additional epochs on targets included in previous DRs.
    The majority of the targets are in the Milky Way galaxy, but DR17 also contains observations of stars in the Large and Small Magellanic Clouds and eight dwarf spheroidal satellites as well as integrated light observations of both M33 and M31. 
    Notably, DR17 contains 408,118 new spectra taken with the APOGEE-S spectrograph at LCO; this brings the total APOGEE-2S observations to 671,379 spectra of 204,193 unique stars. 
    DR17 also includes all previously released APOGEE and APOGEE-2 spectra for a cumulative total of 2,659,178 individual spectra, all of which have been re-reduced with the latest version of the APOGEE data reduction and analysis pipeline (J.~Holtzman et al. in prep.). 
    In addition to the reduced spectra, element abundances and stellar parameters are included in this data release. 
    APOGEE-2 is also releasing a number of VACs, which are listed in Table \ref{table:vac}.
\item  MaNGA and MaStar are releasing all scientific data products
    from the now-completed surveys.  This contains a substantial
    number of new galaxy and star observations respectively, along
    with updated products for all observations previously released in
    DR15 and before.  These updated data products include
    modifications to achieve sub-percent accuracy in the spectral
    line-spread function, revised flux calibration, and Data Analysis
    Pipeline (DAP) products that now use stellar templates constructed from
    the MaStar observations to model the MaNGA galaxy stellar
    continuum throughout the full optical and near-infrared (NIR) wavelength range. MaNGA
    reached its target goal of observing more than 10,000 nearby
    galaxies, as well as a small number of non-galaxy targets, while
    bright time observations enable MaStar to collect spectra for
    almost 30,000 stars through the MaNGA instrument.  MaNGA is
    also releasing a number of VACs (Table \ref{table:vac}).
\item There is no change in the main survey eBOSS data released since
    DR16, when a total of 1.4 million eBOSS spectra were
    released, completing its main survey goals. However, a number of Value Added Catalogs (VACs) useful for cosmological and other applications are released in DR17. The TDSS survey also released its complete dataset in DR16. However, on-going eBOSS-like observations of X-ray sources under the SPIDERS program and continued monitoring of quasars under the reverberation mapping program (SDSS-RM) are released in DR17. 
\item DR17 also includes data
    from all previous SDSS data releases. All MaNGA, BOSS, eBOSS, APOGEE and APOGEE-2 spectra that were previously released have all been reprocessed with the latest reduction and analysis pipelines. eBOSS main survey data were last released in DR16 \citep{DR16}, SDSS-III MARVELS spectra were finalized in DR12 \citep{2015ApJS..219...12A}. SDSS Legacy Spectra were released in its final form in DR8 \citep{2011ApJS..193...29A}, and the SEGUE-1 and SEGUE-2 surveys had their final reductions released with DR9 \citep{2012ApJS..203...21A}. The SDSS imaging had its most recent release in DR13 \citep{2017ApJS..233...25A}, when it was recalibrated for eBOSS imaging purposes.

\end{enumerate}

A numerical overview of the total content of DR17 is given in Table \ref{tab:scope}. An overview of the value-added catalogs that are new or updated in DR17 can be found in Table \ref{table:vac}; adding these to the VACs previously released in SDSS, the total number of VACs in SDSS as of DR17 is now \revision{63} (DR17 updates \revision{14} existing VACs and introduces \revision{11} new ones). 
\revision{DR17 also contains the VACs that were first published in the mini-data release DR16+ on 20 June 2020. DR16+ did not contain any new spectra, and consisted of VACs only. Most of the VACs in DR16+ were based on the final eBOSS DR16 spectra, and these include large scale structure and quasar catalogs. In addition, DR16+ contained three VACs based on DR15 MaNGA sample. The DR16+ VACs can be found in Table \ref{table:vac}, and are described in more detail in the sections listed there.}

\begin{deluxetable*}{lll}
\tablecaption{New or Updated Value Added Catalogs (DR16+ where noted, otherwise new or updated for DR17) \label{table:vac}}
\tablehead{\colhead{Name (see Section for Acronym definitions)} & \colhead{Section} &  \colhead{Reference(s)}}
\startdata
\multicolumn{3}{l}{APOGEE-2} \\ 
Open Cluster Chemical Abundances and Mapping catalog & \S \ref{vac:apogeeobjects} & \citet{2013ApJ...777L...1F,Donor2018,2020AJ....159..199D}, \\
 (OCCAM)             & & N.~Myers et al. (in prep.)  \\
Red-Clump (RC) Catalog & \S \ref{vac:apogeeobjects} & \citet{2014ApJ...790..127B}\\
APOGEE-Joker & \S \ref{vac:apogeeobjects} & A. Price-Whelan et al. (in prep.) \\
Double lined spectroscopic binaries in APOGEE spectra & \S \ref{vac:apogeeobjects} & \citet{Kounkel2021}\\
StarHorse for APOGEE DR17 + \Gaia EDR3 & \S \ref{vac:apogeedist}  & \citet{Queiroz2020}\\
AstroNN &  \S \ref{vac:apogeedist} & \citet{2019MNRAS.483.3255L,2019arXiv190208634L,2019arXiv190104502M} \\
APOGEE Net: a unified spectral model & \S \ref{vac:apogeeprocess} & \citet{olney2020,sprague2022} \\
APOGEE on FIRE Simulation Mocks & \S \ref{vac:apogeefire} & \citet{Sanderson_2020}, \citet{Nikakhtar_2021} \\
\tableline
\multicolumn{3}{l}{MaNGA} \\ 
NSA Images (DR16+) & \S \ref{vac:nsa} & \citet{blanton2011,Wake2017} \\
SWIFT VAC (DR16+) & \S \ref{vac:nsa} &  \citet{swiftVAC}\\
Galaxy Zoo: 3D & \S \ref{vac:morph} & \citet{Masters2021} \\
Updated Galaxy Zoo Morphologies (SDSS, UKIDSS and DESI) & \S \ref{vac:morph} & \citet{Hart16,Walmsley21} \\
Visual Morphologies from SDSS + DESI images (DR16+)  & \S \ref{vac:morph} & \citet{VazquezMata2021}\\
PyMorph DR17 photometric catalog & \S \ref{vac:morph} & \citet{DominguezSanchez2021}\\
Morphology Deep Learning DR17 catalog & \S \ref{vac:morph} &  \citet{DominguezSanchez2021}\\
PCA VAC (DR17) &  \S \ref{vac:stellarpops}   &   \citet{mangapca_1_pace19a,mangapca_2_pace19b}. \\
\textsc{Firefly} Stellar Populations & \S \ref{vac:stellarpops} &  \citet{Goddard2017}, Neumann et al. (in prep.)\\
Pipe3D & \S \ref{vac:stellarpops} & \citet{Sanchez2016,Sanchez2018}\\
HI-MaNGA DR3 & \S \ref{vac:himanga}  & \citet{Masters2019,Stark2021} \\
The MaNGA AGN Catalog & \S \ref{vac:mangaagn} & \citet{Comerford2020}\\
Galaxy Environment for MaNGA (GEMA) & \S \ref{vac:gema} & \citet{2015AA...578A.110A} \\
Spectroscopic Redshifts for DR17 & \S \ref{vac:mangaz} & \citet{2018MNRAS.477..195T},  M. Talbot et al. (in prep.)\\
Strong Gravitational Lens Candidate Catalog & \S \ref{vac:gravlens} & M. Talbot et al. (in prep.) \\
\tableline
\multicolumn{3}{l}{MaStar} \\ 
Photometry Crossmatch & \S \ref{vac:mastarphot} & R. Yan et al. (in prep.)\\
Stellar Parameters & \S \ref{vac:mastar} & R. Yan et al. (in prep.)\\
\tableline
\multicolumn{3}{l}{eBOSS} \\ 
 ELG (DR16+) & \S \ref{vac:lss} &  \citet{raichoor17a,raichoor19a} \\
 LRG  (DR16+) & \S \ref{vac:lss} & \citet{prakash16a,ross20a} \\
 QSO  (DR16+) & \S \ref{vac:lss} & \citet{myers15a,ross20a} \\
 DR16 Large-scale structure multi-tracer EZmock catalogs & \S \ref{vac:mock} &  \citet{EZmock_eBOSS2021} \\
 DR16Q catalog  (DR16+) & \S \ref{vac:qso} & \citet{lyke20a} \\
 Ly$\alpha$ catalog  (DR16+) & \S \ref{vac:lya} & \citet{2019duMasdesBourbouxH}   \\
 Strong Gravitational Lens Catalog  (DR16+) & \S \ref{vac:lenses} & \citet{2021MNRAS.502.4617T} \\
 ELG-LAE Strong Lens Catalog & \S \ref{vac:ebossELGLAE} & \citet{Shu16} \\
Cosmic Web Environmental Densities from MCPM & \S \ref{vac:ebossMCPM} & \citet{Burchett:2020_slime} \\
\enddata
\end{deluxetable*}

%% file: dataaccess.tex
\label{sec:access}

There are various ways to access the SDSS DR17 data products, and an overview of all these methods is available on the SDSS website \url{https://www.sdss.org/dr17/data_access/}, and in Table \ref{table:dataaccess}. In general, the best way to access a data product will depend on the particular data product and what the data product will be used for. We give an overview of all different access methods below, and also refer to tutorials and examples on data access available on this website: \url{https://www.sdss.org/dr17/tutorials/}. 

\begin{deluxetable*}{ll}
\tablecaption{Summary of Methods for Accessing SDSS Data \label{table:dataaccess}}
\tablehead{\colhead{Name} & \colhead{Brief Description}}
\startdata
SAS & Science Archive Server - direct access to reduced images and spectra, and downloadable catalog files \\
SAW & Science Archive Webservers - for visualisation of images and 1D spectra\\
CAS & Catalog Archive Server - for optimized access to searchable catalog data from a database management system \\
SkyServer & web app providing visual browsing and synchronous query access to the CAS\\
Explore & a visual browsing tool in SkyServer to examine individual objects\\
Quicklook & a more succinct version of the Explore tool in SkyServer \\
CasJobs & batch (asynchronous) query access to the CAS\\
SciServer & science platform for server-side analysis. Includes browser-based and Jupyter notebook access to SkyServer, CasJobs and \texttt{Marvin} \\
\texttt{Marvin} & a webapp and python package to access MaNGA data\\
SpecDash & a SciServer tool to visualize 1D spectra with standalone and Jupyter notebook access\\
Voyages & an immersive introduction to data and access tools for K-12 education purposes\\
\enddata
\end{deluxetable*}

For those users interested in the reduced images and spectra of the SDSS, we recommend that they access these data products through the SDSS Science Archive Server (SAS, \url{https://data.sdss.org/sas/}). These data products were all derived through the official SDSS data reduction pipelines, which are also publicly available through SVN or GitHub (\url{https://www.sdss.org/dr17/software/}). The SAS also contains the VACs that science team members have contributed to the data releases (see Table \ref{table:vac}), as well as raw and intermediate data products. All files available through the SAS have a data model that explains their content (\url{https://data.sdss.org/datamodel/}). Data products can be downloaded from the SAS either directly through browsing, or by using methods such as wget, rsync and Globus Online (see  \url{https://www.sdss.org/dr17/data_access/bulk}, for more details and examples). For large data downloads, we recommend the use of Globus Online. Since SDSS data releases are cumulative, in that data products released in earlier data releases are included in DR17, and will have been processed by the latest available pipelines, we reiterate that users should always use the latest data release, as pipelines have often been updated to improve their output and fix previously known bugs.

The Science Archive Webservers (SAW) provides visualisations of most of the reduced images and data products available on the SAS. The SAW offers the option to display spectra with their model fits, and to search spectra based on a variety of parameters (e.g. observing program, redshift, coordinates). These searches can be saved as permalinks, so that they can be consulted again in the future and be shared with collaborators. All SAW webapps are available from \url{https://dr17.sdss.org/}, and allow for displaying and searching of images (SDSS-I/II), optical single-fiber spectra (SDSS-I/II, SEGUE, BOSS and eBOSS), infrared spectra (APOGEE-1 and APOGEE-2), and MaStar stellar library spectra. Images and spectra can be downloaded through the SAW, and previous data releases are available back to DR8. The SAW also offers direct links to SkyServer Explore pages (see below).

The MaNGA integral-field data is not incorporated in the SAW due to its more complex data structure, and can instead be accessed through \texttt{Marvin} (\url{https://dr17.sdss.org/marvin/}; \citealt{cherinka19}). \texttt{Marvin} offers not only visualisation options through its web interface, but also allows the user to query the data and analyze data products remotely through a suite of Python tools. \texttt{Marvin} also offers access to various MaNGA value added catalogs, as described in \S \ref{sec:mangavacs}. \texttt{Marvin}'s Python tools are available through pip-install, and installation instructions as well as tutorials and examples are available here: \url{https://sdss-marvin.readthedocs.io/en/stable/}. No installation is required to use \texttt{Marvin}'s Python tools in SciServer, as described later in this section and in \S \ref{sec:manga.marvin}. 

Catalogs of derived data products are available on the SAS, but can be accessed more directly through the Catalog Archive Server \citep[CAS,][]{2008CSE....10...30T}. These include photometric and spectroscopic properties, as well as some value added catalogs. The SkyServer webapp (\url{https://skyserver.sdss.org}) allows for visual inspection of objects using e.g. the QuickLook and Explore tools, and is also suitable for synchronous SQL queries in the browser. Tutorials and examples explaining the SQL syntax and how to query in SkyServer are available at \url{http://skyserver.sdss.org/en/help/docs/docshome.aspx}. For DR17, the SkyServer underwent a significant upgrade, which includes a completely redesigned user interface as well as migration of the back end to a platform independent, modular architecture. Although SkyServer is optimal for smaller queries that can run in the browser, for larger ones we recommend using CASJobs (\url{https://skyserver.sdss.org/casjobs}). CASJobs allows for asynchronous queries in batch mode, and offers the user free storage space for query results in a personal database (MyDB) for server-side analysis that minimizes data movement \citep{2008CSE....10...18L}. 

SkyServer and CASJobs are now part of the SciServer science platform \citep[][\url{https://www.sciserver.org}]{TAGHIZADEHPOPP2020100412}, which is accessible with free registration on a single-sign-on portal, and offers server-side analysis with Jupyter notebooks in both interactive and batch mode, via SciServer Compute. SciServer is fully integrated with the CAS, and users will be able to access the data and store their notebooks in their personal account (shared with CASJobs). SciServer offers data and resource sharing via its Groups functionality that greatly facilitates its use in the classroom, to organize classes with student, teacher and teaching assistant privileges. Several SciServer Jupyter notebooks with use cases of SDSS data are available through the SDSS education webpages (\url{https://www.sdss.org/education/}), some of which have been used by SDSS members in college-level based courses as an introduction to working with astronomical data. SciServer has prominently featured in the ``SDSS in the Classroom" workshops at AAS meetings.

Users can now analyze the MaNGA DR17 data in SciServer, using the \texttt{Marvin} suite of Python tools.  SciServer integration enables users to use the access and analysis capabilities of \texttt{Marvin} without having a local installation.  In the SciServer Compute system\footnote{\url{https://www.sciserver.org/about/compute/}}, the MaNGA dataset is available as an attachable MaNGA Data Volume, with the \texttt{Marvin} toolkit available as a loadable \texttt{Marvin} Compute Image.   Once loaded, the \texttt{Marvin} package along with a set of \texttt{Marvin} Jupyter example notebooks and tutorials are available on the compute platform.

With DR17, we are also releasing in SciServer a new feature called SpecDash  \citep{manuchehr_taghizadeh_popp_2021_5083750} to interactively analyze and visualize \revision{one-dimensional optical spectra from SDSS Legacy and eBOSS surveys, and soon from APOGEE as well}. SpecDash is available both as stand-alone website\footnote{\url{https://specdash.idies.jhu.edu/}}, and as a Jupyter notebook widget in SciServer.

 Users can load and compare multiple spectra at the same time, smooth them with several kernels, overlay error bars, spectral masks and lines, and show individual exposure frames, sky background and model spectra.
For analysis and modeling, spectral regions can be interactively selected for fitting the continuum or spectral lines with several predefined models.
All spectra and models shown in SpecDash can be downloaded, shared, and then uploaded again for subsequent analysis and reproducibility. Although the web-based version shares the same functionality as the Jupyter widget version, the latter has the advantage that users can use the SpecDash python library to programmatically load any kind of 1-D spectra, and analyze or model them using their own models and kernels.

All tools and data access points described above are designed to serve a wide range of users from undergraduate level to expert users with significant programming experience. In addition, Voyages (\url{https://voyages.sdss.org/}) provides an introduction to astronomical concepts and the SDSS data for less experienced users, and can also be used by teachers in a classroom setting. The Voyages activities were specifically developed around pointers to K-12 US science standards, and a Spanish language version of the site is available at \url{https://voyages.sdss.org/es/}.

%% file: apogee.tex
The central goal of APOGEE is to map the chemodynamics of all structural components of the Milky Way Galaxy via near-twin, multiplexed NIR high-resolution spectrographs operating simultaneously in both hemispheres \citep[APOGEE-N and APOGEE-S spectrographs respectively; both described in][]{Wilson2019}. 
DR17 constitutes the sixth release of data from APOGEE, which has run in two phases (APOGEE-1 and APOGEE-2) spanning both SDSS-III and SDSS-IV.  
As part of SDSS-III, the APOGEE-1 survey operated for approximately 3 years from August 2011 to July 2014 using the 2.5-m Sloan Foundation Telescope at APO.
At the start of of SDSS-IV, APOGEE-2 continued its operations in the Northern Hemisphere by initiating a $\sim$6-year survey (APOGEE-2N). 
Thanks to unanticipated on-sky efficiency, APOGEE-2N operations concluded in November 2020 with an effective $\sim$7.5 years of bright time observations, with many programs expanded from their original 6-year baseline. 
In April 2017, operations began with the newly built APOGEE-S spectrograph and associated fiber plugplate infrastructure on the 2.5-m
Ir\'en\'ee du Pont Telescope at LCO; APOGEE-2S observations concluded in January 2021. 
A full overview of the APOGEE-1 scientific portfolio and operations was given in \citet{Majewski2017} and a parallel overview for APOGEE-2 is forthcoming (S.~Majewski et al., in prep.). 

The APOGEE data in DR17 encompass all SDSS-III APOGEE-1 and SDSS-IV APOGEE-2 observations acquired with both instruments from the start of operations at APO in SDSS-III (September 2011) through the conclusion of SDSS-IV operations at APO and LCO (in November 2020 and January 2021, respectively). 
Compared to the previous APOGEE data release (DR16), DR17 contains roughly two additional years of observations in both hemispheres; this doubles the number of targets observed from APOGEE-2S (see Table \ref{tab:scope}).

DR17 contains APOGEE data and information for 657,135 unique targets, with 372,458 of these (57\%) as part of the main red star sample that uses a simple selection function based on de-reddened colors and magnitudes \citep[for more details see][]{Zasowski_2013_apogeetargeting,Zasowski_2017_apogee2targeting}.
The primary data products are: 
    (1) reduced visit and visit-combined spectra, 
    (2) radial velocity measurements, 
    (3) atmospheric parameters (eight in total), and 
    (4) individual element abundances (up to 20 species). 
Approximately 2.6 million individual visit spectra are included in DR17; 399,505 sources have three or more visits (54\%) and 35,009 sources (5\%) have ten or more visits.  

\begin{figure*}
\centering
\includegraphics[angle=0,width=15cm]{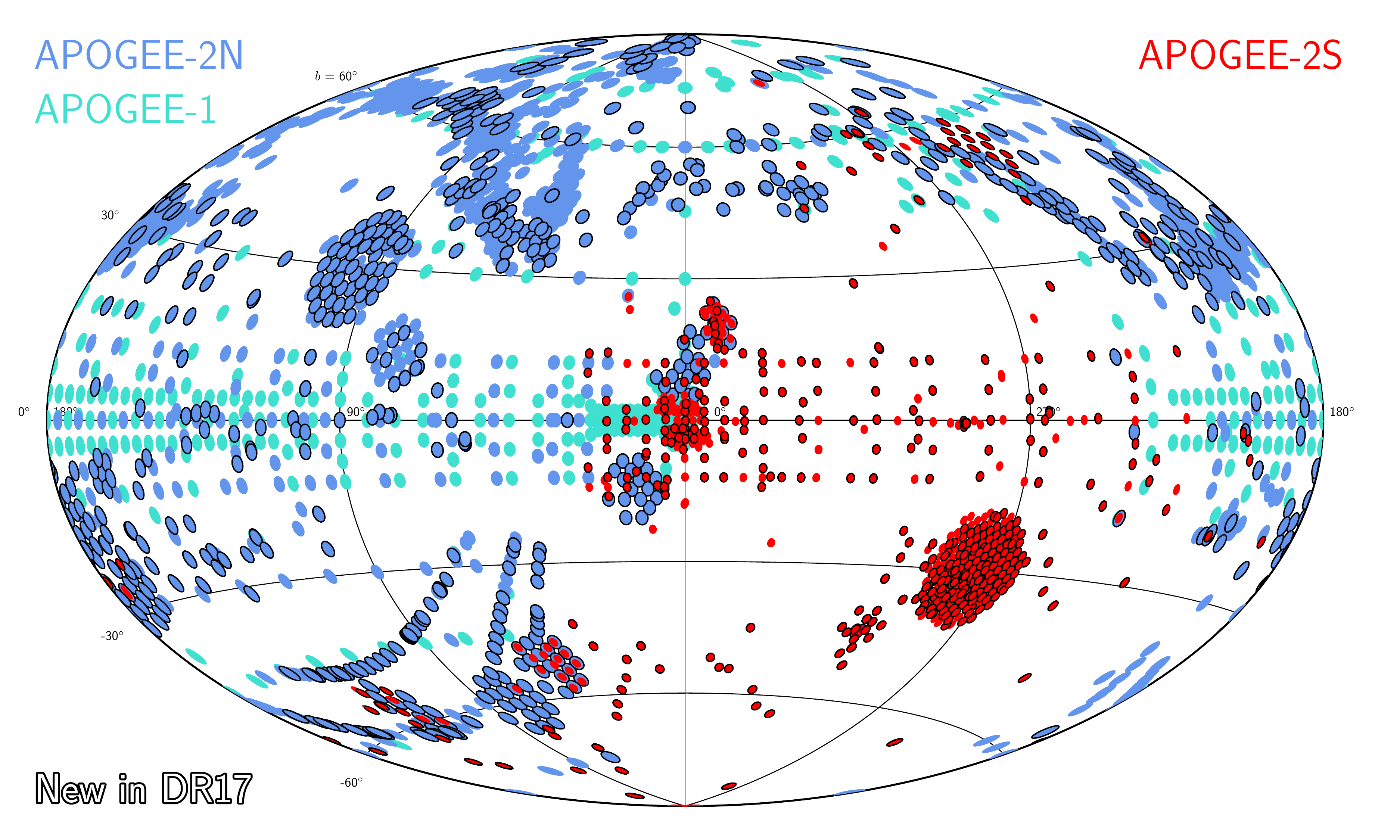}
\caption{
    The DR17 final APOGEE sky coverage shown in Galactic coordinates with fields color-coded by the survey phase in which the field was observed: APOGEE-1 (cyan), APOGEE-2N (blue), and APOGEE-2S (red). 
    The fiber plugplates used with the APOGEE-N spectrograph have a 7 square degree field-of-view while those used with the APOGEE-S spectrograph have a 2.8 square degree field of view. 
    Those fields with any new observations in DR17 are highlighted with a black outline. 
}
\label{fig:apogeedr17new}
\end{figure*}

The final APOGEE survey map is shown in Figure \ref{fig:apogeedr17new}, where each circle represents a single ``field'' that is color-coded by survey phase: APOGEE-1 (cyan), APOGEE-2N (blue), or APOGEE-2S (red). 
The difference in field-of-view between APOGEE-N and APOGEE-S is visible by the size of the symbol, with each APOGEE-S field spanning 2.8~deg$^{2}$ and APOGEE-N spanning 7~deg$^{2}$ \citep[for the instrument descriptions, see][]{Wilson2019}.
Those fields with any new data in DR17 are encircled in black; new data can either be fields observed for the first time or fields receiving additional epochs. 
The irregular high Galactic latitude coverage is largely due to piggyback ``co-observing'' with MaNGA during dark time.
Notably, these cooperative operations resulted in observations of an additional 162,817 targets, or 22\% of the total DR17 targets ($\sim$30\% of targets in APOGEE-2), which is a comparable number of targets as were observed in all of APOGEE-1. 


\begin{figure*}
\centering
\includegraphics[angle=0,width=15cm]{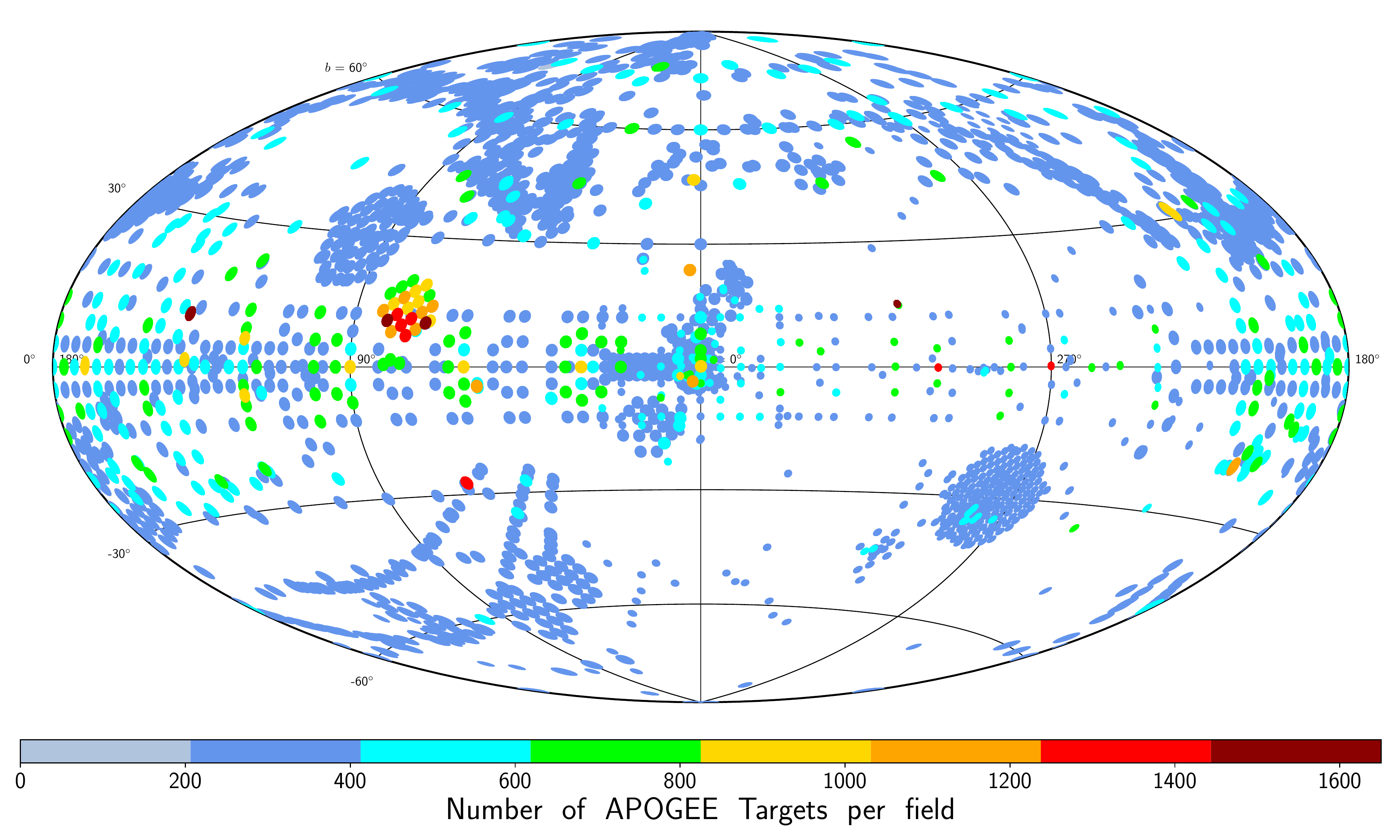}
\caption{
    A sky map in Galactic coordinates showing the number of stars per APOGEE field.
    The disk is targeted with a more or less systematic grid of pointings within $|b| < 15\deg$.
    For $\ell < 30 \deg$ there is more dense coverage of the bulge and inner Galaxy. 
    The circle sizes reflect the different field-of-view of APOGEE-N and APOGEE-S.
    The dense coverage at the North Galactic Cap is due to co-observing with the MaNGA survey, which contributed 22\% of the targets in DR17. 
    }
\label{fig:apogeenstars}
\end{figure*}

A different visualization of the final field plan is given in \autoref{fig:apogeenstars}, where now each field is color-coded by the number of unique stars targeted in each field. 
APOGEE plates have 300 fibers, but APOGEE targeting uses a ``cohorting'' strategy by which exposure is accumulated over many visits for the faintest targets in a field while brighter targets are swapped in and out over time \citep[for a schematic see][Figure 1 therein]{Zasowski_2013_apogeetargeting}. 
Moreover, some fields were included in multiple programs, like those in the \emph{Kepler} footprint, and as many as 1600 unique targets were accommodated in a single 7~deg$^{2}$ APOGEE-2N field over the full span of the APOGEE-1 and APOGEE-2 observing programs.

Extensive descriptions of the target selection and strategy are found in \citet{Zasowski_2013_apogeetargeting} for APOGEE-1 and in \citet{Zasowski_2017_apogee2targeting} for APOGEE-2. 
Details about the final target selection schemes used for APOGEE-2N and APOGEE-2S, which evolved over time, are presented in \citet{Beaton2021} and \citet{Santana2021}, respectively.

\subsection{DR17 Sample Highlights}

DR17 represents the culmination of the APOGEE-2 program (and, indeed, all of APOGEE) and presents a number of large, focused subsamples that are worth noting briefly.
DR17 contains over 18,000 targets in the TESS Northern Continuous Viewing Zone (CVZ) and over 35,000 targets in the TESS Southern CVZ \citep{Ricker2016}.
In DR17, there are over 35,000 targets which are part of 13 of the \emph{Kepler} K2 Campaigns and over 20,000 in the primary \emph{Kepler} field. 
In total, over 100,000 targets are also found in high-cadence, space-based photometry programs. 
Among all scientific targeting programs, there are more than 13,000 targets that have more than 18 individual epochs, spanning all parts of the Galaxy.

DR17 includes extensive APOGEE coverage for numerous star clusters, including 29 open clusters, 35 globular clusters, and 18 young clusters. 
However, detailed membership characterization identifies at least one possible member in as many as 126 open clusters and 48 globular clusters, after accounting for targets in Contributed and Ancillary Science programs (N. Myers et al., in prep, R. Schiavon et al., in prep.).
Thus, some observations exist in DR17 for approximately 200 star clusters spanning a range of ages and physical properties. 

In addition, DR17 contains measurements of resolved stars from ten dwarf satellite galaxies of the Milky Way (including the dwarf spheroidal systems Bo\"otes I, Sextans, Carina, Fornax, Sculptor, Sagittarius, Draco, and Ursa Minor, as well as the Large and Small Magellanic Clouds); 14,000 of the over 20,000 targets toward dwarf satellites are in the Magellanic System.
In addition, DR17 contains integrated light observations of star clusters in Fornax, M31, and M33 and of the central regions of M31 and of its highest--surface brightness dwarf satellites.

\subsection{APOGEE DR17 Data Products}
The basic procedure for processing and analysis of APOGEE data is similar to that from previous data releases  \citep{2018ApJS..235...42A,Holtzman2018,Jonsson_2020AJ_apogeedr16}, but a few notable differences are highlighted here. 
More details are presented in J.~Holtzman et al. (in prep.).

\subsubsection{Spectral Reduction and Radial Velocity Determinations}

\citet{2015AJ....150..173N} describes the original reduction 
procedure for APOGEE data, and the various APOGEE Data Release
papers present updates \citep[][J.~Holtzman et al. in prep.]{2018ApJS..235...42A,Holtzman2018,Jonsson_2020AJ_apogeedr16}. 
For DR17, at the visit reduction level, a small change was made to the criteria by which pixels are 
flagged as  being potentially affected by poor sky subtraction.

The routines for combination of the individual visit spectra 
were rewritten for DR17 to incorporate a new radial velocity
analysis, called Doppler \citep{Nidever_doppler}. 
Doppler performs a least squares fit to a set of visit spectra, solving simultaneously for basic stellar parameters (\teff, \logg, and [M/H]) and the radial velocity for each visit. 
The fitting is accomplished by using a series of
Cannon \citep{Ness_2015,Casey_2016} models to generate spectra for arbitrary choices of stellar parameters across the Hertzsprung-Russell diagram (from 3500 K to 20,000 K in \teff); the Cannon models were trained on a grid of spectra produced using Synspec \citep[e.g.,][]{Hubeny_2017,Hubeny_2021} with 
Kurucz model atmospheres \citep[][]{Kurucz_1979,Castelli_2003,Munari_2005}. 
\revision{
The primary output of Doppler are the radial velocities; while the stellar parameters from Doppler are stored, 
they are not adopted as the final values (see ASPCAP, \S \ref{details} below).
The Doppler routine produces slightly better results for radial velocities in most cases, as judged by scatter across repeated visits of stars. 
Details will be given in J.~Holtzman et al. (in prep), but, for example, for $\sim$ 85,000 stars that have more than 3 visits, 
\texttt{VSCATTER}$<$ 1 km/s, \texttt{TEFF}$<$ 6000 K, and no additional data since DR17, the median \texttt{VSCATTER} is reduced from 128 m/s to 96 m/s.
}

In addition to the new methodology, the radial velocities for
faint stars were improved. This was accomplished by making an
initial combination of the visit spectra using only the 
barycentric correction. This initial combination provided a 
combined spectrum from which a radial velocity was 
determined. The radial velocity for each individual visit was 
then determined separately, but was required to be within
50 km/s of the original estimate. This yielded a higher fraction
of successful radial velocities for faint stars, as judged by
looking at targets in nearby dwarf spheroidal galaxies.

\subsubsection{Atmospheric Parameter and Element Abundance Derivations}\label{details}

Stellar parameters and abundances are determined using the
APOGEE Stellar Parameters and Chemical Abundance Pipeline (ASPCAP,
\citealt{garciaperez2016}) 
\revision{that relies on the FERRE optimization code \citep{2006ApJ...636..804A}}.\footnote{\url{https://github.com/sdss/apogee}}

The basic methodology of ASPCAP remained the same for DR17 as in previous releases, but 
new synthetic spectral grids were created. 
These took advantage of new, non-local thermodynamic equilibrium (NLTE) population calculations by \citet{Osorio2020} for four elements: Na, Mg, K, and Ca; 
\revision{as discussed in \citet{Osorio2020} the H-band abundance differences between LTE and NLTE were always less than 0.1 dex}. 
Adopting these calculations, however, required the adoption of a different spectral
synthesis code from that used in the last several APOGEE data 
releases: for DR17, the Synspec code \citep[e.g.,][]{Hubeny_2017,Hubeny_2021} was adopted for the 
primary analysis instead of the Turbospectrum code \citep{Alvarez_1998,Plez2012} used in previous releases. 
This was not a straightforward choice because, while 
Synspec allows the NLTE levels to be used, it calculates the 
synthetic spectra under the assumption of plane parallel 
geometry, which becomes less valid for the largest giant stars. 
On the other hand, Turbospectrum can use spherical geometry, but does not accommodate NLTE populations to be specified. 

DR17 uses multiple sub-grids to span from \teff=3000~K (M dwarf) to \teff=20,000~K (BA), with \logg~ ranges from 0 to 5 (3 to 5 for the BA grid). 
The full details of these grids and the reliability of the parameters as a function of stellar type are provided in J.~Holtzman et al. (in prep.). 
Modifications to the linelists used for the syntheses are described in \citet{Smith_2021_apogeelinelist}, 
\revision{which is an augmentation to prior linelist work for APOGEE \citep{2015ApJS..221...24S,2016ApJ...833...81H,2017ApJ...844..145C}.}

\revision{
The ASPCAP results from the new Synspec grid are the primary APOGEE DR17 results and the majority of users will 
likely be satisfied with the results in this catalog; only this primary catalog will be loaded into the CAS. 
However, unlike prior releases, DR17 also includes supplemental analyses constructed using alternate libraries that have different underlying physical assumptions.
The different analyses in DR17 are provided in separate summary files and include:
\begin{enumerate}
    \item the primary library using Synspec including NLTE calculations for Na, Mg, K, and Ca (with files on the SAS under dr17/synspec\_rev1)\footnote{This is a revised version of the dr17/synspec directories, correcting a minor problem with the LSF convolution for a subset of stars observed at LCO, however, since Value Added Catalogs were constructed with the original dr17/synspec we have retained it for completness.};
    \item one created using Synspec, but assuming LTE for all elements (files under dr17/synspec\_lte);
    \item another created using Turbospectrum 20 (files under dr17/turbo20), using spherical geometry for \logg $<=$3;  
    \item one created with Turbospectrum, but with plane parallel geometry (files under dr17/turbo20\_pp) for all stars.
\end{enumerate}
All of the libraries use the same underlying MARCS stellar atmospheres for stars with \teff$<$ 8000 K, computed with spherical geometry for \logg $<=$3. 
A full description of these spectral grids will be presented in J.~Holtzman et al. (in prep.) and a focused discussion on 
the differences between the libraries and the physical implications will be presented in Y.~Osorio et al. (in prep.).
In summary, however, the differences are subtle in most cases.
We encourage those using the APOGEE DR17 results to clearly specify the catalog version that they are using in their analyses\footnote{Users may find the library version in the name of the summary file, as well as in the \texttt{ASPCAP\_ID} tag provided for each source in these files.}. 
}

For DR17, we present 20 elemental abundances: 
 C, C I, N, O, Na, Mg, Al, Si, S, K, Ca, Ti, Ti II, V, Cr, Mn, Fe, Co, Ni, and Ce. 
In DR16, we attempted to measure the abundances of Ge, Rb, and  Yb, but given the poor results for extremely weak lines, we did not attempt these in DR17.  
While we attempted measurements of P, Cu, Nd, and $^{13}$C in DR17, these were judged to be unsuccessful. 
Overall, the spectral windows used to measure the abundances were largely unchanged, but several additional windows were added for Cerium, such that the results for Ce appear to be significantly improved over those in DR16.

As in DR16, both the raw spectroscopic stellar parameters as well as
calibrated parameters and abundances are provided. 
\revision{
Calibrated effective temperatures are determined by a comparison to photometric effective temperatures, as determined from the relations of \citep{GHB2009}, using stars with low reddening.
}
Calibrated surface gravities are provided by comparison to a set of surface gravities from asteroseismology \citep[][M. Pinsonneault et al. in prep.]{Serenelli2017} and isochrones \citep{Berger2020}. 
For DR17, the surface gravity calibration was applied using a neural network, unlike 
previous data releases where separate calibrations were derived and 
applied for different groups (red giants, red clump, and main sequence) of stars. 
\revision{
The new approach eliminates small discontinuities that were previously apparent, and is described in more detail in J.~Holtzman et al. (in prep.).
}
For the elemental abundances, calibration just consists of a zeropoint offset (separately for dwarfs and giants), 
determined by setting the median abundance [X/M] of solar metallicity
stars in the solar neighborhood with thin disk kinematics such that 
[X/M]=0.

Additional details on the ASPCAP changes are described in J.~Holtzman et al. (in prep.). 

\subsubsection{Additional data}

Several other modifications were made for DR17. 
\revision{
\begin{enumerate}
    \item The summary data files for APOGEE that are available on the Science Archive Server now include data from the \Gaia Early Data Release 3 (EDR3) for the APOGEE targets \citep{Gaia2021, GaiaMission_2016}. Positional matches were performed by the APOGEE team. More specifically, the following data are included: 
    \begin{itemize} 
        \item \Gaia EDR3 identifiers \citep{Gaia2021},
        \item \Gaia EDR3 parallaxes and proper motions \citep{Lindegren21},
        \item \Gaia EDR3 photometry \citep{Gaia_EDR3_Photometry},
        \item \Gaia EDR3 RVs \citep{Gaia_EDR3_RVs}, 
        \item Distances and uncertainties following \cite{BailerJones2021}.
    \end{itemize}
    \item Likely membership for a set of open clusters, globular clusters, and dwarf spheroidal galaxies, as determined 
from position, radial velocity, proper motion, and distance, is provided in a \texttt{MEMBERS} column. More specifically, 
initial memberships were computed based on position and literature RVs, and these are then used to determine proper motion 
and distance criteria. Literature RVs were taken from:
    \begin{itemize} 
        \item APOGEE-based mean RVs for the well-sampled ``calibration clusters'' in \citet{Holtzman2018},
        \item mean RVs for globular clusters from \citet{Harris_2010}\footnote{This is the 2010 update to the \citet{Harris_1996} catalog.}, and
        \item mean RVs for dwarf spheroidal galaxies from \citet{McConnachie_2012}.
    \end{itemize}
 Users interested in the properties of the clusters or satellite galaxies are encouraged to do more detailed membership characterization and probabilities \citep[e.g.,][Schiavon et al., in prep., Shetrone et al., in prep.]{2019A&A...622A.191M,2020MNRAS.492.1641M,Hasselquist2021}  
    \item Some spectroscopic binary identification is provided through bits in the \texttt{STARFLAG} and \texttt{ASPCAPFLAG} bitmasks. A more comprehensive analysis of spectroscopic binaries is provided in a VAC (see \S \ref{vac:apogeeobjects} below) .
\end{enumerate}
We encourage those utilizing these data in our summary catalogs to cite the original references as given above.
}

\subsection{Data Quality}  \label{sec:apogee_dq}

The \revision{overall} quality of the DR17 results for radial velocities, stellar parameters, and chemical abundances is similar to that of previous APOGEE data releases \revision{(full evaluation will be provided in Holtzman et al. in prep.).\footnote{The web documentation contains the details of the data model. Morevoer, the documentation communicates how data was flagged, including a brief list of changes relative to prior releases.}} 
\revision{As in DR16,} uncertainties for stellar parameters and abundances are estimated by analyzing the scatter in repeat observations of a set of targets.

Users should be aware that deriving consistent abundances across a wide range of parameter space is challenging, so some systematic features and trends arise. Users should be careful when comparing abundances of stars with significantly different stellar parameters.
Also, the quality of the abundance measurements varies between different elements, across parameter space, and with signal-to-noise. 

Some regions of parameter space present larger challenges than others. 
In particular, it is challenging to model the spectra of the coolest stars and, while abundances are derived for the coolest stars in DR17, there seem to be significant systematic issues for the dwarfs with \teff $<$ 3500 K such that although we provide calibrated results in the PARAM array, we do not populate the ``named tags.'' 
Separately, for warm/hot stars (\teff$>$7000), information on many abundances is lacking in the spectra, and uncertainties in the model grids at these temperatures may lead to systematic issues with the DR17 stellar parameters.

\begin{figure*}
\centering
\includegraphics[width=\textwidth]{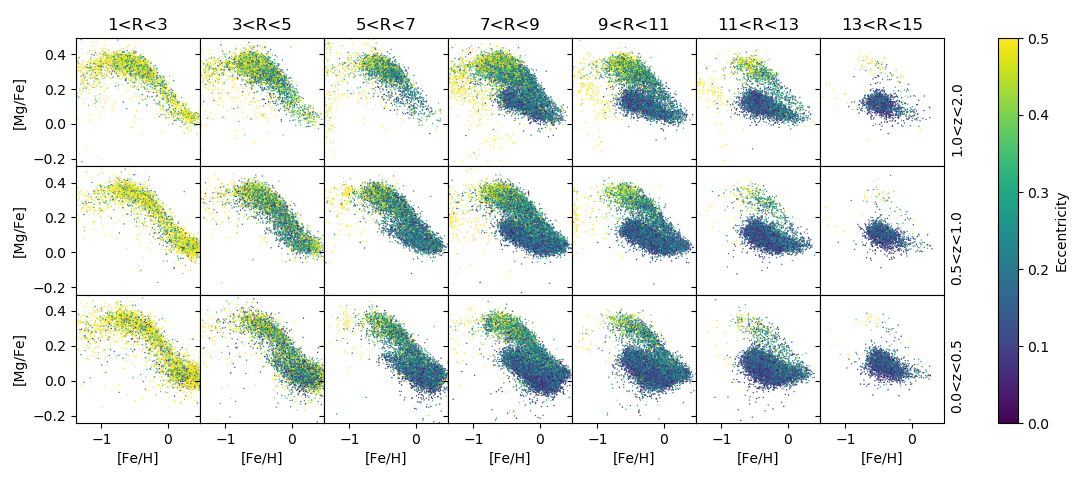}
    \caption{A series of [Mg/Fe] vs [Fe/H] plots from APOGEE DR17 for different zones in the Milky Way. 
    Distances from the DistMass VAC are used to determine Galactocentric radius ($R_G$) and height above the plane ($z$). 
    Points are color-coded by orbital eccentricities as computed with \texttt{GalPy} \citep{2015ApJS..216...29B}
    using \Gaia EDR3 proper motions and APOGEE radial velocities.
}
\label{fig:alphafe}
\end{figure*}

As a demonstration of the quality and scientific potential of the data, Figure \ref{fig:alphafe} shows a set of [Mg/Fe] versus [Fe/H] diagrams for different three-dimensional spatial zones within the disk of the Milky Way, restricted to giant-stars with 1 $<$ \logg $<$ 2.5 to minimize \revision{potential} systematics \revision{or sampling bias}. 
\revision{
Spectrophotometric distances to individual stars are determined from Value Added Catalogs\footnote{In this visualization, from the DistMass VAC to be released in 2022 that uses a Neural Net at the parameter level to determine spectroscopic distances.} and then are used with stellar positions to determine the Galactocentric radius ($R_G$) and height above the plane ($z$) for each individual star; this highlights the scientific potential enabled via the analyses in the Value Added Catalogs.
The color coding indicates the orbital eccentricity based on calculations from \texttt{GalPy} \citep{2015ApJS..216...29B} using \Gaia EDR3 proper motions \citep{Gaia2021} and APOGEE DR17 radial velocities.
Figure \ref{fig:alphafe} is a merging of similar visualizations previously presented in \citet{Hayden_2015} and \citet{Mackereth_2019}, such that the spatial zones of the former are merged with the dynamical inference of the latter.
The stars of the solar neighborhood (middle panel, $7<R_G<9$) show two distinct chemical sequences, commonly referred to the the low- and high- [$\alpha$/Fe] sequences that are also somewhat dynamically distinct (apparent in the color-coding by orbital eccentricity). 
The inner Galaxy, however, is dominated both by high-eccentricity (bulge-like orbits) stars on the high-[$\alpha$/Fe] sequence just as the outer galaxy is dominated by low-eccentricity (near circular orbits) stars on the low-[$\alpha$/Fe] sequence, with some slight dependence on $z$. 
The relative contributions of low-eccentricity versus high-eccentricity and low-[$\alpha$/Fe] versus high-[$\alpha$/Fe] sequences shift throughout the Galaxy.
These spatial, chemical, and dynamical groupings provide evidence for various disk-formation and disk-evolution scenarios \citep[e.g., as discussed in][among others]{Hayden_2015,Mackereth_2019} that add complexity and nuance to the canonical schemes.
} .
 
\subsection{APOGEE Value Added Catalogs}\label{vac:apogee}

There are a large number of APOGEE-associated VACs in DR17. In what follows we provide brief descriptions of each VAC along with references where the reader can find more detail. Broadly speaking, APOGEE VACs can be split into characterising special subsamples, like binary stars, open clusters, and photometric variables, those which calculate stellar or orbital parameters for all (or most) APOGEE target stars (e.g. Starhorse, APOGEEnet and others). We also document the release of a mock catalog of APOGEE based on a hydrodynamical simulation. 

\subsubsection{VACs Describing Categories of Objects in APOGEE}\label{vac:apogeeobjects}
The first set of APOGEE VACs describe special categories of objects in APOGEE data and in most cases provide additional information/characteristics for these objects. They are: 

\begin{enumerate}
\item{\it Open Cluster Chemical Abundances and Mapping catalog (OCCAM)}: The goal of OCCAM is to leverage the APOGEE survey to create a large, {\it uniform} catalog of open cluster chemical abundances and use these clusters to study Galactic chemical evolution. The catalog contains average chemical abundances for each cluster and membership probability estimates for APOGEE stars in the cluster area. We combine proper motion (PM) and radial velocity (RV) measurements from \Gaia EDR3 \citep{Gaia2021} with RV and metallicity measurements from APOGEE to establish cluster membership probabilities for each star observed by APOGEE.  The VAC includes 26,699 stars in the areas of 153 cataloged disk clusters. Detailed descriptions of the OCCAM survey, including targeting and the methodology for membership determinations, are presented in \citet{2013ApJ...777L...1F}, \citet{Donor2018}, and  \citet{2020AJ....159..199D}. This third catalog from the OCCAM survey includes 44 new open clusters, including many in the Southern hemisphere and those targeted specifically in GC size ($R_{GC}$) ranges with little coverage in the DR16 catalog (specific targeting described in \citealt{Beaton2021,Santana2021}). Average RV, PM, and abundances for reliable ASPCAP elements are provided for each cluster, along with the visual quality determination. Membership probabilities based individually upon PM, RV, and [Fe/H] are provided for each star, stars are considered 3$\sigma$ members if they have probability $> 0.01$ in all three membership dimensions \footnote{However, some stars near the main sequence turn-off may ``fail" the [Fe/H] cut due to evolutionary diffusion effects \citep{2018ApJ...857...14S,2019ApJ...874...97S}}. The results and caveats from this VAC will be discussed thoroughly in N. Myers et al. (in prep.). 
  
\item{\it APOGEE Red-Clump (RC) Catalog}: DR17 contains an updated version of the APOGEE red-clump (APOGEE-RC) catalog. This catalog is created in the same way as the previous DR14 and DR16 versions of the catalog, with a more stringent $\log g$ cut compared to the original version of the catalog \citep{2014ApJ...790..127B}. The catalog contains 50,837 unique stars, about 30\% more than in DR16. 
\revision{
The catalog is created using a spectrophotometric technique first presented in \citet{2014ApJ...790..127B} that results in a rather pure sample of red-clump stars (e.g., minimal contamination from red-giant-branch, secondary-red-clump, and asymptotic-giant-branch stars that have similar CMD and H-R positions). \citeauthor{2014ApJ...790..127B} estimated a purity of $\sim$95\%. The narrowness of the RC locus in color-metallicity-luminosity space allows distances to the stars to be assigned with an accuracy of 5\%-10\%, which exceeds the precision of spectrophotometric distances in other parts of the H-R diagram. We recommend users adopt the most recent catalog (DR17) for their analyses; additional discussion on how to use the catalog is given in \citet{2014ApJ...790..127B}.  While the overall datamodel is similar to previous versions of the catalog, the proper motions are from \Gaia EDR3 \citep{Gaia2021,Gaia_EDR3_Astrometry}.
}
  
\item {\it APOGEE-Joker}: The APOGEE-Joker VAC contains posterior samples for binary-star orbital parameters (Keplerian orbital elements) for 358,350 sources with three or more APOGEE visit spectra that pass a set of quality cuts as described in A. Price-Whelan et al. (in prep.). The posterior samples are generated using \textit{The Joker}, a custom Monte Carlo sampler designed to handle the multi-modal likelihood functions that arise when inferring orbital parameters with sparsely-sampled or noisy radial velocity time data \citep{Price-Whelan:2017}. This VAC deprecates the previous iterations of the catalog \citep{Price-Whelan:2018, Price-Whelan:2020}.

For 2,819 stars, the orbital parameters are well constrained, and the returned samples are effectively unimodal in period. For these cases, we use the sample(s) returned from \textit{The Joker} to initialize standard MCMC sampling of the Keplerian parameters using the time-series optimized MCMC code known as  \textit{exoplanet}\footnote{\url{https://docs.exoplanet.codes/en/latest/}} \citep{exoplanet} and provide these MCMC samples. For all stars, we provide a catalog containing metadata about the samplings, such as the maximum \textit{a posteriori} (MAP) parameter values and sample statistics for the MAP sample. A. Price-Whelan et al. (in prep.) describes the data analysis procedure in more detail, and defines and analyzes a catalog of $\gtrsim$40,000 binary star systems selected using the raw orbital parameter samples released in this VAC.

\item {\it Double lined spectroscopic binaries in APOGEE spectra}: Generally, APOGEE fibers capture a spectrum of single stars. Sometimes, however, there may be multiple stars of comparable brightness with the sky separations closer than the fiber radius whose individual spectra are captured by the same recorded spectrum. Most often, these stars are double-lined spectroscopic binaries or higher order multiples (SBs), but on an occasion they may also be chance line-of-sight alignments of random field stars (most often observed towards the Galactic center). Through analyzing the cross-correlation function (CCF) of the APOGEE spectra, \citet{Kounkel2021} have developed a routine to automatically identify these SBs using Gaussian deconvolution of the CCFs \citep{Kounkel2021_code}\footnote{\url{https://github.com/mkounkel/apogeesb2}}, and to measure RVs of the individual stars. The catalog of these sources and the sub-component RVs are presented here as a VAC. For the subset of sources that had a sufficient number of measurements to fully characterize the motion of both stars, the orbit is also constructed.

The data obtained though April/May 2020 were processed with the DR16 version of the APOGEE radial velocity pipeline and this processing was made available internally to the collaboration as an intermediate data release. All of the SBs identified in this internal data release have undergone rigorous visual vetting to ensure that every component that can be detected is included and that spurious detections have been removed. However, the final DR17 radial velocity pipeline is distinct from that used for DR16 (summarized above; J. Holtzman et al. in prep.) and the reductions are sufficiently different that they introduce minor discrepancies within the catalog. In comparison to DR16, the DR17 pipeline limits the span of the CCF for some stars to a velocity range around the mean radial velocity to ensure a more stable overall set of RV measurements; on the other hand the DR16 pipeline itself may fail on a larger number of individual visit spectra and thus not produce a full set of outputs. For the sources that have both good parameters and a complete CCF coverage for both DR16 and DR17, the widely resolved components of SBs are generally consistent with one another; close companions that have only small RV separations are not always identified in both datasets. For this reason, SBs that could be identified in both the DR16 and DR17 reductions are kept as separate entries in the catalog. Visual vetting was limited only to the data processed with the DR16 pipeline (e.g., data through April/May 2020); the full automatic deconvolutions of the DR17 CCFs are presented as-is.

\end{enumerate}

\subsubsection{VACs of Distances and other parameters} \label{vac:apogeedist} 
VACs providing distances and other properties (mostly related to orbital parameters) are released (or re-released): 

\begin{enumerate}
\item {\it StarHorse distances, extinctions,  and stellar parameters for APOGEE DR17 + \Gaia EDR3:} We combine high-resolution spectroscopic data from APOGEE DR17 with broad-band photometric data from 2MASS, unWISE and PanSTARRS-1, as well as parallaxes from \Gaia EDR3. Using the Bayesian isochrone-fitting code StarHorse \citep{Santiago2016, Queiroz2018}, we derive distances, extinctions, and astrophysical parameters. We achieve typical distance uncertainties of $\sim$ 5 \% and extinction uncertainties in V-band amount to $\sim$ 0.05 mag for stars with available PanSTARRS-1 photometry, and $\sim$ 0.17 mag for stars with only infra-red photometry. The estimated StarHorse parameters are robust to changes in the Galactic priors assumed and corrections for \Gaia parallax zero-point offset. This work represents an update of DR16-based results presented in \citet{Queiroz2020}.

\item{\it APOGEE-\texttt{astroNN}:} The APOGEE-\texttt{astroNN} value-added catalog holds the results from applying the \texttt{astroNN} deep-learning code to APOGEE spectra to determine stellar parameters, individual stellar abundances \citep{2019MNRAS.483.3255L}, distances \citep{2019arXiv190208634L}, and ages \citep{2019arXiv190104502M}. For DR17, we have re-trained all neural networks using the latest data, i.e., APOGEE DR17 results for the abundances, \Gaia EDR3 parallax measurements, and an intermediate APOKASC data set with stellar ages (v6.6.1, March 2020 using DR16 ASPCAP). Additionally,  we augmented the APOKASC age data with low-metallicity asteroseismic ages from \citet{2021NatAs.tmp...90M} to improve the accuracy of ages at low metallicities; the \citet{2021NatAs.tmp...90M} analysis is similar to that of APOKASC, but performed by an independent team. As in DR16, we correct for systematic differences between spectra taken at LCO and APO by applying the median difference between stars observed at both observatories. In addition to abundances, distances, and ages, properties of the orbits in the Milky Way (and their uncertainties) for all stars are computed using the fast method of \citet{2018PASP..130k4501M} assuming the \texttt{MWPotential2014} gravitational potential from \citet{2015ApJS..216...29B}. Typical uncertainties in the parameters are 35 K in \teff, 0.1 dex in \logg, 0.05 dex in elemental abundances, 5\,\% in distance, and 30\,\% in age. Orbital properties such as the eccentricity, maximum height above the mid-plane, radial, and vertical action are typically precise to 4 to 8\,\%.  

\end{enumerate} 

\subsubsection{APOGEE Net: a unified spectral model} \label{vac:apogeeprocess}
A number of different pipelines are available for extracting spectral parameters from the APOGEE spectra. These pipelines generally manage to achieve optimal performance for red giants and, increasingly, G \& K dwarfs, which compose the bulk of the stars in the catalog. However, the APOGEE2 catalog contains a number of parameter spaces that are often not well characterized by the primary pipelines. Such parameter spaces include pre-main sequence stars and low mass stars, with their measured parameters showing systematic $T_{\rm eff}$ \& $\log g$ deviations making them inconsistent from the isochrones and the main sequence. OBA stars are also less well constrained and in prior data releases many were classified as F dwarfs (due to grid-edge effects) and have their $T_{\rm eff}$ underestimated in the formal results. By using data-driven techniques, we attempt to fill in those gaps to construct a unified model of APOGEE spectra. In the past, we have developed a neural network, APOGEE Net \citep{olney2020}, which was shown to perform well to extract $T_{\rm eff}$, $\log g$, \& [Fe/H]  on all stars with $T_{\rm eff}<$6,500 K, including pre-main sequence stars. We now expand these efforts to also characterize hotter stars with 6,500$<T_{\rm eff}<$50,000 K. APOGEE NET II is described in \citet{sprague2022}.

\subsubsection{APOGEE FIRE VAC} \label{vac:apogeefire}

Mock catalogs made by making simulated observations of sophisticated galaxy simulations provide unique opportunities for observational projects, in particular, the ability to test for or constrain the impact of selection functions, field plans, and algorithms on scientific inferences.
One of the most realistic galaxy simulations to date is the Latte simulation suite, which uses FIRE-2 \citep{Hopkins_2018} to produce galaxies in Milky Way-mass halos in a cosmological framework \citep{Wetzel_2016}. 
\citet{Sanderson_2020} translated three of the simulations into realistic mock catalogs (using three solar locations, resulting in nine catalogs), known as the Ananke simulations\footnote{For data access see: \url{ https://fire.northwestern.edu/ananke/\#dm}}. 
Ananke contains key \Gaia measureables for the star particles in the simulations and these include radial velocity, proper motion,  parallax, and photometry in the \Gaia bands as well as chemistry (10 chemical elements are tracked in the simulation), and other stellar properties. 
Because the input physics and the global structure of the model galaxy are known, these mock catalogs provide an experimental laboratory to make connections between the resolved stellar populations and global galaxy studies. 

In this VAC, Ananke is expanded to permit APOGEE-style sampling of the mock-catalogs. 
For all observed quantities both the intrinsic, e.g., error-free, and the observed values are reported; the observed values are the intrinsic values convolved with an error-model derived from observational data for similar object types. 
As described in \citet{Nikakhtar_2021}, Ananke mock-catalogs now contain: 
    (i) 2MASS ($JHK_s$) photometry and reddening, 
    (ii) abundance uncertainties following APOGEE DR16 performance \citep[following][]{Poovelil_2020,Jonsson_2020AJ_apogeedr16}, and 
    (iii) a column that applies a basic survey map \citep{Zasowski_2013_apogeetargeting,Zasowski_2017_apogee2targeting,Beaton2021,Santana2021}. 
The full mock-catalogs are released such that users can impose their own selection function to constructs a mock APOGEE survey in the simulation. 
Mock-surveys can then be used to test the performance of methods and algorithms to recover the true underlying galactic physics as demonstrated in \citet{Nikakhtar_2021}.

%% file: manga.tex
The MaNGA survey \citep{2015ApJ...798....7B} uses a custom-built set of hexagonal integral field unit (IFU) fiber bundles \citep{Drory2015} to feed spectroscopic fibers into the BOSS spectrograph \citep{Smee2013}. Over its operational lifetime, MaNGA has successfully met its goal of obtaining integral field spectroscopy for $\sim$ 10,000 nearby galaxies \citep{Law2015, Yan2016survey} at redshift $z \sim 0.03$ with a nearly flat distribution in stellar mass \citep{Wake2017}.

DR17 contains all MaNGA observations taken throughout SDSS-IV, and more than doubles the sample size of fully reduced galaxy data products previously released in DR15 \citep{2019ApJS..240...23A}.
These data products include raw data, intermediate reductions such as flux-calibrated spectra from individual exposures, and final calibrated data cubes and row-stacked spectra (RSS) produced using the MaNGA Data Reduction Pipeline \citep[DRP;][]{law16,law21,yan16calibration}.

DR17 includes DRP data products (see \S \ref{sec:manga.drp}) for 11,273 MaNGA cubes distributed amongst 674 plates.  10,296 of these data cubes are for ``traditional" MaNGA type galaxies, and 977 represent data cubes associated with non-standard ancillary programs (targeting a variety of objects including globular clusters, faint galaxies and intracluster light in the Coma cluster, background reference sky, and also tiling of the large nearby galaxies M31 and IC342; see \S \ref{sec:manga_ancillary} for more details).  Of the 10,296 galaxy cubes, 10,145 have the highest data quality with no warning flags indicating significant issues with the data reduction process.  These 10,145 data cubes correspond to 10,010 unique targets (as identified via their {\tt MANGAID}) with a small number of repeat observations taken for cross-calibration purposes (each has an individual plate-ifu code, {\tt MANGAID} needs to be used to identify unique galaxies). 
As in previous releases, DR17 also includes the release of derived spectroscopic products (e.g., stellar kinematics, emission-line diagnostic maps, etc.) from the MaNGA Data Analysis Pipeline \citep[DAP;][]{belfiore19,westfall19}; see \S \ref{sec:manga.dap}. Additionally, DR17 contains the final data release for the MaNGA Stellar Library \citep[MaStar;][and \S \ref{sec:mastar}]{Yan2019}, which includes calibrated 1D spectra for 28,124 unique stars spanning a wide range of stellar types. 

\begin{figure*}
\epsscale{1.0}
\plotone{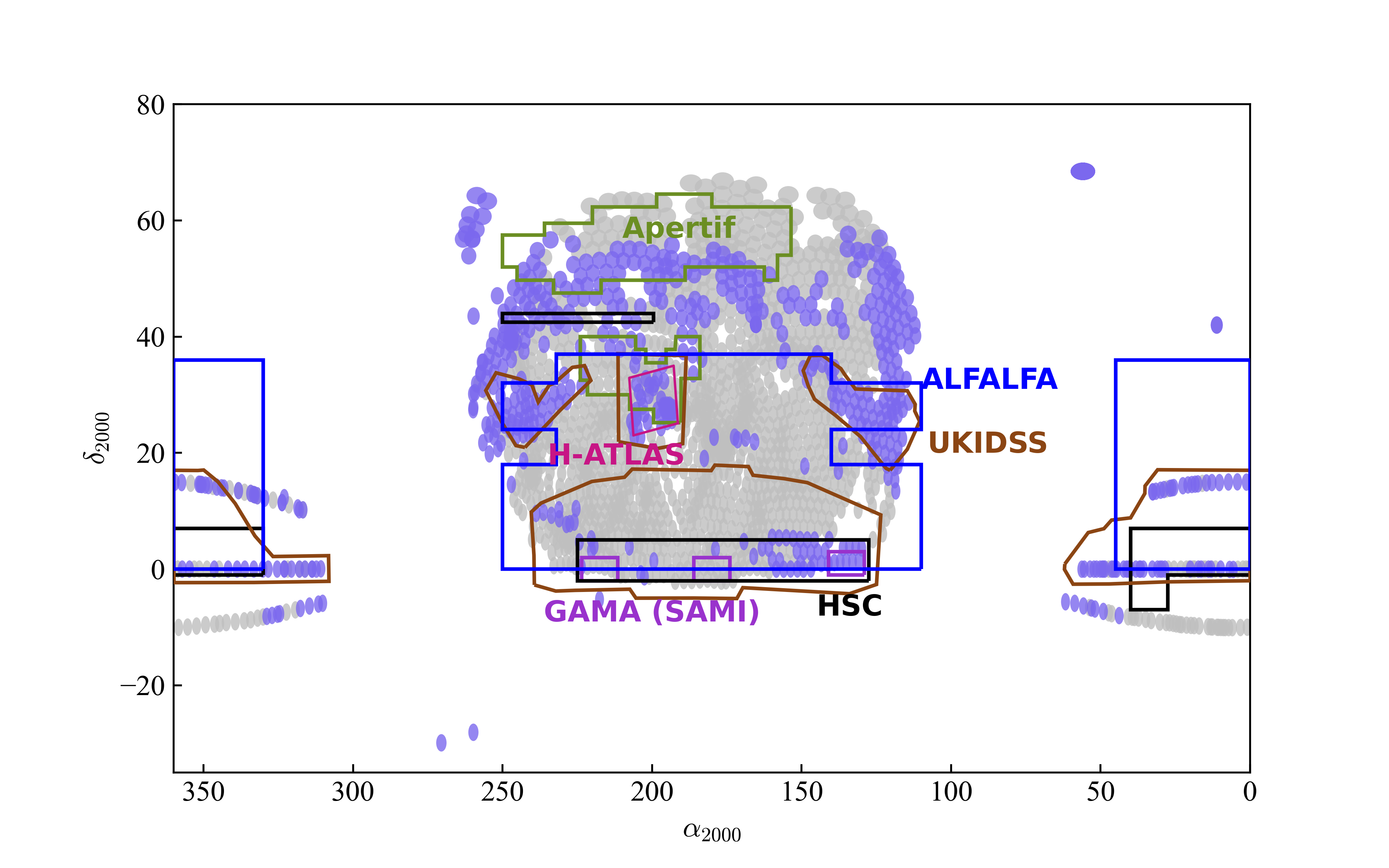}
\caption{DR17 final MaNGA survey area; blue tiles indicate observed fields (plates), grey tiles indicate potential fields from which the MaNGA final sample was drawn.  Colored boxes indicate the regions observed by a variety of other surveys as described in the text.}
\label{fig:manga_sample}
\end{figure*}

We illustrate the sky footprint of MaNGA galaxies released in DR17 in Figure \ref{fig:manga_sample}, along with colored boxes indicating the locations of a selection of other galaxy surveys, namely the HI surveys Apertif (K. Hess et al. in prep) and \revision{ALFALFA (or Arecibo Legacy Fast ALFA}, \citealt{haynes2018}; also see \S \ref{vac:himanga} for more HI followup); IR surveys like Herschel-ATLAS, (H-ATLAS, \citealt{hatlas}), the UKIRT Infrared Deep Sky Survey, (UKIDSS, \citealt{Lawrence2007}), and other optical surveys, like Galaxy and Mass Assembly Survey (GAMA, \citealt{GAMA}), the footprint of which includes most of the SAMI IFU observations, \citep[][in total, 74 galaxies are observed by both MaNGA and SAMI]{Croom2021} and Hyper Suprime-Cam (HSC, \citealt{Aihara2019PASJ...71..114A}). In some cases the prioritization of which MaNGA plates to observe was driven by the availability of these ancillary data (e.g. note how observed plates fill in parts of the UKIDSS footprint). MaNGA plates in an earlier projected footprint of Apertif were also prioritized but changes in Apertif observation plans has significantly reduced the final overlap. 

\subsection{MaNGA Data Reduction Pipeline and Products}
\label{sec:manga.drp}


The MaNGA DRP has evolved substantially throughout the survey across a variety of both
public (DR) and internal (``MaNGA Product Launch", or MPL) data releases.  A summary of these various DRP versions and
the number of unique galaxies in each is given by \citet[][see their Table 1]{law21}.
These authors also provide a detailed description of the differences in the DRP for DR17
compared to previous releases.\footnote{Strictly \citet{law21} describe the team-internal
data release MPL-10, but these data are practically identical to the final 
public data release 
DR17 (which is the team internal release MPL-11) in everything except the total number of galaxies.}  In brief, changes in the DR17 data products compared
to DR15 include:

\begin{enumerate}
    \item Updated spectral line-spread function (LSF):  Many stages of the pipeline have been rewritten to further improve the accuracy of the LSF estimate, which is now good to better than 1\%.  As demonstrated by \citet{law21} by comparison against observations with higher-resolution spectrographs, this allows MaNGA emission-line velocity dispersions to be reliable down to 20 \kms\ at signal-to-noise ratio (SNR) above 50, which is well below the 70 \kms\ instrumental resolution.
    \item Multiple pipeline changes have affected the overall MaNGA survey flux calibration.  The most significant changes included adoption of a different extinction model for the calibration standard stars and correction for a few-percent scale error in lab measurements of the MaNGA fiber bundle metrology using on-sky self calibrations \citep[see][their Appendix A]{law21}.
    \item New data quality flags have been defined to better identify potential reduction problems.  These include a new {\tt UNUSUAL} data quality bit to identify cubes that are different from ordinary data quality but still useful for many analyzes (e.g., that may be missing a fraction of the field of view due to hardware problems).  These are distinct from the previously-defined {\tt CRITICAL} data quality bit that indicates data with significant problems that should preclude it from most scientific analyzes ($<1$\% of the total sample).
    \item Introduction of a new processing step to detect and subtract bright electronic artifacts (dubbed the ``blowtorch") arising from a persistent electronic artifact within the Charge-coupled devices (CCDs) in one of the red cameras during the final year of survey operations \citep[see][their Appendix B]{law21}.
\end{enumerate}


\subsection{MaNGA Data Analysis Pipeline and Products}
\label{sec:manga.dap}

In this section we describe two specific changes to the DAP analysis
between MaNGA data released in DR15 and DR17. The first is a change in the stellar continuum
templates used for the emission line measurements; this change only
affects emission line measurements and does not affect stellar
kinematic measurements. The second is the addition of new spectral
index measurements more appropriate for stacking analyzes and
coaddition of spaxels; the previously existing spectral index
measurements are not affected by this addition.

The MaNGA Data Analysis Pipeline (DAP) as a whole is discussed
extensively in the DR15 paper \citep{2019ApJS..240...23A} and
in \citet{westfall19}, \citet{belfiore19}, and \citet{law21}. The last
provides a summary of other improvements made to the DAP since DR15.

The SDSS data release
website (\url{https://www.sdss.org/}) provides
information on data access and changes to the DAP data models in DR17
for its major output products. Further information can be found in the
documentation of the code
base.\footnote{\url{https://sdss-mangadap.readthedocs.io/en/latest/}}

\subsubsection{Stellar Continuum Templates}

In DR17, we use different spectral templates to model the galaxy
continuum for emission line measurements than we use for stellar
kinematics measurements. In DR15, we used the same templates in both
cases, but as discussed by \citet{law21}, these template sets
diverged starting with our ninth internal data set (MPL-9; between
DR15 and DR17). For the emission line measurements, the new templates
are based on the MaStar survey, allowing us to take advantage of the
full MaNGA spectral range \revision{(3600-10000 \AA)} and, e.g., model the
[\ion{S}{3}]$\lambda\lambda$9071,9533\AA ~doublet and some of the blue
Paschen lines.  For the stellar kinematics measurements, we have
continued to use the same templates used in DR15,
the \texttt{MILES-HC} library, taking advantage of its modestly higher
spectral resolution than MaStar. Since MILES only spans between 3575 to 7400 \AA, this means MaNGA stellar kinematics do not include, e.g., contributions from the calcium near-infrared triplet near 8600 \AA.

In DR17, we provide DAP emission line measurements based on two
different continuum template sets, both based on the MaStar
Survey \citep[][and \S \ref{sec:mastar}]{Yan2019}, and referred to
as \texttt{MASTARSSP} and \texttt{MASTARHC2}. There are four different
analysis approaches, indicated by \texttt{DAPTYPE}. Three
use \texttt{MASTARSSP}, with three different spatial binning
approaches, and the fourth uses \texttt{MASTARHC2}.

The template set referred to as the \texttt{MASTARSSP} library by the
DAP are a subset of simple-stellar-population (SSP) models provided
by \citet{Maraston2020}.  Largely to decrease execution time, we
down-selected templates from the larger library provided
by \citet{Maraston2020} to only those spectra with a Salpeter Initial Mass Function (IMF) and the following grid in
SSP age and metallicity, for a total of 54 spectra:
\begin{enumerate}
    \item Age/[1 Gyr] = 0.003, 0.01, 0.03, 0.1, 0.3, 1, 3, 9, 14
    \item $\log(Z/Z_\odot)$ = -1.35, -1., -0.7, -0.33, 0, 0.35.
\end{enumerate}
\revision{Extensive testing was done to check differences in stellar-continuum fits based on this choice; small differences that were found are well within the limits described by \citet{belfiore19}. Section 5.3 of \citet{lawbpt} show further analysis, including a direct comparison of results for the BPT emission-line diagnostics plots when using either the \texttt{MASTARHC2} or \texttt{MASTARSSP} templates showing that the templates have a limited effect on their analysis.} Importantly, note that the DAP places no constraints on how these
templates can be combined (e.g., unlike methods which use the \revision{Penalized PiXel-Fitting, or} pPXF; \revision{\citealt{2004PASP..116..138C,
2017MNRAS.466..798C},} implementation of regularized weights), and the weight applied to each
template is not used to construct luminosity-weighted ages or
metallicities for the fitted spectra.  The use of the SSP models, as
opposed to spectra of single stars, is meant only to impose a
physically relevant prior on the best-fitting continua, even if
minimally so compared to more sophisticated stellar-population
modeling.

The template set referred to as the \texttt{MASTARHC2}\footnote{\texttt{MASTARHC2} was the second of two
library versions \revision{based on hierarchical clustering (HC) of MaStar spectra}.  \texttt{MASTARHC1} is also available from the DAP code
repository, but it was only used in the processing for MPL-9.} library
by the DAP is a set of 65 hierarchically clustered templates based on
$\sim$2800 MaStar spectra from MPL-10.  Only one of the
four \texttt{DAPTYPE}s provided in DR17 uses these templates; however,
we note that the results based on these templates are the primary data
sets used by \citet{lawbpt, law21} to improve the DRP (see above).
The approach used to build the \texttt{MASTARHC2} library is inspired
by, but different in many details, from the hierarchical clustering
method used to build the \texttt{MILESHC} library \citep[cf.,][Section
5]{westfall19}, as described below.

The principles of the hierarchical clustering approach used
by \citet{westfall19} to construct the \texttt{MILESHC} library are
maintained, except we perform the clustering for
the \texttt{MASTARHC2} library in two steps.  The first step clusters
spectra based on their low-order continuum differences, leading to a
set of ``base clusters.''  We use pPXF \citep{2004PASP..116..138C,
2017MNRAS.466..798C} to perform a least-squares fit of each spectrum
using every other spectrum; however, we do not include Gaussian kernel
terms or polynomial continuum optimization, meaning the least-squares
fit simply optimizes the scaling between the two spectra.  We use the
$rms$ difference between the best-fit spectra as the clustering
``distance,'' and the distance matrix is used to construct eight base
clusters. \revision{The choice of eight clusters was based on an a qualitative assessment of the appropriate number which separated MaStar spectra into distinct types.} The second step uses pPXF to fit each spectrum using every
other spectrum \textit{within its base cluster}.  In this step, we
modestly degrade the resolution of the template being fit with $\sigma
= 1$ pixel, and then our pPXF fit includes a freely fit Gaussian
kernel with bounds of a $\pm 1$ pixel shift and a $0.1-2$ pixel
broadening.  \revision{This was done in the same way across all parts of the spectra.} We also include a multiplicative Legendre polynomial of
order 100 to optimize the continuum match between the two templates.
The very high-order fit \revision{(the choice of the exact number of 100 was arbitrary)} acts like a high-pass filter on the
differences between the two spectra, ensuring that the optimized $rms$
difference between the two spectra is driven by the high-order (line)
structure differences.  The spectra within each base cluster are
organized into ``template clusters'' and visually inspected.  The
visual inspection leads to iterations on the number of template
clusters in each base cluster, as well as removing some of the spectra
from the analysis.  The number of template clusters per base cluster
ranged from 6 to 16, depending on a by-eye assessment of the spectra
in each template cluster.  The final assignment of each MaStar
spectrum (identified by its {\tt MANGAID}) to a template cluster is
provided in the DAP code
repository.\footnote{\url{https://github.com/sdss/mangadap/blob/master/mangadap/data/spectral\_templates/mastarhc\_v2/README}}
Note that 34 of 99 clusters were not included in
the \texttt{MASTARHC2} library because they were either composed of
single stars, resulted in noisy spectral stacks, contained isolated
specific data-reduction artifacts, or contained a set of spectra that
were considered too disparate for a single cluster.  For the vetted
set of 65 template clusters, the median number of spectra per cluster
is 14, but the range is from 2 to more than 300.

\begin{figure}
\epsscale{1.15}
\plotone{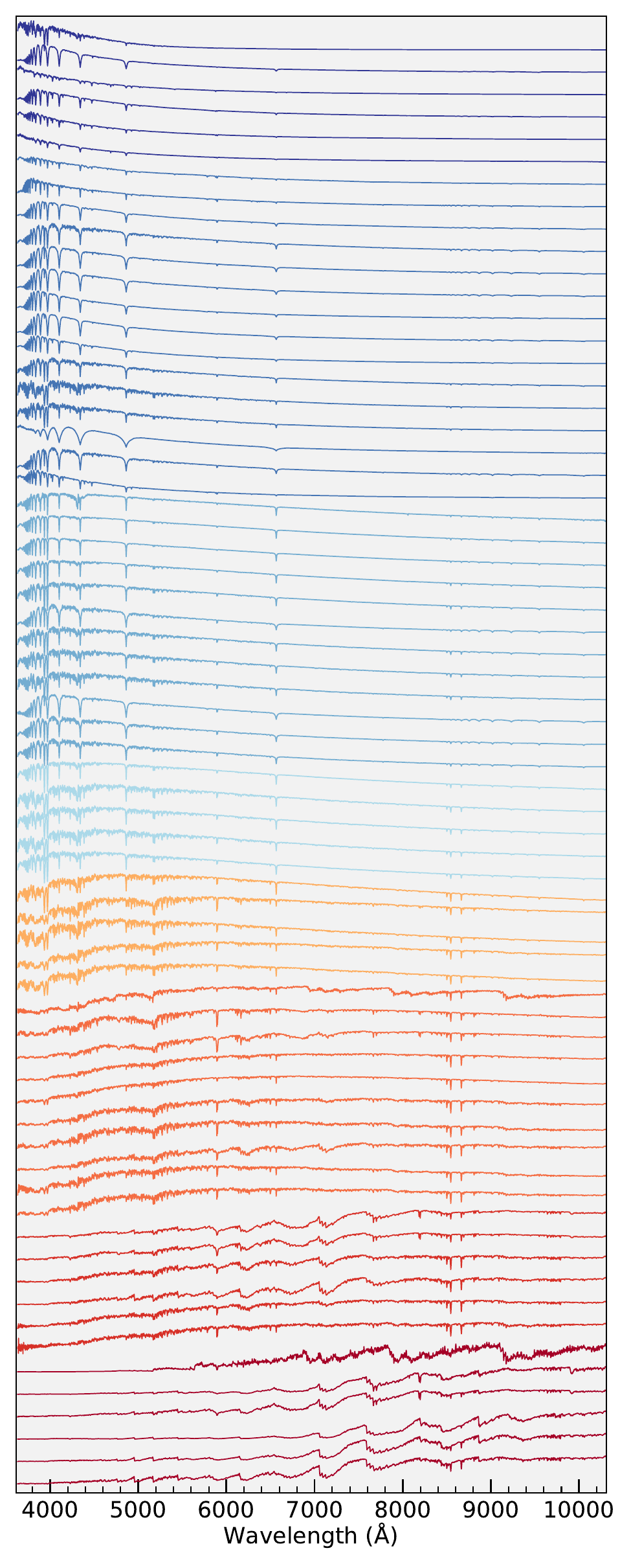}
\caption{Spectra in the \texttt{MASTARHC2} template library.  Spectra are arranged and colored according to the membership in one of eight base clusters (the first clustering step used in the process to generate the template library from individual MaStar spectra).}
\label{fig:mastarhc2}
\end{figure}

With the assignments in hand, we combine spectra in each template
cluster as follows.  We first scale each spectrum by their median flux
and create an initial stack, weighting each spectrum by its median
SNR.  We then calculate the ratio of each spectrum to the stacked
spectrum and fit this with an order-14 Legendre polynomial, which
provides a low-order correction function to the continuum shape of
each spectrum.  \revision{The specific choice of order 14 was driven by a desire to match the choice made in the DAP fitting of galaxy spectra, which was justified in \citet{westfall19}} We constrain the correction function to be no more
than a factor of 2, which is particularly important to the stacks of
late-type stars with very little flux toward the blue end of MaStar's
spectral range.  The low-order correction function is then applied to
each spectrum in the template cluster before the final S/N weighted
stack.  The error vector for each stack is the quadrature sum of the
propagated error from the stacking operations and the, typically much
more significant, standard deviation measured for the spectra in the
stack.  The final spectra in the \texttt{MASTARHC2} library are shown
in Figure \ref{fig:mastarhc2}.

\subsubsection{Spectral Index Measurements}

In DR17, we have added spectral index measurements that are more
suited to stacking analyzes and coaddition of spectra among spaxels, such as
those based on definitions of \citet{burstein84} and \citet{faber85}.
These measurements are particularly useful for low SNR spaxels.

The motivation for this change emerges from the fact that in DAP's
hybrid
binning scheme, the spectral index measurements are performed on
individual spaxels, which can have very low S/N 
\citep[cf.][Section 9]{westfall19}.  \citet{westfall19} recommend
improving the precision of the spectral index measurements using
specific aggregation calculations that closely match the results
obtained by performing the measurements on stacked spectra over the
same spatial regions (specifically, see their Section 10.3.3).
However, the comparison between an aggregated index and an index
measured using a stacked spectrum is not mathematically identical for
the index definitions used by \citet{westfall19}.  Motivated by the
analysis of \citet{swiftVAC}, the DAP calculates the spectral indices
(specifically the absorption line indices) using two definitions for
DR17: (1) those definitions provided by \citet{worthey94}
and \citet{trager98} and (2) earlier definitions provided
by \citet{burstein84} and \citet{faber85}.  The advantage of the
definitions provided by \citet{burstein84} and \citet{faber85} is that
they allow for a mathematically rigorous aggregation of spectral
indices, as we derive below.

Following the derivation by \citet{westfall19}, we define a utility function, which is a sum of pixel values, multiplied by pixel width, 
\begin{equation}
S(y) \equiv \int_{\lambda_1}^{\lambda_2} y\ d\lambda \approx \sum_i y_i\ {\rm d}p_i\ {\rm d}\lambda_i,
\end{equation}
where $y$ is usually, but not always a function describing the flux in the spectrum, $f(\lambda)$, ${\rm d}p_i$ is the fraction of spectral pixel $i$ (with width ${\rm
d}\lambda_i$) in the passband defined by $\lambda_1 < \lambda
< \lambda_2$.  Note that masked pixels in the passband are excluded by
setting ${\rm d}p_i = 0$, and $S(1) = \Delta\lambda \equiv \lambda_2
- \lambda_1$ if no pixels are masked.  We can then define a linear
continuum between two sidebands, referred to as the blue and red
sidebands, as
\begin{equation}
C(\lambda) = (\langle f\rangle_{\rm red} - \langle f\rangle_{\rm blue})\ \frac{\lambda - \lambda_{\rm blue}}{\lambda_{\rm red}-\lambda_{\rm blue}} + \langle f\rangle_{\rm blue},
\end{equation}
where $f$ is the spectrum flux density, $\lambda_{\rm blue}$ and
$\lambda_{\rm red}$ are the wavelengths at the center of the two
sidebands, and $\langle f\rangle = S(f)/S(1)$.

The absorption-line index definitions used by \citet{worthey94}
and \citet{trager98} are:
\begin{equation}
{\mathcal I}_{\rm WT} = \left\{
\begin{array}{ll}
S(1 - f/C), & \mbox{for \AA\ units} \\[3pt]
-2.5\log\left[\langle f/C \rangle\right], & \mbox{for magnitude units}
\end{array}\right.,
\label{eq:wtabsindex}
\end{equation}
where the measurements are made on a rest-wavelength
spectrum.\footnote{Note the subtle difference between
Equation \ref{eq:wtabsindex} and Equation 22 from \citet{westfall19};
the latter has an error in the expression for the index in magnitudes
units.}  Under this definition, the integration is performed over
the \textit{ratio} of the flux to a linear continuum, which means that
the sum of, say, two index measurements is not identical to a single
index measurement made using the sum of two spectra.  In
contrast, \citet{burstein84} and \citet{faber85} define:
\begin{equation}
{\mathcal I}_{\rm BF} = \left\{
\begin{array}{ll}
S(1) - S(f)/C_0, & \mbox{for \AA\ units} \\[3pt]
-2.5\log\left[\langle f\rangle/C_0 \right], & \mbox{for magnitude units}
\end{array}\right.,
\label{eq:bfabsindex}
\end{equation}
where $C_0$ is the value of the continuum, $C(\lambda)$, at the center of the main passband.  Note that, given that $C(\lambda)$ is linear and assuming no pixels are masked, \\ $S(C) = C_0\ \Delta\lambda$. Using the definition in Equation \ref{eq:bfabsindex}, we can calculate a weighted sum of indices using the value \revision{of the continuum,} $C_0$, for each index as the weight to obtain
\begin{equation}
\frac{\sum_i C_{0,i} {\mathcal I}_{\rm BF}}{\sum_i C_{0,i}}
        = \Delta\lambda - \frac{\sum_i S(f)_i}{\sum_i C_{0,i}},
\label{eq:siangsum}        
\end{equation}
assuming no pixels are masked such that $S(1) = \Delta\lambda$. That is, the weighted sum of the individual indices is mathematically identical (to within the limits of how error affects the construction of the linear continuum) to the index measured for the sum (or mean) of the individual spectra. Similarly, for the indices in magnitude units, we find:
\begin{equation}
    -2.5\log\left[\frac{\sum_i C_{0,i} 10^{-0.4 {\mathcal I}_{\rm BF}}}{\sum_i C_{0,i}}\right]
         = -2.5\log\left[\frac{\sum_i \langle f\rangle_i}{ \sum_i C_{0,i}}\right].
\label{eq:simagsum}        
\end{equation}
Given the ease with which one can combine indices in the latter
definition, we provide both ${\mathcal I}_{\rm BF}$ \revision{(in the {\tt SPECINDEX\_BF} extension of the DAP MAPS file)} and $C_0$ \revision{(in {\tt SPECINDEX\_WGT})} for all absorption-line indices in DR17, along with the original definitions (${\mathcal I}_{\rm WT}$; {\tt SPECINDEX}) provided in DR15/DR16.


\subsection{Marvin Visualization and Analysis Tools}
\label{sec:manga.marvin}

\begin{figure*}
\epsscale{1.2}
\plotone{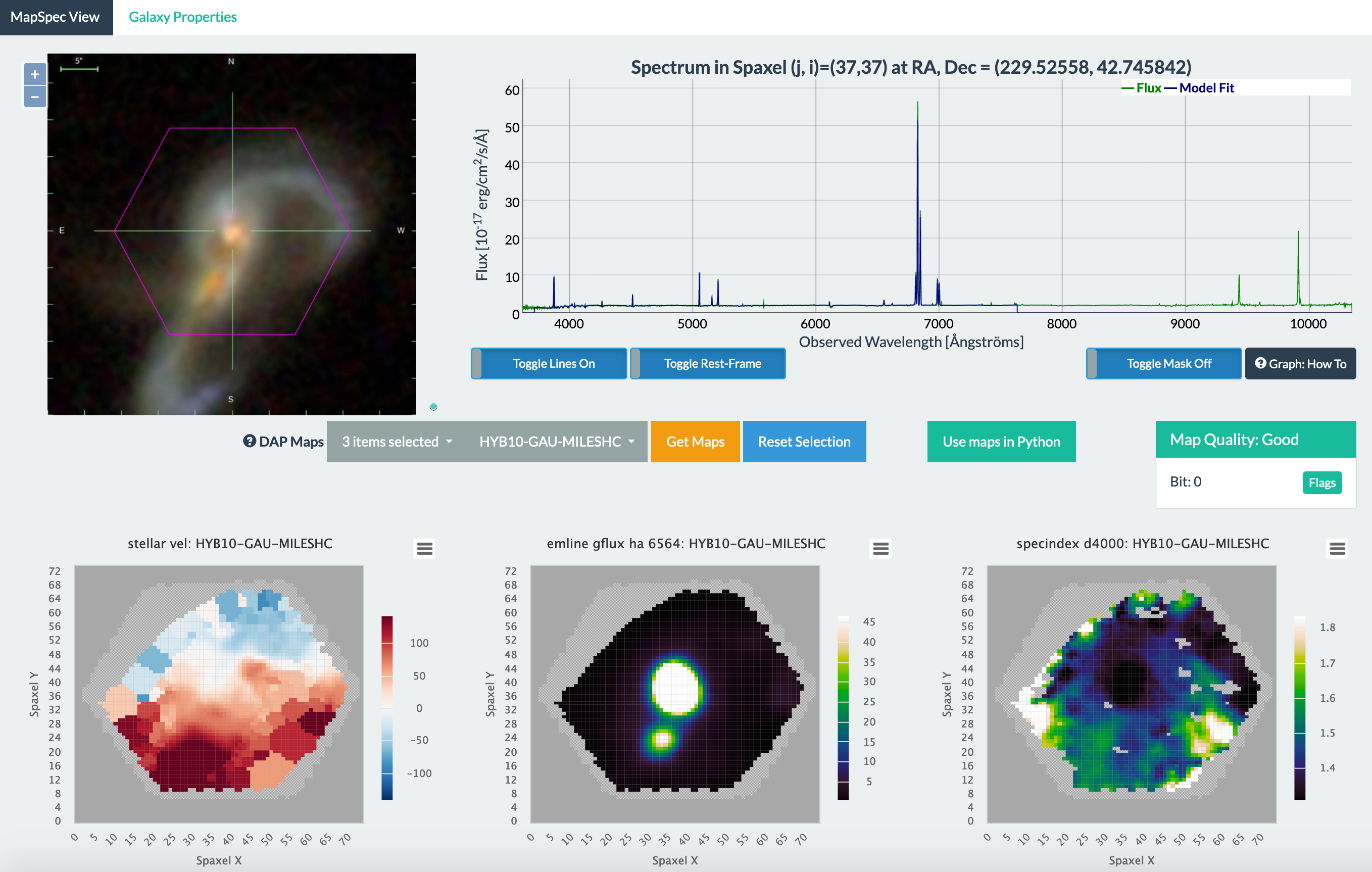}
\caption{A screenshot of the galaxy maps view of the \texttt{Marvin} Web for the MaNGA galaxy 12-193481 (Mrk 848). The SDSS three-color image of the galaxy is shown in the top left part of the figure. The upper right panel shows the spectrum of the spaxel at the position (37,37), which corresponds to the center of the bundle. The maps show: (lower left) stellar kinematics; (lower middle) H$\alpha$ emission line flux; and (lower right) D4000 spectral index for this galaxy based on its ``hybrid-binned" spectral data cube from the MaNGA DAP.}
\label{fig:marvin_web}
\end{figure*}

\texttt{Marvin} \citep{cherinka19} was developed as the tool for streamlined access to the MaNGA data, optimized for overcoming the challenges of searching, accessing, and visualizing the complexity of the MaNGA dataset.  Besides patches and internal optimizations, \revision{the DR17 updates to \texttt{Marvin} include} several enhancements such as querying targets by MaNGA quality and target bitmasks and values, full support for installation on Windows machines, as well as updates to the web interface.  The \texttt{Marvin} Web Galaxy Page (Figure \ref{fig:marvin_web}) now includes data quality indicators for the DAP Maps, as well as toggle-able features for the spectrum display.  For how to use the web or Python tools, see the \texttt{Marvin} documentation\footnote{\url{https://sdss-marvin.rtfd.io/en/latest/index.html}}.  See the \texttt{Marvin} Changelog for a complete list of what has changed since the last released version.  Contributions to \texttt{Marvin} are welcome and encouraged.  Please see the contribution guidelines\footnote{\url{https://sdss-marvin.rtfd.io/en/latest/contributing/contributing.html}} for more details. 

\texttt{Marvin} now includes access for many of the MaNGA VACs\footnote{\url{https://sdss-marvin.rtfd.io/en/latest/tools/vacs.html}}, that have been integrated into the \texttt{Marvin} ecosystem.  Each integrated VAC is accessible either as the full catalog through the new \texttt{Marvin} VACs Tool, or on a per-target basis through the existing \texttt{Marvin} Tools, e.g. Cube or Maps, via a new ``vacs" attribute attached to each \texttt{Marvin} Tool instance.  Check the VACs section of the DR17 Datamodel in the \texttt{Marvin} documentation to see which VACs are available in this release.


\subsection{Ancillary Programs}\label{sec:manga_ancillary}

As described in detail in \citet{Wake2017}, the MaNGA galaxy sample is comprised of a Primary sample covering galaxies to 1.5 $r_e$ (where $r_e$ is the effective radius; the radius containing 50\% of the light), a Secondary sample covering galaxies to 2.5 $r_e$, and a color-enhanced sample designed to fill
in underrepresented locations in the galaxy color-magnitude plane. However the number density of these main sample targets was not uniform on the sky, and in regions of lower-than-average target density, not all of the IFUs on a plate could be assigned to core target categories. In order to fill the remaining $\sim$5\% of MaNGA bundles, MaNGA held two competitive calls for ancillary targets (in July 2014 and January 2017), and a variety of ancillary programs targeting assorted kinds of galaxies or other targets were selected. 

We document here the final set of ancillary targets, with detail on how to identify them for use, or to exclude them from studies of the primary, secondary and/or color-enhanced samples. We provide in Table \ref{table:ancillary} an updated list of the number of bundles available in each documented sample, along with the binary bit mask digit stored in {\tt MANGA\_TARGET3} (sometimes {\tt MNGTARG3}) which can be used to identify the sample.\footnote{For advice on using bitmasks see \url{https://www.sdss.org/dr17/algorithms/bitmasks}}

 Ancillary programs in general were designed to increase the numbers of specific types of galaxies in the MaNGA sample. 
  We provide a short summary of all programs here (also see {\url{http://www.sdss.org/dr17/manga/manga-target-selection/ancillary-targets}): 
 \begin{enumerate}
 \item {\it Luminous AGN:} various luminous active galactic nuclear (AGN) samples were targeted, either selected from Swift BAT ({\tt AGN\_BAT}), [OIII] emission selected \citep[][{\tt AGN\_OIII}]{Mullaney2013}, Wide-field Infrared Survey Explorer \citep[WISE; ][]{WISE} colors ({\tt AGN\_WISE}), \revision{or other AGN from the Palomar survey \citep[{\tt AGN\_PALOMAR},][]{Ho1995}}. The goal of this program was to increase the range of luminosities of AGN observed by MaNGA. 
\item {\it Void Galaxies:} \revision{({\tt VOID})} this program targeted rare void galaxies located in low-density large scale environments. Targets were selected from the Void Galaxy Survey (VGS; \citealt{Kreckel2011}).
\item {\it Edge-On SF Galaxies:} a set of edge-on star forming (SF) galaxies were selected, using WISE data to estimate star-formation rates (SFRs), and S\'ersic axial ratios ($b/a$) from the NASA Sloan Atlas (NSA; \citealt{blanton2011}) to estimate inclinations. \revision{The {\tt BITNAME} is {\tt EDGE\_ON\_WINDS}.}
\item {\it Close Pairs and Mergers: } a set of close pairs and/or mergers were observed. These were either selected to be in larger bundles than typical \revision{({\tt PAIR\_ENLARGE})}, re-centered bundles (i.e. not centered on one of the pair, but somewhere in the middle, \revision{({\tt PAIR\_RECENTER})}), or in overlapping tiles, sometimes two bundles are assigned - one to each of the pair  \revision{({\tt PAIR\_2IFU})}. \revision{In addition one bundle was assigned to a merger simulated by the {\it Galaxy Zoo: Mergers} program \citep[{\tt PAIR\_SIM},][]{GZmergers}.}
\item {\it Writing MaNGA:} two bundles were assigned to an education/public outreach (EPO) program to obtain MaNGA data for galaxies in the shape of letters in the word MaNGA (selected from the Galaxy Zoo ``Alphabet"\footnote{\url{https://writing.galaxyzoo.org/}}). These are an ``M" (plate-ifu = 8721-6102) and a ``g" (9499-9102). \revision{See {\tt LETTERS}.}
\item {\it Massive Nearby Galaxies:}  very massive nearby galaxies are underrepresented in MaNGA as they are too large in angular size to fit in the bundles. This program targeted bundles at the central regions of very massive nearby galaxies \revision{({\tt MASSIVE})}.  
\item {\it Milky Way Analogs:}  two different sets of Milky Way Analog galaxies are included. These galaxies are identified using the method described in \citet{Licquia2015}: one set matched on stellar mass and SFR ({\tt MWA}), the other matched on stellar mass and bulge-to-total ratio ({\tt MW\_ANALOG}). 
\item {\it Dwarf Galaxies:} a sample of dwarf galaxies selected from the \citet{Geha2012} catalog \revision{({\tt DWARF})}.
\item {\it ETGs with Radio Jets:} a sample of early-type galaxies (ETGs) with radio jets and evidence of suppressed SF \citep[{\tt RADIO\_JETS}][]{Lin2010}.
\item {\it DiskMass Sample:} a sample of face-on disc galaxies which had previously been observed in the DiskMass survey \citep{Bershady2011} with the goal of providing a cross-calibration set \revision{({\tt DISKMASS})}.
\item {\it Brightest Cluster Galaxies:} a sample of Brightest Cluster Galaxies (BCGs; \revision{{\tt BCG}}) from the \citet{2007ApJ...671..153Y} catalog. This type of galaxy is otherwise underrepresented in MaNGA.
\item {\it Resolved Stellar Populations:} observations of very nearby galaxies in the ACS Nearby Galaxy Survey Treasury \citep[ANGST, {\tt ANGST}; ][]{Dalcanton2009} survey, as well as in M31 \revision{({\tt M31})}, to facilitate detailed studies of the resolved stellar populations. 
\item {\it Coma Plates:} a set of very deep observations of the Coma cluster \revision{({\tt DEEP\_COMA})}. Each dedicated plate used for this program observes the central massive cD galaxies (varying placement between the central regions and the galaxy outskirts), a selection of ordinary galaxies, 3 ultrafaint galaxies, and 3 regions of intracluster light (ICL). The goal was to provide very high quality spectra to enable detailed stellar population analysis.
\item {\it IC 342:}  a mosaic of 49 MaNGA plates, covering $5\arcmin \times 5\arcmin$ (5 kpc $\times$ 5 kpc) across the disk of the nearby galaxy IC342 with $\sim$30 pc spatial resolution \revision{({\tt IC342})}. This project provides test data for the Local Volume Mapper (LVM) in SDSS-V (see \S\ref{sec:future}).
\item {\it SN Hosts:} observations of the host galaxies of known Supernova, both SN Type 1a under {\tt SN1A\_HOST} and other types of SN under {\tt SN\_ENV}.
\item {\it Giant LSB galaxies:} a set of giant low surface brightness (GLSB,  \revision{{\tt GLSB}}) galaxies identified in the NSA.
\item {\it Globular clusters:} a set of \revision{dithered observations of the cores of eight} globular clusters (GCs) \revision{and 19 bulge/background fields around 3 GCs (NGC6316, NGC6522, and NGC6528)} taken to help with MaStar \revision{({\tt GLOBULAR\_CLUSTER}}).
 \end{enumerate}
 
 \begin{deluxetable*}{lcrc}
\tablecaption{Summary of MaNGA Ancillary Programs and Targeting Bits. \revision{See \S \ref{sec:manga_ancillary} for an explanation of each program and the {\tt BITNAMES}} \label{table:ancillary}}
\tablehead{\colhead{Ancillary Program} & \colhead{Number of bundles observed} & \colhead{{\tt BITNAME}} & \colhead{Binary digit}} 
\startdata
Luminous AGN & 6& {\tt AGN\_BAT} & 1 \\
 & 37 & {\tt AGN\_OIII} & 2 \\
 & 23 & {\tt AGN\_WISE} & 3 \\
 & 5 & {\tt AGN\_PALOMAR} & 4 \\
Void Galaxies & 4 & {\tt VOID} & 5\\
Edge-On SF Galaxies & 58 & {\tt EDGE\_ON\_WINDS} & 6 \\
Close Pairs and Mergers & 56 & {\tt PAIR\_ENLARGE} & 7 \\
& 38 & {\tt PAIR\_RECENTER} & 8 \\
& 1 & {\tt PAIR\_SIM} & 9 \\
& 22 & {\tt PAIR\_2IFU} & 10 \\
Writing MaNGA & 2 & {\tt LETTERS} & 11\\
Massive Nearby Galaxies  & 70 & {\tt MASSIVE} & 12 \\
Milky Way Analogs & 38 & {\tt MWA} & 13 \\
& 40 & {\tt MW\_ANALOG} & 23\\
Dwarf Galaxies & 31 & {\tt DWARF} & 14 \\
ETGs with Radio Jets & 10 & {\tt RADIO\_JETS} & 15 \\
DiskMass Sample & 7 & {\tt DISKMASS} & 16 \\
Brightest Cluster Galaxies & 55 & {\tt BCG} & 17 \\
Resolved Stellar Pops. & 3 & {\tt ANGST} & 18 \\
 & 18 & {\tt M31} & 21 \\
Coma & 85 & {\tt DEEP\_COMA} & 19 \\
IC 342 (LVM like observations) & 810 & {\tt IC342} & 20 \\
SN Hosts & 19 & {\tt SN1A\_HOST} & 26\\
& 30 & {\tt SN\_ENV} & 22\\ 
Post-starburst galaxies &24 &{\tt POST-STARBURST} &24\\
Giant LSB galaxies & 3 & {\tt GLSB} &25 \\
Globular clusters & 27\tablenotemark{1} & {\tt GLOBULAR\_CLUSTER} &  27\\
\tablenotetext{1}{8 GC targets, plus 19 bulge background fields} 
\enddata
\end{deluxetable*}


\subsection{MaNGA Related VACs}\label{sec:mangavacs}

A large number of MaNGA related VACs are presented in DR17, and will be summarized in brief below. 

\subsubsection{DR16+ VACs} \label{vac:nsa}
Two MaNGA related VACs were released in DR16+ (a mini-data release which happened in July 2020). In addition a version of the ``Visual Morphology from DECaLS Images" VAC, which is updated for DR17, was also released in DR16+. We document those first. 

\begin{enumerate}
\item {\it NASA Sloan Atlas Images and Image Analysis}:
This VAC contains the underlying image and image analysis for the NASA
Sloan Atlas (NSA). The methods used are described in
\citet{blanton2011} and \citet{Wake2017}. Briefly, for a set of nearby
galaxies of known redshift ($z< 0.15$) within the SDSS imaging area,
we have created and analyzed GALEX \citep{2007ApJS..173..682M} and SDSS images. This analysis
forms the basis for the MaNGA targeting, and resulted in the {\tt
  v1\_0\_1} NSA catalog released originally with DR14. We are now
releasing the images which were analyzed to create those parameters. The data set includes the original
catalogs from which the NSA sample was drawn, the mosaic images and
inverse variance images that were analyzed, the deblending results for
each object, the curve-of-growth and aperture corrections for each
object, and other intermediate outputs. We expect that this data set
may be useful for reanalysis of the GALEX or SDSS imaging. The full
data set is large (15 terabytes) and therefore any users interested in
using a large fraction of it should transfer the data through Globus (see \S \ref{sec:access} for details on how to use Globus\footnote{and \url{https://www.sdss.org/dr16/data_access/bulk/\#GlobusOnline}}).

\item {\it MaNGA SWIFT VAC}:\label{vac:swift}
The {\it Swift}+MaNGA (SwiM) value added catalog comprises 150 galaxies with both SDSS-IV/MaNGA IFU spectroscopy and archival {\it Swift} Ultraviolet Optical Telescope (UVOT) near-ultraviolet (NUV) images, and is presented in \citet{swiftVAC}. The similar angular resolution ($\sim3\arcsec$) between the {\it Swift}/UVOT three NUV filters and the MaNGA IFU maps allows for spatially-resolved comparisons of optical and NUV star formation indicators, which is crucial for constraining attenuation and star formation quenching in the local universe. The UVOT NUV images, SDSS optical images, and MaNGA emission line and spectral index maps have been spatially matched and re-projected so that all of the data match the pixel sampling, resolution and coordinate system of the UVOT uvw2 image for each galaxy. The spectral index maps utilize the definition given in \citet{burstein84}, which allows users to more easily compute spectral indices when binning the maps. Spatial covariance is properly accounted for in the propagated uncertainties. In addition to the spatially-matched maps, \citet{swiftVAC} also provides a catalog with PSF-matched aperture photometry for the SDSS optical and {\it Swift}/UVOT NUV bands.
\end{enumerate}

\subsubsection{Galaxy Morphology VACs} \label{vac:morph}

A variety of galaxy morphology catalogs are provided as VACs, with analysis done in a variety of ways, using a variety of images. We provide a short summary of each here - for more details please see the appropriate paper. 

\begin{enumerate}
\item {\it Galaxy Zoo: 3D}: (GZ:3D; \citealt{Masters2021}) provides crowdsourced spaxel masks locating galaxy centers, foreground stars, bars and spirals in the SDSS images of MaNGA target galaxies. Available for use within \texttt{Marvin}, these masks can be used to pick out spectra, or map quantities likely associated with the different structures \citep[see][for example use cases]{Peterken2019Density,Peterken2019TimeSlice,FraserMcKelvie2019,FraserMcKelvie2020,Greener2020,Krishnarao2020}.

\item {\it Galaxy Zoo Morphologies from SDSS, DECaLS and UKIDSS}: The Galaxy Zoo method, which involves combining classifications from a large number of classifiers collected via an online interface, has been applied to a variety of images, including the original SDSS images \citep{GZ2,Hart16}, the UK Infrared Telescope Infrared Deep Sky Survey \citep[UKIDSS;][]{Lawrence2007,Galloway}  and most recently the Dark Energy Camera Legacy Survey \citep[DECaLS;][]{Dey2019,Walmsley21}. This latter analysis combines Machine Learning (ML) methods with crowdsourcing in an active loop (for details see \citealt{Walmsley21}). We collect together all these crowdsourced morphologies for as many MaNGA galaxies as possible in this VAC. 

\item {\it Visual Morphology from DECaLS Images:} 
This VAC contains a direct visual morphological classification, based on the inspection of image mosaics generated from a combination of SDSS and DECaLs \citep[][]{Dey2019} images, for the MaNGA galaxies. The DR16+ version contains the classification for the first half of MaNGA galaxies (4600, MaNGA DR15) while the DR17 version contains the classification for the full MaNGA DR17 with unique {\tt MaNGAID}. Through a digital image post-processing, we exploit the advantages of this combination of images to identify inner structures, as well as external low surface brightness features for an homogeneous classification, following an empirical implementation of the methods in \citet{HernandezToledo2010} and \citet{Cheng2011}. The visual morphological classification is carried out by two classifiers inspecting three-panel image mosaics, containing a gray logarithmic-scaled $r-$band image, a filter-enhanced $r-$band image and the corresponding RGB color composite image from SDSS and a similar mosaic using DECaLS images \revision{incorporating the residual image after subtraction of a best surface brightness model from the DESI \emph{legacypipeline} \footnote{https://github.com/legacysurvey/legacypipe}.} The catalog contains the T-Type morphology, a variety of visual morphological attributes (bars, bar families, tidal debris, etc.) and our estimate of the non-parametric structural, Concentration, Asymmetry and Clumpiness \citep[CAS;][]{Conselice2003} parameters from the DECaLS images. For more detail in see \citet{VazquezMata2021}. \revision{An updated version including morphologies for all DR17 MaNGA galaxies in being prepared, see V{\'a}zquez-Mata et al. (in prep.).}

\item {\it MaNGA PyMorph DR17 photometric catalog:} (MPP-VAC, see \citealt{Fischer2019,DominguezSanchez2021}  for details) provides photometric parameters obtained from S{\`e}rsic and S{\`e}rsic+Exponential fits to the 2D surface brightness profiles of the final MaNGA DR17 galaxy sample (e.g. total fluxes, half light radii, bulge-disk fractions, ellipticities, position angles, etc.). It extends the MaNGA PyMorph DR15 photometric VAC to now include all MaNGA galaxies in DR17. 

\item {\it MaNGA Morphology Deep Learning DR17 catalog:} (MDLM-VAC, see \citealt{DominguezSanchez2021} for details) provides morphological classifications for the final MaNGA DR17 galaxy sample using Convolutional Neural Networks (CNN). The catalog provides a T-Type value (trained in regression mode) plus four binary classifications: P$_{\rm LTG}$  \revision{(separates early type galaxies, or ETGs, from late types, or LTGs)},  P$_{\rm S0}$  (separates ellipticals from S0) ,  P$_{\rm edge-on}$  (identifies edge-on galaxies) ,  P$_{\rm bar}$  (identifies barred galaxies). It extends the ``MaNGA  Deep Learning Morphology DR15 VAC'' \citep{Fischer2019} to now include galaxies which were added to make the final DR17. There are some differences with respect to the previous version -  namely, the low-end of the T-Types are better recovered in this new version. In addition, the P$_{LTG}$ classification separates \revision{ETGs from LTGs} in a cleaner way, especially at the intermediate types ($-1 < $T-Type $< 2$), where the T-Type values show a large scatter. Moreover, the value provided in the catalog is the average of 10 models trained with k-folding for each classification task (15 for the T-Type classification). The standard deviation, which can be used as a proxy for the uncertainty in the classification, is also reported.
\end{enumerate}

\subsubsection{Stellar Population Modeling VACs}  \label{vac:stellarpops}
There are a variety of stellar population modeling based VACs released. 

\begin{enumerate}
\item {\it  Principle Component Analysis (PCA) VAC (DR17):} this VAC includes measurements of resolved and integrated galaxy stellar masses, obtained using a low-dimensional, PCA-derived fit to the stellar continuum and subsequent matches to simulated star-formation histories (SFHs). The general methodology for obtaining the principal component basis set, the stellar continuum fitting routine, and the process of inferring stellar population properties such as mass-to-light ratio are discussed in \citet{mangapca_1_pace19a}. The aggregation of pixel-based mass estimates and adopted aperture-correction procedure are described in \citet{mangapca_2_pace19b}. This procedure yields estimates of galaxy-wide, integrated stellar masses also provided as part of this VAC. Key VAC characteristics remain unchanged in comparison to DR16 \citep{DR16}, where a holistic description of the VAC can be found. The principal enhancement in this release is in the sample size: the number of galaxies has been expanded to include all MaNGA galaxies to which the analysis could be readily applied, a total of 10223 unique plate-ifu designations (this number differs from unique galaxy counts, as some MaNGA galaxies were observed multiple times, so have multiple plate-ifus, each of which are analyzed separately in this VAC).

\item {\it \textsc{Firefly} Stellar Populations:} This VAC provides measurements of spatially resolved stellar population properties of MaNGA galaxies employing the \textsc{Firefly}\footnote{\url{https://www.icg.port.ac.uk/firefly/}} \citep{Wilkinson2017} full spectral fitting code. For DR17, \textsc{Firefly} v1.0.1 was run over all 10735 datacubes that had been processed by both the DRP and the DAP\footnote{This number differs from the total number of cubes listed in Table \ref{tab:scope}, as 538 data cubes did not run through the DAP for various reasons}. The major difference to the DR15 VAC is that we now provide the catalog in two versions. The first employs the stellar population models of \citet{Maraston2011} based on the MILES stellar library \citep{Sanchez-Blazquez2006}. The second version uses new MaStar models described in \citet{Maraston2020}. Both model libraries assume a \citet{Kroupa2001} IMF. Compared to the \textsc{Firefly} VAC in DR15, the radius (stored in {\tt HDU4} in the file) is now given in elliptical coordinates and the azimuth is added. Masses ({\tt HDU11} and {\tt HDU12}) are given per spaxel and per Voronoi cell. We do not provide absorption index measurements anymore. Each version of the VAC is offered as a single FITS file ($\sim$ 6 GB) comprising the whole catalog of global and spatially resolved parameters and, additionally, as a small version ($\sim$ 3 MB) that contains only global galaxy stellar population parameters. A detailed description can be found in J. Neumann et al. (in prep.) and \citet{Goddard2017}.

\item {\it Pipe3D}: This VAC containes the Pipe3D \citep{Sanchez2016} analysis of the full MaNGA dataset comprises the main properties of the stellar populations and emission lines for more than 10,000 galaxies, both spatial resolved and integrated across the entire field-of-view (FoV) of the IFUs. The content of the released distribution was originally described in \citet{Sanchez2018}, and updated in S. F. S\'anchez et al in prep. The new releases include considerable modifications from the previous ones, the most important ones being (i) the use of an updated version of the code fully transcribed to python (E. Lacerda et al. in prep.), (ii) the use of a new stellar population library based on the MaStar stellar library (A. Mejia-Narvaez et al. in prep.), and (iii) an update on the list of analyzed emission lines. 
\end{enumerate}

\subsubsection{HI-MaNGA DR3} \label{vac:himanga}
HI-MaNGA is a HI 21cm line followup program, to provide estimates of total atomic hydrogen content for galaxies in the MaNGA survey \citep{Masters2019}. It makes use of both previously published HI data (primarily from the ALFALFA survey; \citealt{haynes2018}) and new observations using the Robert C. Bryd Green Bank Telescope (GBT; to date under observing codes {\tt GBT16A\_095}, {\tt GBT17A\_012}, {\tt GBT19A\_127}, {\tt GBT20B\_033} and {\tt GBT21B\_130}). This VAC comprises the third data release (DR3) of HI 21cm detections or upper limits for \revision{6358} galaxies in the MaNGA sample. In some cases both GBT and ALFALFA data exist for a single galaxy, and we provide both observations separately, \revision{so the total number of rows is 6632, with 3358 coming from our GBT observations}. The observation and reduction strategy are documented in \citet{Masters2019,Stark2021}.  As part of this program a 20\% offset between actual, and estimated L-band calibration at GBT was noticed (see \citealt{Goddy2020}). \citet{Stark2021} provide guidance on dealing with confusion, and including upper limits into statistical analysis.  Observations are ongoing (under proposal code {\tt GBT21B\_130}), with the program on track to observe, or homogenise HI data for all MaNGA galaxies at $z<0.05$, with no pre-selection on color or morphology. This is expected to result in a final HI-MaNGA sample size of around 7000 MaNGA galaxies, with over \revision{6800} already having at least some data in hand. 


\subsubsection{The MaNGA AGN Catalog} \label{vac:mangaagn}
The MaNGA AGN Catalog presents AGN in the DR15 sample of MaNGA that are identified via mid-infrared WISE colors, Swift/BAT ultrahard X-ray detections, NRAO VLA Sky Survey (NVSS) and Faint Images of the Radio Sky at Twenty-cm (FIRST) radio observations, and broad emission lines in SDSS spectra. The catalog further divides the radio AGN into quasar-mode and radio-mode subpopulations, and provides estimates of the AGN bolometric luminosities. Full details of the AGN selection and luminosity measurements are \revision{described} in \citet{Comerford2020}. It is intended that this will be updated to include all MaNGA galaxies in \revision{the future. }

\subsubsection{GEMA-VAC: Galaxy Environment for MaNGA Value Added Catalog}\label{vac:gema}
The Galaxy Environment for MaNGA (GEMA) VAC (M. Argudo-Fern\'adez et al. in prep.) provides a variety of different measures of environment for galaxies in MaNGA. The combination of mass-dependent and mass-independent parameters provided in the catalog can be used to explore the effects of the local and large-scale environments on the spatial distribution of star formation enhancement/quenching, in the interaction of AGN with galaxies, or in the connection with kinematics or galaxy morphology, for instance, in an homogenous way allowing comparisons between different studies. In DR17, we present the final and updated version of GEMA-VAC for the final MaNGA sample. The quantifications of the environments are based on the methods described in \citet{2015AA...578A.110A} to estimate tidal strengths and projected number densities, as well as that in \citet{2015MNRAS.451..660E} to estimate overdensity-corrected local densities (MaNGA galaxies in the SDSS-DR15 only), and \citet{2016ApJ...831..164W} for an estimation of the cosmic web environment.  To better explore the environment of galaxies located in dense local environments, for instance, galaxies in compact groups or with strong interactions (close paris/mergers), but not necessarily a high density environment at larger scale, we also provide these same quantifications considering MaNGA galaxies in groups, according to an updated version of the catalog of groups compiled by \citet{2007ApJ...671..153Y}; and MaNGA galaxies in close pairs, according to the sample used in \citet{2019ApJ...881..119P}.  

\subsubsection{MaNGA Spectroscopic Redshifts for DR17}\label{vac:mangaz}
We present a catalog of precise spectroscopic redshifts M. Talbot et al. (in prep.) for the RSS and spaxels in MaNGA, updating the previous version of this catalog \citep{2018MNRAS.477..195T} to include the completed sample of MaNGA galaxies.  These spectroscopic redshifts are computed using the {\sc spec1d -- zfind} code from the publicly available BOSS pipeline~\citep[][]{bolton2012}, in which the NSA catalog provides the initial redshift.  Once spectroscopic redshifts were determined for the high signal-to-noise region within the galaxy half-light radius, a second pass attempted to determine the remaining redshifts in the low signal-to-noise spectra using the mean spectroscopic redshift as a prior.   The spectroscopic redshifts and a foreground model are presented for each spectrum with sufficient SNR to model, in this VAC.

\subsubsection{MaNGA Strong Gravitational Lens Candidate catalog}\label{vac:gravlens}
We present six likely, 12 probable, and 74 possible candidate strong galaxy-galaxy scale gravitational lenses found within the completed MaNGA survey. The lens candidates are found by the Spectroscopic Identification of Lensing Object program (SILO; \citealt[][M. Talbot et al. in prep]{2018MNRAS.477..195T}) , which was adapted from the BELLS \citep{2012ApJ...744...41B} spectroscopic detection method to find background emission-lines within co-added foreground-subtracted row-stacked-spectra of the MaNGA IFU, in which the co-added residuals are stacked across exposures from the same fiber at the same dither position.  Visual inspection of any background emission-line detected was performed, including the position of detections in proximity to an estimated Einstein radius.  Narrowband images were constructed from the co-added residuals to search for any lensing features.

%% file: mastar.tex



The MaNGA Stellar Library (MaStar) is a project in SDSS-IV to build a large library of well-calibrated empirical stellar spectra,  covering a wide range in stellar parameter space, roughly from 2,500K to 35,000K in effective temperature ($T_{\rm eff}$), from -1 to 5.5 in surface gravity ($\log g$), and from -2.5 to 0.5 in metallicity ([Fe/H]). It is conducted using the same instrument as MaNGA but during bright time \citep{Yan2019}. Most of the observations were done by piggybacking on APOGEE-2N, in the sense that the field centers of those plates, the time spent on the field, and the number of visits were determined according to the science need of APOGEE-2N. Only in a small number of fields were observational parameters determined by the science needs of MaStar. DR17 presents data for all of the stars observed in the MaStar program, along with complementary analysis of all the standard stars targeted on MaNGA-led plates. 

The MaStar targets were selected to cover a wide range in the 4-dimensional stellar parameter space ($T_{\rm eff}$, $\log g$, [Fe/H], [$\alpha$/Fe].) In the part of parameter space covered by APOGEE \citep{Majewski2017}, APOGEE-2 (\S \ref{sec:apogee}), SEGUE \citep{2009AJ....137.4377Y}, and the Large-sky Area Multi-Object fiber Spectroscopic Telescope (LAMOST \citealt{LAMOST}), we make use of the stellar parameters derived from these surveys to select targets, aiming to evenly sample the parameter space. However, due to availability of stars of certain parameters, the constraints of the fields selected by APOGEE-2N and the evolving field choices, the stellar parameter space coverage cannot be completely even. In addition to these selections, we further use photometry data from the Panoramic Survey Telescope and Rapid Response System (PanSTARRS-1; \citealt{Kaiser_etal_2010}) and the American Association of Variable Star Observers Photometric All-Sky Survey (APASS; \citealt{APASS}) to select stars that are more likely to have extreme temperatures, either very hot or very cool. \revision{Further details of the MaStar target selection have been described in \citet{Yan2019}.} Once {\it Gaia} Data Release 2 \citep{GaiaDR2,GaiaMission_2016} was available, we made use of {\it Gaia} color and absolute magnitudes (derived using distances from \citealt{BailerJones2018}) to select stars to fill up the parts of the color-magnitude space that were not sufficiently sampled, including hot main sequence stars, blue supergiants, yellow supergiants, stars at the tip of red giant branch, and red supergiants, Carbon stars and other asymptotic giant branch (AGB) stars, white dwarfs (WD), extreme horizontal branch stars, metal poor dwarfs, and late M-dwarfs. \revision{These recent changes to the target selection involving Gaia photometry will be described by R. Yan et al. (in prep.).}

Within the APOGEE-2 bright time extension ancillary call, the MaStar project was given a small number of hours to observe stars that could not be targeted by piggybacking on APOGEE-2N. During these times, we targeted a number of star-forming fields with a large number of hot main sequence stars, blue and red supergiants, a number of fields with known metal-poor late M dwarfs, and a number of open cluster and globular cluster fields. For the dedicated globular cluster fields, we conducted dithered observations to obtain integrated spectra for the core regions of the globular clusters with some fibers targeting relatively isolated stars in the outskirts of globular clusters. 

\subsection{MaStar-specific Changes to the MaNGA DRP}

The MaStar data are obtained using the same MaNGA fiber feed system and the BOSS spectrographs as the main MaNGA survey. The data reductions for MaStar are done with the MaNGA DRP through its 2D phase. The details of this were described by \citet{law16} with DR17 updates described in \S\ref{sec:manga.drp}. In the 2D phase, the only difference is in how we correct for the extinction of the standard stars in the flux calibration module, which we describe below. The reduction for the 3D phase is done differently from MaNGA. The basics of the data reduction were described by \citet{Yan2019}. We briefly describe the updates since DR15/16 below. More details of these will be presented in R. Yan et al. (in prep.).  

\begin{itemize} 
\item Flux calibration for both MaNGA and MaStar plates are done using a set of 12 standard stars observed simultaneously with the science targets. By comparing the spectra of these 12 stars with the theoretical spectra we determine the throughput ratio between the observed spectra and the expected spectra above the atmosphere. This ratio is then applied to all the spectra from all the fibers to determine the per-fiber spectra. The theoretical spectra used in the comparison need to have galactic extinction applied. For MaNGA-led plates, we use the values from \cite{SFD} dust map as the standards in MaNGA-led fields are at high galactic latitude and at a far enough distance to be beyond most of the dust in the Milky Way in those line of sights. But this is not always the case for MaStar plates. On MaStar plates, we use the spectra themselves (relative to their respective models) to estimate the relative extinction difference between different standard stars. Then we use the broadband colors to estimate the absolute extinction level for all the stars. With that,  we can then determine the combined throughput curve of the atmosphere, the telescope, and the instrument. This throughput curve is then applied to calibrate spectra from all science fibers.   

\item Standard stars targeted with mini-bundles on MaNGA-led plates are also treated like other MaStar targets. This adds a significant number of F stars to the library. 

\item We have updated the template set used in the determination of stellar radial velocity search, which is a selected subset from the BOSZ templates \citep{bohlin17}. We expanded the subset to include templates with temperatures between 3500K and 35000K, with surface gravity between 1 and 5\ in log (g cm$^{-1}$ s$^2$), and with two different [$\alpha$/Fe] settings (0 and 0.5). We also included the Koester white dwarf templates for DA-type white dwarfs \citep{Koester2010}.

\item We changed the method used to select the fiber on which the final spectrum is based, for stars that saturate the central fiber in a bundle. We also changed the criteria of determining whether to combine spectra of multiple fibers together in each exposure. When combining spectra, the risk of ``red upturn" is evaluated and taken into consideration. The ``red upturn'' refers to the artificial extra flux at the extreme red wavelengths in some of the spectra, which is likely introduced by crosstalk between adjacent spectral traces and imperfect 1D extraction at the extreme wavelengths. This will be discussed in more detail in R. Yan et al. (in prep.).

\item For some MaStar plates (usually those done in APOGEE-2 time and therefore led by APOGEE-2), we adopted exposure times much shorter than 900s in order to target bright stars. Three exposures time settings were adopted: 28s, 83s, and 250s. For those exposures shorter than 180s, the flexure-compensation algorithm adopted by DRP for MaNGA-length exposures no longer works due to the faintness of sky emission lines. In these cases, we measure the radial velocities for the standard stars, separately for the blue and red cameras, to figure out the median relative shifts between the two cameras. This is then used to adjust the blue cameras' wavelength solutions to match those of the red cameras'. The flexure could also cause the wavelength solution to differ from exposure to exposure. In this case, we use the radial velocities derived for all stars (both science targets and standard stars) to shift the wavelength solutions of all exposures to be consistent with the first exposure on a given visit, which is closest in time to the arc calibration frame. 
\end{itemize}

We also added many quality checks and quality flagging in this data release. 
\begin{itemize}
\item We added checks in the DRP to indicate the risk of red upturns. A subset of per-exposure spectra with significant upturn or downturn risks were visually inspected. The results were stored in a metadata file and read in by the pipeline to flag those per-exposure spectra. We exclude those per-exposure spectra with {\tt UPTURNRISK}, {\tt REDUPTURN}, or {\tt REDDOWNTURN} set in their quality bitmask, if possible, when producing the per-visit spectra. 

\item We run all MaStar spectra through an emission-line measurement code. The results were used to select a subset for visual inspection. The visual inspection results were combined with the automated measurements to decide which spectra to be flagged as having emission lines. Spectra with H$\alpha$ equivalent widths (EW) greater than 0.6 Angstrom are flagged with the {\tt EMLINE} bit in {\tt EXPQUAL} and {\tt MJDQUAL}. We note that usual convention has positive EW for absorption not emission, but for this context (MaStar emission-line identification), we define emission lines to be positive in EW.
 
\item We evaluated the quality of flux calibration \revision{for both the spectra from individual exposures and the combined spectra per visit. For individual exposure spectra, we flag them according to the chi-square produced when fitting for the flux ratios between the central fiber and the surrounding fibers. When the chi-square} is greater than 50, we flag \revision{the} {\tt BADFLUX} \revision{bit of the quality bitmask of this exposure ({\tt EXPQUAL}). This threshold corresponds to an uncertainty larger than 0.15 mag ($\sim14\%$) uncertainty in the relative flux calibration between the BP and RP bands (evaluated by comparing the synthetic color with Gaia photometry). This only flags the 1\% worst cases on a per-exposure per-star basis.}. 

\item \revision{When we combine spectra from multiple exposures together to construct the combined exposure per visit, the quality of flux calibration is going to significantly improve due to averaging and due to the dominance by spectra with higher signal-to-noise ratio which tend to have better calibration. The evaluation of flux calibration quality for the combined spectra per-visit is based on uncertainty determined through Jackknife resampling technique. If the uncertainty on the synthetic BP-RP color is more than 0.05 mag, which corresponds to 5\% relative calibration error, we flag the {\tt BADFLUX} bit of the quality bitmask for this visit ({\tt MJDQUAL}).}

\item We exclude those per-exposure spectra with {\tt BADSKYSUB}, {\tt POORCAL}, or {\tt SEVERBT} set, if possible, when producing the per-visit spectra. 
\end{itemize} 

All of these changes and updates will be discussed in more detail by R. Yan et al. (in prep.).

\subsection{Changes to the MaStar Post-processing Pipeline}

The MaStar post-processing pipeline (mastarproc) is updated for several purposes. It processes the result from a preliminary DRP run to identify candidates with emission line risks and red upturn/downturn risks for visual inspection. It is also updated to give more information in the MaStar summary files. We evaluate the variations of heliocentric radial velocities among all visits of a star (same {\tt MaNGAID}) and provide both the median velocity per visit and that across all visits. The significance of the variation is provided and if it is more than 3$\sigma$, we flag the \texttt{VELVARFLAG} column in the summary file. We also added several useful columns to the summary files to indicate the SNR, bad pixel fraction, etc. See more details in R. Yan et al. (in prep.).

\subsection{MaStar Summary files}

In this data release, we provide several summary files. The {\tt mastarall} file contains only metadata information about the stars and visits, but no spectra. It has four extensions containing four tables: \texttt{GOODSTARS}, \texttt{GOODVISITS}, \texttt{ALLSTARS}, and \texttt{ALLVISITS}. The \texttt{GOODSTARS} table contains the summary information for all stars with at least one good quality visit spectrum, which we define as good stars. It has one entry per unique {\tt MaNGAID}. The \texttt{GOODVISITS} table lists out all of the good quality visits for the good stars. It has one entry per unique visit. The \texttt{ALLSTARS} table contains the summary information for all of the stars observed in MaStar, regardless of the quality of the visits, with one entry per unique {\tt MaNGAID}. The \texttt{ALLVISITS} table contains the information for all of the visits of all of the stars. 

The {\tt mastar-goodspec} file contains all of the good quality visit-spectra. It matches row-to-row to the \texttt{GOODVISITS} table in the \texttt{mastarall} file. The file {\tt mastar-badspec-v3\_1\_1-v1\_7\_7.fits.gz} contains all the other visit-spectra, that are excluded from the \texttt{GOODVISITS} table. 

In addition, we also provide two sets of files that contain spectra with unified spectral resolution curves. Because each spectrum in \texttt{mastar-goodspec} files can have different spectral resolution curves or line spread function curves, it could be cumbersome for the users. We thus defined four resolution curves based on the distribution of the spectral resolution at each wavelength among all good visit-spectra. For each resolution curve, we select visit-spectra that have higher resolution at all wavelengths and broadened their line spread function by convolution to match the common resolution curve. We provide four files containing four subsets of visit-spectra convolved to these four uniform resolution curves, respectively. 

With resolution curves unified, we could easily combine multiple visit-spectra for the same star to improve signal-to-noise. These combined spectra are also provided corresponding to the four subsets defined by the four resolution curves.

The detailed data models for these files can be found in R. Yan et al. (in prep.) or on the SDSS data release website (see \S \ref{sec:access}). 

\subsection{Photometry crossmatch} \label{vac:mastarphot}

With this data release, we also provide a value-added catalog giving crossmatch information between MaStar and a few other catalogs. We crossmatch it with \Gaia DR2 \citep{GaiaDR2,GaiaMission_2016}, \Gaia EDR3 \citep{Gaia2021}, PanSTARRS-1 \citep{flewelling2016}, 2MASS \citep{2mass_paper}, and Simbad \citep{simbad_2000}. 

\revision{The crossmatches between MaStar and \Gaia (DR2 or EDR3) are performed in a few steps. For each MaStar target, given its coordinates and epoch of the coordinates, we select all \Gaia targets within 40\arcsec~ of the star, apply proper motion correction to shift them to the epoch for which the MaStar coordinates were given. Then we search for the corresponding match within 3\arcsec. If there is only one match, the match is considered to be the correct one. If there is more than one candidate within the search radius, we compute a positional matching probability and a photometry-matching probability for all candidates. The photometry-matching probability is the average among probabilities in multiple bands available for the MaStar targets, computed using empirical relationships we established between Gaia photometry bands and the photometry bands of the MaStar targets. Both probabilities take into account their respective uncertainties. The product of the two probabilities are used to determine the best match among the multiple matches within 3\arcsec. Nearly all MaStar good stars have a match with Gaia. Among 24,290 good stars with unique MaNGAID, all but one have a match in \Gaia EDR3; all but 14 have a match in \Gaia DR2.}

\revision{Sometimes, a single MaStar target based on ground-based photometry is resolved by Gaia into multiple sources. Our algorithm tends to choose the brighter and more dominant source as the match. In such cases, the MaStar spectra could also be affected in two ways. First, the spectrum would contain light from both stars. Second, if the separation between the two stars is large enough to make the combined image non-circular, then the fiber-aperture correction could be significantly affected resulting in a poor flux calibration for the final spectrum. These can be identified or excluded by checking the {\tt gaia\_cleanmatch} column in the two tables. Sources with {\tt gaia\_cleanmatch}=1\footnote{This corresponds to a value lower than 0.0084 in the ``contamination" column, whose meaning is defined in Appendix D.8 of \citet{Yan2019}. The threshold adopted here is 3 times larger than that in \citet{Yan2019}. The threshold adopted by \citet{Yan2019} is more conservative for targeting purposes.} are considered cleanly isolated, for which the flux correction should be sufficiently accurate, according to Gaia astrometry and photometry.}

\revision{With most stars crossmatched with \Gaia, the crossmatch with PanSTARRS-1 is done through \Gaia astrometry since it has more accurate coordinates and epoch information. Applying Gaia proper motion, we shift the \gaia-provided coordinates for MaStar targets to Epoch 2012.3 which is the approximate average epoch of the PanSTARRS-1 photometry catalog we used. Then we search for crossmatches in the PanSTARRS-1 catalog using a search radius of 2\arcsec. All sources with a unique match within 1\arcsec ~are considered a secure match. For those with multiple candidates within 2\arcsec, we choose the candidate with the largest product of the positional-matching probability and the photometry-matching probability. }

\revision{The Simbad catalog \citep{simbad_2000} contains useful spectral type and object type information for a small fraction of our targets. For crossmatching to Simbad, we shift the coordinates of all our targets to epoch 2000.0, then use a search radius of 3\arcsec. Sometimes, the same source appears as multiple entries in Simbad with different object types. In these cases, we choose the one that is most relevant or more informative for the star. About 13.3\% of all the good stars and 23.4\% of the science targets have a Simbad crossmatch with object type information. About 6.1\% of all good stars and 9.9\% of the science targets have a Simbad match with spectral type information.}

 For 2MASS, the crossmatch is done through the crossmatch table provided by \Gaia \revision{\citep{Marrese19}, using the astrometry solution provided by \Gaia Data Processing and Analysis Consortium  \citep[DPAC][]{Lindegren18, Lindegren21}}. We also derive extinction-corrected absolute magnitude and colors based on the \Gaia photometry using a 3D dust map \citep{Green2019} and Bailer-Jones distances (\citealt{BailerJones2018} for DR2 and \citealt{BailerJones2021} for EDR3). We also include spectral type and object type information available from Simbad. We provide two files. Both contain information from MaStar, PanSTARRS-1, 2MASS, and Simbad. The only difference is that one file is based on \Gaia DR2, while the other file is based on \Gaia EDR3.

\subsection{MaStar Stellar Parameters VAC} \label{vac:mastar}

Accurate stellar parameter labeling of the stars is essential for a stellar library. Although a significant fraction of the stars targeted in our library have been observed by other surveys with parameters derived, a large fraction still lack such information. Furthermore, the parameter derivations from previous surveys were inhomogeneous and some were based on data with poorer quality than we have. Thus, we initiated multiple parallel efforts to determine the stellar parameters for MaStar based on MaStar spectra themselves. Within DR17, we include a VAC giving four sets of stellar parameter measurements based on different methods along with the median values of them when available and deemed robust. 

The four sets of parameters are described below. More details can be found in the respective papers. A comparison between the parameters will be presented in R. Yan et al. (in prep.).

\begin{description}
\item[DL] This parameter set (D. Lazarz et al. in prep.) is derived using full-spectrum fitting with an Markov Chain Monte Carlo (MCMC) sampler using interpolated BOSZ model spectra with continuum shape information included in the chi-square calculation. Extinction is fitted as a by-product. No photometry prior is used.

\item[JI] This parameter set \citep{Imig2021} is derived using a neural network which models flux as a function of labels and is trained on a combination of empirical MaStar spectra with parameters from the APOGEE Stellar Parameters and Chemical Abundance Pipeline (ASPCAP, see \S \ref{details} below) and the model spectra produced by \citet{Allende2018}. 

\item[LH] This parameter set \citep{Hill2021} is derived using full-spectrum, single-template, pPXF fitting with an MCMC sampler, using interpolated BOSZ and MARCS model spectra, with a flat prior based on \Gaia color-magnitude diagram. The continuum is modeled with a multiplicative polynomial. 

\item[YC] This parameter set (Y. Chen et al in prep.) is derived using full-spectrum fitting using both the BOSZ and MARCS model spectra without interpolation, with the result produced by Bayesian average and a flat prior based on \Gaia color-magnitude diagram. The continuum is modeled with a multiplicative polynomial. 
\end{description}
\revision{All four methods provide \teff, \logg, and \feh.  On top of that, the methods by DL, LH and JI also provide \afe, and the method by JI additionally provides micro turbulence velocity ($v_{\rm micro}$). All methods have been applied to all spectra with quality control applied differently for different methods. Each method flags the spectra for which the parameters are considered invalid due to poor fitting quality. When we take the median, we only take the median among those methods that provide a valid measurement for a given spectrum. The uncertainties of the median values are also computed accordingly. Which methods are used in the median calculation are indicated by the {\tt INPUT\_GROUPS} and {\tt INPUT\_GROUPS\_NAME} columns for \teff, \logg, and \feh. The quality control is more strict for \afe, for which the contributing methods are indicated by the {\tt INPUT\_ALPHA\_GROUPS} and {\tt INPUT\_ALPHA\_GROUPS\_NAME} columns in the VAC. The details of these will be provided in R. Yan et al. in prep.}

Metallicity measurements are crucial for assigning the right library spectra to the right metallicity bin when building stellar population models. We found some of the parameter sets could have slight systematic bias in metallicity measurements. Therefore, we calibrated the metallicity measurements for three of the four sets against APOGEE ASPCAP \feh~ measurements. In the VAC, in addition to the straight median among the four sets, we also provide a calibrated metallicity for each set and the median calibrated one among the four. We consider this to be a more accurate representation of the true metallicities of the stars. \revision{The stellar parameters are more reliable when at least two of the four groups have valid measurements for the given star. The users could choose to apply similar cuts to select a set of stars with more reliable parameters.}

\revision{In the right panel of Figure \ref{fig:mastar_hr}, we show the median effective temperature vs. median surface gravity for the subset of good science stars in MaStar with at least two of four groups providing valid measurements. This includes 91\% of all good science stars. }

\subsection{MaStar Sample Statistics}

In total, the MaStar library includes 24,130 unique good quality stars with 59,266 good quality visits. Among these, 11,817 unique stars were targeted as science targets and 12,345 unique stars were targeted as flux standards, with some overlap between the two categories. The 24,130 unique stars correspond to 24,290 unique {\tt MaNGAID}s as some stars correspond to more than one {\tt MaNGAID}s when taken from different source catalogs.

\begin{figure*}
\begin{center}
\includegraphics[width=0.45\textwidth]{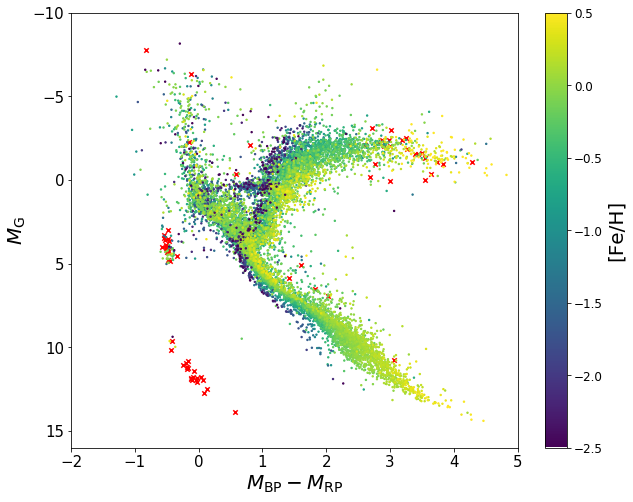}
\includegraphics[width=0.43\textwidth]{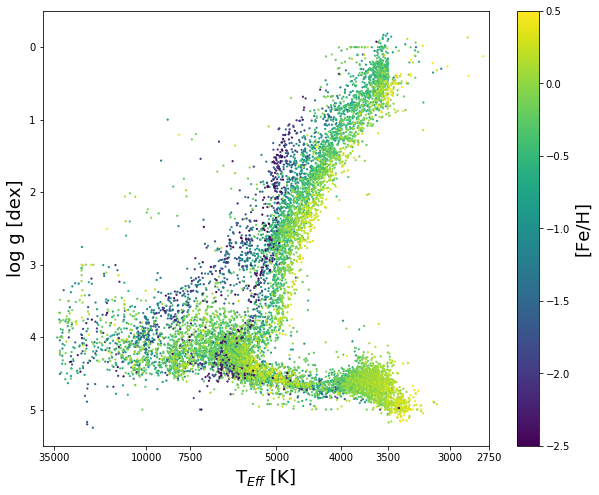}
\caption{\revision{Left: Extinction-corrected G-band absolute magnitude vs. BP-RP color for all good science targets in the MaStar library. The color coding indicates the median calibrated metallicity. Stars without valid metallicity measurements are marked as red crosses. Right: Median effective temperature vs. median surface gravity for MaStar science targets with valid measurements from at least two of the four groups. Standard stars are also included in the library but are not shown in these plots.}}
\label{fig:mastar_hr}
\end{center}
\end{figure*}

In Figure~\ref{fig:mastar_hr}, we show a Hertzsprung-Russell diagram for all good stars in MaStar, based on \Gaia EDR3 photometry after correcting for dust using the 3D dust map. The points are color-coded according to the median calibrated metallicity. This illustrates the comprehensive stellar parameter coverage of our library.

%% file: eboss.tex
eBOSS \citep{Dawson16} concluded its observations of galaxies and quasars as tracers of
large-scale structure on March 1, 2019.  The goal of these
measurements was to measure the distance-redshift relation with the baryon acoustic oscillation (BAO)
feature that appears at a scale of roughly 150 Mpc.  These data were also used to measure the growth of structure through redshift space distortions
\citep[RSD;][]{kaiser87a}.

The final eBOSS cosmology measurements were presented in a series of papers submitted in July, 2020.  These results included consensus measurements of BAO and RSD for luminous red galaxies \citep[LRG;][]{LRG_corr,gil-marin19a} over $0.6<z<1.0$, emission line galaxies \citep[ELG;][]{tamone19a,demattia19a}
over $0.6<z<1.1$, and quasars \citep{hou19a,neveux19a} over $0.8<z<2.2$.  In addition, measurements of BAO were performed at
$z>2.1$ using clustering in the Lyman-$\alpha$ forest and cross-correlations between quasars and the forest \citep{2019duMasdesBourbouxH}.
These measurements were combined with the final SDSS and BOSS \citep{Dawson13a} BAO and RSD measurements spanning redshifts $0.07<z<0.6$
\citep{ross15,howlett15,boss17} to form a final sample of distinct
clustering measurements over roughly ten billion years.   The aggregate precision of the expansion history measurements is 0.70\% at redshifts $z < 1$ and 1.19\% at redshifts $z > 1$, while the aggregate precision of the growth measurements is 4.78\% over the redshift interval $0 < z < 1.5$.  Using the BAO technique by itself, with no other constraints, SDSS has built up a clear picture of the distance-redshift relationship revealing a clear need for dark energy with a detection significance of $8\sigma$ \revision{\citep{ebosscosmo21}}.

The full cosmological interpretation of these data \revision{are described in \citet{ebosscosmo21} and demonstrate} the power of BAO for constraining curvature and providing robust estimates
of $H_0$  The analysis also demonstrates the ability of RSD data to complement weak lensing and cosmic microwave background measurements in providing
independent evidence for a flat cosmological model with dark energy described by a cosmological constant.
The combined BAO and RSD measurements indicate $\sigma_8=0.85 \pm0.03$,
implying a growth rate that is consistent with predictions from $\it Planck$ temperature and polarization data \citep{planck20} and with General Relativity.  
Combining these results with $\it Planck$, Pantheon Type Ia supernovae \citep[SNe~Ia;][]{scolnic18a},
and weak lensing and clustering measurements \citep{troxel18a} from the Dark Energy Survey (DES) leads to significant advances in cosmological
constraints relative to the prior generation of experiments.
Each of the three parameters, $\Omega_\Lambda$, $H_0$, and $\sigma_8$ is constrained at roughly 1\% precision, even in a model that allows
free curvature and a time-evolving equation of state for dark energy.
In total, the data are best described by a flat $\Lambda CDM$ model
with
$H_0= 68.18 \pm 0.79 \, {\rm km\, s^{-1} Mpc^{-1}}$ \revision{(for full details of this analysis see \citealt{ebosscosmo21}).}
The 
Dark Energy Task Force Figure of Merit \citep{albrecht06a} of these
data sets together is 94\footnote{The measurements and cosmological impact are summarised in two web pages \url{https://www.sdss.org/science/final-bao-and-rsd-measurements/} and \url{https://www.sdss.org/science/cosmology-results-from-eboss/}}.

New value-added catalogs derived from eBOSS data were released publicly at the same time as the cosmology results.  These catalogs contain the redshifts and
weights for each of the LRG, ELG, quasar, and Ly-$\alpha$ forest samples, as well as the properties of all quasars observed during
the four generations of SDSS.  In addition, the mock catalogs
used to characterize covariance in the clustering measurements are being released in coordination with DR17.  A description of each of these
cosmology value-added catalogs is as follows:

\subsubsection{eBOSS Large Scale Structure Catalogs}\label{vac:lss}

DR16 included full reductions of the completed set of observed eBOSS spectra. \revision{An additional series of redshift estimates for the eBOSS galaxy samples was produced by an algorithm known as {\tt redrock}\footnote{\url{https://github.com/desihub/redrock}}. The galaxy redshifts derived from {\tt redrock} are described in Section 4 of \citet{ross20a}.  An additional series of classifications and redshift estimates was also performed for the BOSS and eBOSS quasar samples.  The origin of quasar redshift estimates is described in \citet{lyke20a}, while a summary of how those redshifts were used in the clustering measurements is also found in Section 4 of \citet{ross20a}.  From these updated redshift estimates} large-scale structure (LSS) VACs are created, which, together map the three-dimensional structure of the Universe using galaxies and quasars over redshifts $0.6<z<2.2$.
These maps are carefully constructed with corrections for observational systematic errors and random positions that sample the survey selection function to allow unbiased cosmological inference.  Three distinct samples were observed by SDSS-IV and used to produce LSS catalogs: LRG \citep{prakash16a}; ELG \citep{raichoor17a}; and quasars \citep{myers15a}. A value-added catalog for each of these samples was released in July 2020.  The LSS catalogs for the LRG and quasar samples are described in \citet{ross20a} while the LSS catalog for the ELG sample is described in \citet{raichoor19a}.

\subsubsection{eBOSS DR16 Large-scale structure multi-tracer EZmock catalogs}\label{vac:mock}
We present 1000 realizations of multi-tracer EZmock catalogs, with redshift evolution and observational systematics, for each sample of the DR16 LSS data. These mock catalogs are generated using the EZmock method \citep{EZmock_method2015}, and applied the survey footprints and redshift distributions extracted from the corresponding data. They accurately reproduce the two- and three-point clustering statistics of the DR16 data, including cross correlations between different tracers, down to the scale of a few $h^{-1}\,{\rm Mpc}$, and provide reliable estimates of covariance matrices and analyzes on the robustness of the cosmological results. Details on the construction and clustering properties of the EZmock catalogs are presented in \citet{EZmock_eBOSS2021}.

\subsubsection{eBOSS Quasar Catalog}\label{vac:qso}

Beginning with SDSS-I, SDSS has maintained a tradition of releasing a visually-inspected quasar (or quasi-stellar object; QSO)  catalog alongside major data releases. 
The new SDSS-DR16Q catalog (DR16Q;~\citealt{lyke20a}) represents the most recent, and largest, catalog of known unique quasars within SDSS. To ensure completeness, quasars from previous catalog releases (DR7Q;~\citealt{dr7q_paper}, DR12Q;~\citealt{dr12q_paper}) have been combined with observations from eBOSS in SDSS-IV. The catalog contains data for more than 750,000 unique quasars, including redshifts from visual inspections, principle component analysis (PCA), and the SDSS automated pipeline. 
Additionally, the catalog is the first from SDSS to contain both the  \citet{dr6q_hw_paper} DR6 redshift estimates and the \citet{Shen_etal_2011} DR7 redshift estimates that are based on the \citet{dr6q_hw_paper} algorithm. Where applicable, the catalog also contains information about broad absorption line (BAL) troughs,
damped Lyman-$\alpha$ (DLA) absorbers, and emission line redshifts (via PCA). As in previous releases, DR16Q also contains properties for each quasar from GALEX~\citep{galex_paper}, UKIDSS~\citep{Lawrence2007}, WISE~\citep{WISE}, FIRST~\citep{first_paper}, 2MASS~\citep{2mass_paper}, ROSAT/2RXS~\citep{Boller16}, XMM-Newton~\citep{Rosen2016}, and \Gaia~\citep{GaiaDR2}, when available. To facilitate analyzes of pipeline accuracy and automated classification, a superset was also released.  This sample contained $\sim 1.4$ million unique observations for objects targeted as quasars from SDSS-I/II/III/IV. 

\subsubsection{Lyman-$\alpha$ Forest Transmission VAC}\label{vac:lya}

This VAC contains the estimated fluctuations of transmitted flux fraction in the pixels across
the Lyman-$\alpha$ and Lyman-$\beta$ spectra region of DR16Q quasars.
In total, 211,375 line-of-sights contribute to the Lyman-$\alpha$ spectral regions and 70,626 to the Lyman-$\beta$ one.
This VAC contains everything needed to compute the three-dimensional auto-correlation of Lyman-$\alpha$ absorption in two different spectral regions as in \citet{2019duMasdesBourbouxH}. 
When combined with the DR16Q quasar catalog, this VAC also provides the information to compute the
three-dimensional quasar $\times$ Lyman-$\alpha$ cross-correlation. These two measurements are used to measure
the location of the BAO as reported in \citet{2019duMasdesBourbouxH}.

\subsection{Other VACs based on Single-Fiber Optical Spectra}

The final eBOSS data sample contains more than one million spectra of stars, galaxies,
and quasars obtained during SDSS-IV.  This catalog has been used for a range of studies of astrophysical processes beyond the
BAO and RSD measurements described above.  Demonstrating the impact of these data for additional studies, this release includes
a value-added catalog of strong lensing systems (originally released in July 2020 in a mini data release), a new catalog of lensed Lyman-$\alpha$
emitting (LAE) galaxies, a new catalog of the cosmic web, and a catalog of metal absorbers.
A description of these value-added catalogs is as follows:

\subsubsection{eBOSS Strong Gravitational Lens Detection Catalog}\label{vac:lenses}
A value-added catalog of 838 likely, 448 probable, and 265 possible candidate strong 
galaxy gravitational lens systems was released along with the batch of cosmology results and
value-added catalogs in July 2020 \revision{(aka DR16+)}.  These systems were discovered by the presence of higher 
redshift background emission-lines in DR16 eBOSS galaxy spectra. \revision{The methodology, including quantitative explanations of the ``likely", ``probably" and ``possible" categorisation is described in full in \citet{2021MNRAS.502.4617T}; also see \citet{2018MNRAS.477..195T}}.  \revision{This Spectroscopic Identification of Lensing Objects (SILO)} 
program extends the method of the BOSS Emission-Line Lens 
Survey~\citep[BELLS;][]{2012ApJ...744...41B} and Sloan Lens 
ACS~\citep[SLACS;][]{2006ApJ...638..703B} survey to higher redshift, and 
has recently been applied to the spectroscopic discovery of strongly 
lensed galaxies in MaNGA~\cite[SILO;][also see \S \ref{vac:gravlens}]{2018MNRAS.477..195T}. Although 
these candidates have not been studied through a dedicated imaging 
program, an analysis of existing imaging from the SDSS 
Legacy Survey and the DESI Legacy 
Surveys~\citep[][]{Dey2019} provides additional information on these systems.
This catalog includes the results of a 
manual inspection process, including grades and comments for each 
candidate, consideration of sky contamination, low 
signal-to-noise emission-lines,
improper calibration, weak target emission-lines, systematic errors, 
Gaussian modeling, and potential lensing features visually identified 
within the available imaging.

\subsubsection{eBOSS ELG-LAE Strong Lens Catalog} \label{vac:ebossELGLAE}
The eBOSS ELG-LAE Strong Lens Catalog contains roughly $150$ candidate strong lens systems selected from the ELG sample of the eBOSS survey released by DR17 using the method presented in \citet{Shu16}. By construction, the lensing galaxies in this catalog are ELGs up to $z \simeq 1.1$, and the source galaxies are Ly$\alpha$ emitters at $z > 2$. A full description of the catalog will be presented in A. Filipp et al. (in prep.). This catalog is complementary to the eBOSS Strong Gravitational Lens Detection Catalog constructed by \citet[][\S \ref{vac:lenses} above]{2021MNRAS.502.4617T}, in which the source galaxies are all [O\textsc{ii}] emitters at $z < 1.7$. 

\subsubsection{Cosmic Web Environmental Densities from Monte Carlo Physarum Machine} \label{vac:ebossMCPM}
The Monte Carlo Physarum Machine (MCPM) cosmic web reconstruction algorithm \citep{Burchett:2020_slime,Elek2021} was employed to characterize the matter density field from galaxies spectroscopically observed in SDSS, including those previously included in the NSA \citep{blanton2011} and LOWZ and eBOSS LRG catalogs.  The MCPM framework, which is particularly sensitive to the filamentary structure of the cosmic web, requires as inputs galaxy coordinates, redshifts, and halo masses, and we leverage stellar mass measurements included in VACs from previous Data Releases, the NASA-Sloan Atlas itself and the eBOSS \textsc{Firefly} Value-Added Catalog \citep{Comparat2017}.  Halo masses were then derived by the halo abundance matching relations from \citet{Moster:2013lr}. Using the Polyphorm software \citep{Elek:2020_polyphorm}, in which MCPM is implemented, reconstructions were produced in various redshift ranges corresponding to the varying depth/completeness of the galaxy samples.  MCPM then provides a proxy metric (known as the trace) for the environmental density at each point in the fitted volume.  These arbitrary density values are then calibrated to the cosmological matter density relative to the mean matter density by performing MCPM fits to the Bolshoi-Planck \citep{Klypin:2016aa} dark matter-only cosmological simulation's halo catalog \citep{Behroozi:2013aa} and producing a mapping between the MCPM trace and simulation 3D matter density field.  The data products are: (1) galaxy catalogs with local environmental density values evaluated at their locations from the full 3D density field and (2) the 3D density field itself, which users may in turn query for density values at locations away from known galaxy positions.

%% file: spiders.tex
Within eBOSS, the SPectroscopic IDentification of ERosita Sources (SPIDERS) program, dedicated to the 
characterization of the X-ray sources, was the largest and most complete spectroscopic follow up of the best
all-sky X-ray surveys available at the time (mainly the ROSAT all-sky survey; \citealt{Clerc2016, Dwelly17, Salvato2018}). 
As the acronym suggests, the original motivation and plans for SPIDERS were driven by the capabilities of the {\it eROSITA} X-ray telescope
(on board the Spectrum Roentgen Gamma, or SRG, satellite), which launched on July 13, 2019 \citep{Predehl2021AA...647A...1P}, significantly later than the planned date at the start of SDSS-IV. While waiting for {\it eROSITA} to collect data, the SPIDERS program targeted ROSAT/1RXS and XMM-Newton Slew Survey sources within the eBOSS footprint. These targets are expected to make up the bright X-ray flux end of the {\it eROSITA} population.
Much of the SDSS spectroscopy from the SPIDERS program was released as part of SDSS DR16 \citep{DR16}, including the largest spectroscopic redshift catalogs of X-ray selected AGN and clusters \citep{Comparat2020AA...636A..97C, Clerc2020MNRAS.497.3976C, Kirkpatrick2021MNRAS.503.5763K}.  

Below, we highlight some recent pre-{\it eROSITA} SPIDERS results
that are based on the SDSS DR16 dataset. Then, in \S \ref{sec:efeds} we present
a new SPIDERS dataset (to be released as part of SDSS DR17\footnote{\url{https://data.sdss.org/sas/dr17/eboss/spectro/redux/v5_13_2/}}), dedicated
to the spectroscopic follow-up of X-ray sources discovered in early
SRG/{\it eROSITA} performance verification observations. 
In \S \ref{sec:efeds_s5} we discuss future plans for {\it eROSITA} followup in SDSS-V (also see \S \ref{sec:future}). 

\subsection{SPIDERS Galaxy Cluster Highlights}

In this section we describe the SPIDERS cluster sample, the largest catalog of spectroscopically confirmed X-ray detected clusters to date. While no new data is released in DR17, several new results have come out since DR16. We refer the reader to the below papers for details of the ancillary data created and how to access it. 

The bulk of the galaxy cluster population targeted by SPIDERS up to and including DR16 originated from the ROSAT-based CODEX survey \citep{Finoguenov2020AA...638A.114F}, with additions of low-mass systems from the XMM-Newton-based X-CLASS catalog. 
The initial SPIDERS CODEX and X-CLASS targeting plans and survey strategy were described by \citet{Clerc2016}. \citet{Clerc2020MNRAS.497.3976C} provide complementary details of targeting updates that occurred during the course of the project.  \citet{Clerc2020MNRAS.497.3976C} also describe the full sample of acquired targets, the achieved spectral quality, as well as the redshift measurement success rate (nearly 98\%) and precision (20 km\,s$^{-1}$ at $z=0.2$ after averaging all galaxies in a cluster). In particular, \citet{Clerc2020MNRAS.497.3976C} highlight the very homogeneous nature of targets observed throughout the program; this specific feature supports a minimally biased spectroscopic follow-up of clusters.

Galaxy clusters confirmed and validated with spectroscopy from SPIDERS are presented in \citet[][2740 CODEX clusters]{Kirkpatrick2021MNRAS.503.5763K} and \citet[][124 X-CLASS clusters]{Koulouridis2021arXiv210406617K}. These two catalog papers list confirmed clusters up to $z\lesssim 0.6$, and present derived X-ray properties that exploit high precision cluster redshifts, obtained by averaging multiple SPIDERS member galaxy spectra per cluster. \citet{Kirkpatrick2021MNRAS.503.5763K} also provide velocity dispersion estimates for those clusters with more than 15 members, confirming correlations with X-ray luminosity and optical richness found by other studies. A galaxy cluster number count analysis performed in the velocity dispersion--redshift parameter space yields constraints on cosmological parameters $\Omega_{\rm{m}}$ and $\sigma_8$ consistent with the main cosmological analysis of the sample. The main cosmological analysis of the SPIDERS clusters program is presented by \citet{IderChitham2020MNRAS.499.4768I}; it involves the modeling of the redshift--richness relation of galaxy clusters spectroscopically confirmed by SPIDERS. The best-fit parameters are found to be $\Omega_{\rm{m}}=0.34^{+0.09}_{-0.05}$ and $\sigma_8 = 0.73 \pm 0.03$. As a central ingredient in those cosmological studies, the calibration of the mass--observable relations is presented in two papers: \citet{Capasso2019MNRAS.486.1594C} focus on the mass--richness relation and \citet{Capasso2020MNRAS.494.2736C} on the mass--X-ray luminosity relation. The large scale clustering of clusters -- highly biased tracers of the underlying dark matter density field -- is presented and discussed by \citet{Lindholm2020arXiv201200090L}. Taking advantage of the dense and precise spectroscopic sampling of cluster members, \citet{Mpetha2021MNRAS.503..669M} extract the gravitational redshift signal from stacks of SPIDERS confirmed galaxy clusters. Brightest SPIDERS cluster galaxies are studied by \citet{Furnell2018MNRAS.478.4952F} and \citet{Erfanianfar2019AA...631A.175E}, in particular, the scaling of their stellar mass with the host halo properties is shown to lack evolution since $z\sim 0.6$.

\subsection{The eROSITA Final Equatorial-Depth Survey}\label{sec:efeds}

During the SRG performance verification phase (November 2019), {\it eROSITA} was used to survey a contiguous region of $\sim$140\,deg$^2$. This {\it eROSITA} Final Equatorial-Depth Survey \citep[eFEDS;][]{Brunner2021,LiuTeng_Data2021} is intended as an early representative demonstration of the capabilities of the {\it eROSITA} all-sky survey, which will not be completed until late 2023. 
In order to exploit the availability of these X-ray data, we allocated a dedicated set of 12 SDSS-IV/SPIDERS plates to follow up counterparts of eFEDS X-ray sources, to be observed at APO during the first quarter of 2020. 
These special eFEDS plates, documented here, are released as part of SDSS DR17. 
Unfortunately, because of the COVID19-related closure of APO (see \S \ref{sec:intro}), only a small fraction of the originally designed plates were observed, most to shallower than expected depth. Nevertheless, this program did collect useful data, and the experience aqcuired has greatly helped the preparation of the {\it eROSITA} followup component of the SDSS-V Black Hole Mapper program \citep[][also see \S \ref{sec:future}]{2017arXiv171103234K}. 

The target selection procedure for eFEDS broadly followed the steps carried out for follow up of ROSAT sources in SPIDERS \citep{Clerc2016, Dwelly17}. First, the X-ray source detection process \citep[described in detail by][]{Brunner2021}, provides a catalog, including parameters that describe the detection likelihood, extent likelihood (to distinguish point sources from extended ones), and astrometrically corrected X-ray position and positional uncertainty. For the SDSS-IV targeting of eFEDS sources, we used the preliminary data processing software version available at that date (late 2019, `c940'), and only considered eFEDS sources with detection likelihood $>8$, so that the fraction of spurious detections is kept to a relatively low rate. Based on simulations \citep[][]{Comparat2019MNRAS.487.2005C, Comparat2020OJAp....3E..13C, LiuTeng_Sim2021}, we expect only around 2\% of these X-ray detections to be spurious. Optical/IR counterparts were associated with the X-ray sources using the methods described below. Finally, each of these counterparts is identified as being in one or more target classes (recorded using flags in the targeting bitmask, Sec. \ref{subsubsec:maskbits}), and then the final target list is created which includes only objects suited for spectroscopic observations with the BOSS spectrographs. 

\subsubsection{eFEDS: Optical counterparts to extended sources}
\label{subsubsec:xray:clusters:targets}
The identification of SPIDERS clusters and their membership, as well as the detailed target selection process are described fully by A. Merloni et al. (in prep.) and J. Ider Chitham et al. (in prep.). Here we document the resulting targeting bits that describe the contents of the SPIDERS cluster target catalog:

\begin{enumerate}

\item \texttt{EFEDS\_EROSITA\_CLUS}: These are galaxies associated with extended X-ray emission. A combination of all member galaxies selected using redMaPPer in scanning-mode \citep{Rykoff2014ApJ...785..104R} as well as additional BCGs identified using the multicomponent matched filter cluster confirmation tool \citep{Klein2019MNRAS.488..739K}. 
BCGs are allocated the highest target priority (0), while all other members are given a lower priority (33-130). 
The selection method on eFEDS extended sources is described fully in \citet{LiuAng2021,Klein2021}.  
The redMaPPer based selection uses positional X-ray information to estimate optical properties such as photometric redshifts, centering information and member selection in a method analogous to that of as described by \citet{Clerc2016}, as used in the SPIDERS DR16 data release. The method itself is described by \citet{IderChitham2020MNRAS.499.4768I}, J. Ider Chitham (in prep.). The photometric selections of BCGs and cluster members are based on the eighth data release of the DESI Legacy Imaging Surveys (using $g-r$ and $r-z$ colors) as well as HSC photometric ($grizy$) data \citep{Dey2019,Aihara2019PASJ...71..114A}. 
Cluster BCG photometric redshifts are within $0.0 < z < 1.3$, a total of 85 were observed.

\item \texttt{EFEDS\_SDSS\_REDMAPPER}: Publicly available optically selected cluster member catalog based on SDSS DR8 and CODEX data  \citep{Rykoff2014ApJ...785..104R,Finoguenov2020AA...638A.114F}. 
Cluster photometric redshifts are within $0.0 < z < 0.6$, a total of 1031 potential cluster galaxies were observed.

\item \texttt{EFEDS\_HSC\_REDMAPPER}: Optically selected cluster member catalog \citep{Rykoff2014ApJ...785..104R} based on HSC DR2 data \citep{Aihara2019PASJ...71..114A}. 
Cluster photometric redshifts are within $0.3 < z < 1.1$, a total of 13 cluster galaxies were observed. 

\end{enumerate}

\subsubsection{eFEDS: Optical counterparts to point sources}
\label{subsubsec:xray:PS:targets}
The method used to associate optical/IR counterparts to the eFEDS X-ray point-sources is presented by \citet{Salvato2021}. In short, the optical counterparts to the X-ray sources were identified using a Bayesian association algorithm {\sc{NWAY}} \citep{Salvato2018}, which accounts for both position and photometric properties of the counterpart.  The association algorithm was informed by a large training sample of {\it XMM-Newton} and {\it Chandra} X-ray serendipitous sources with secure optical identifications. Supplementary optical/IR counterparts were also provided via a Likelihood Ratio approach \citep{Sutherland1992,Ruiz18}. Note that at the time of selection of targets for the SDSS-IV special plates, only very early versions of the X-ray catalog and counterpart selection algorithms were available - so the pool of targets is different from the counterparts presented by \citet{Salvato2021}.
The main eFEDS point-like target class is \texttt{EFEDS\_EROSITA\_AGN} with 2149 observed targets. Despite the name, 
we expect this target set to also include a number of stars and (inactive) galaxies.

\subsubsection{eFEDS: additional targets}
To ensure that all fibers available in the designed plates could be allocated to interesting scientific targets, beyond the {\it eROSITA}-selected sources described above, several additional target classes were defined before the arrival of the {\it eROSITA} data. 
The largest of these were special classes of AGN drawn from optical targets selected by a combination of the HSC Subaru Strategic Program (SSP) imaging and other multi-wavelength catalogs. 
Note that due to the high sky density seen in the available {\it eROSITA}/eFEDS source catalog, very few fibers were left for filler targets, and so most of the defined target classes were allocated zero fibers (unless the same astrophysical object was also selected as an {\it eROSITA} target). However, since these target selection classes are defined in the DR17 data products, for completeness of the documentation, we give a brief description here of each target class. 

\begin{enumerate}

\item \texttt{EFEDS\_HSC\_HIZQSO}: Optical/NIR High-redshift QSO candidates selected from HSC photometric data witha drop-out technique targeting $z>4$.

\item \texttt{EFEDS\_HSC\_WERGS}: FIRST-detected bright radio galaxies selected for their extreme red Optical-NIR colors

\item \texttt{EFEDS\_HSC\_REDAGN}: WISE-detected obscured AGN candidates selected on the basis of their extremely red Optical-NIR colors.

\item \texttt{EFEDS\_COSMOQSO}:  UV excess/variability selected QSO - following prototype selection criteria of 4MOST Cosmology Survey.

\item \texttt{EFEDS\_WISE\_AGN}: AGN candidate selected via location in WISE color-magnitude space, following criteria of \citet{Assef2018} - with optical counterparts selected from DESI Legacy Imaging Survey DR8. 

\item \texttt{EFEDS\_WISE\_VARAGN}: AGN candidate selected via variability signature in WISE data - with optical counterparts selected from DESI Legacy Imaging Survey DR8. 

\item \texttt{EFEDS\_XMMATLAS}: Optical counterparts to X-ray sources from the 3XMM/dr8 catalog \citep{Rosen2016} and from the XMM-ATLAS survey field \citep{Ranalli2015AA577A121R}.

\item \texttt{EFEDS\_CSC2}:  X-ray sources from the Chandra-CSC2.0 catalog \citep{Evans2010} - with optical counterparts selected from DESI Legacy Imaging Survey (DR8). 

\item \texttt{EFEDS\_SDSSWISE\_QSO}: QSO candidates from the \citet{Clarke2020} ML classification of SDSS-WISE photometric sources 
\item \texttt{EFEDS\_GAIAWISE\_QSO}: QSO candidate from the \citet{Shu2019} Random Forest classification of \gaia-unWISE sources 

\item \texttt{EFEDS\_KIDS\_QSO}: QSO candidate from the \citet{Nakoneczny2019} classification of KiDS+VIKING DR3 sources

\item \texttt{EFEDS\_GAIA\_WD}:  \gaia-selected WD binary candidate in the region between the zero age main sequence and the WD sequence. 

\item \texttt{EFEDS\_GAIAGALEX\_WD}:  \gaia-GALEX UV excess source - expected to be compact binary harbouring a WD 

\item \texttt{EFEDS\_VAR\_WD}:  WD candidate selected on the basis of its optical variability

\end{enumerate}

Further details of the eFEDS target selection process will be presented in A. Merloni et al. (in prep.).

\subsubsection{Target classes, target bits}
\label{subsubsec:maskbits}

The targets described above for the eFEDS field are assigned to 19 different target classes. 
Table \ref{tab:maskbit:description} describes each of the target classes designed for the 12 plates covering eFEDS. 
The maskbits used range from 19 to 47 and can be found in the  \texttt{EBOSS\_TARGET1} target bitmask. 
For clarity, only the bits for target classes that resulted in observations are reported.

\begin{deluxetable*}{clrl}
\tablecaption{Different target classes created for the SDSS-IV/eFEDS plates. \label{tab:maskbit:description}} 
\tablehead{\colhead{{\tt EBOSS\_TARGET1}} & \colhead{Bit name} & \colhead{N}  & \colhead{Description}}
\startdata
 19 & \texttt{EFEDS\_EROSITA\_AGN}  & 2149  & Point-like {\it eROSITA} source - target expected to be AGN, star or compact object          \\
 20 & \texttt{EFEDS\_EROSITA\_CLUS} & 85  & Extended {\it eROSITA} source - target expected to be galaxy cluster member                  \\
 22 & \texttt{EFEDS\_HSC\_HIZQSO}   & 34     & High redshift QSO candidate from HSC photometry                                        \\
 23 & \texttt{EFEDS\_HSC\_WERGS}    & 154 & WISE bright, optically faint radio galaxy from HSC photometry                          \\
 24 & \texttt{EFEDS\_HSC\_REDAGN}   & 108  & Obscured AGN selected in HSC+WISE and/or HSC+VIKING  photometry                        \\
 25 & \texttt{EFEDS\_COSMOQSO}      & 1056 & UVX/variability QSO selection                                                          \\
 26 & \texttt{EFEDS\_WISE\_AGN}     & 649 & WISE color-selected AGN                                                               \\
 27 & \texttt{EFEDS\_WISE\_VARAGN}  & 508 & WISE variability-selected AGN                                                          \\
 28 & \texttt{EFEDS\_XMMATLAS}     & 88  & X-ray selected target from the XMM-ATLAS coverage in the eFEDS field                   \\
 29 & \texttt{EFEDS\_CSC2}         & 6  & X-ray selected target from the Chandra Source Catalog in the eFEDS field             \\
 35 & \texttt{EFEDS\_SDSSWISE\_QSO} & 395 & QSO candidate from UMAP classification of SDSS+WISE photometric sources    \\ 
 36 & \texttt{EFEDS\_GAIAWISE\_QSO} & 651 & QSO candidate from RF classification of \gaia-unWISE sources                   \\ 
 37 & \texttt{EFEDS\_KIDS\_QSO}     & 565  & QSO candidate from classification of KiDS+VIKING DR3 sources           \\ 
 38 & \texttt{EFEDS\_SDSS\_REDMAPPER}& 1031& RedMaPPER galaxy cluster candidate with SDSS photometry                   \\
 39 & \texttt{EFEDS\_HSC\_REDMAPPER} & 13& RedMaPPER galaxy cluster candidate with HSC photometry                         \\
 45 & \texttt{EFEDS\_GAIA\_WD}       & 55 & \gaia-selected WD binary candidate in the region between the ZAMS and the \\
     &                                         &       & WD sequence.  \\
 46 & \texttt{EFEDS\_GAIAGALEX\_WD}  & 9 & \gaia-GALEX UV excess source - expected to be compact binary harbouring a WD            \\
 47 & \texttt{EFEDS\_VAR\_WD}      & 3 & Optical variability selected WD candidate                             \\   
\tablenotetext{}{These bits are located within the {\tt EBOSS\_TARGET1} targeting bitmask. In the column `N', we report the number of observed targets falling in a category. Note that targets can belong to several categories.} 
\enddata
\end{deluxetable*}

\subsubsection{eFEDS Observations}

Observations were carried out in March 2020, but due to a combination of poor weather and Covid-19 shut down (on MJD 58932), the program could unfortunately not be completed. 
The initial success criteria for a plate to be considered complete were  \texttt{SN2\_G1,2} $> 20$,  \texttt{SN2\_I1,2}~$> 40$, where the  \texttt{SN2} are estimates of the squared SNR, at fiducial 2\arcsec~ fiber magnitudes of $g=21.2$ and $i=20.2$, averaged over targets in each of the four BOSS spectrograph arms (G1, I1, G2, I2).
These  \texttt{SN2} thresholds are approximately double those adopted for regular eBOSS observations \citep{Dawson16}. 
This guarantees a highly complete redshift measurements for faint AGN, and allows the measurement of the properties of the AGN and of the host galaxy. 

Table \ref{table:spiders:efeds:sdss4} details the plate numbers and the meta data related to their observations. Only seven out of 12 plates were actually observed in the period MJD 58928--58932; of these seven, only one plate reached nominal depth, three received significant exposure beyond the typical eBOSS limit, and three were only partially exposed.  In Table~\ref{table:spiders:efeds:sdss4} these are marked as good (G), fair (F) and bad (B), respectively. 

The redshifts from eFEDS special plates are part of the catalog that describes the multiwavelength properties of the {\it eROSITA} point sources \citep{Salvato2021}, currently available at \url{https://erosita.mpe.mpg.de/edr/eROSITAObservations/Catalogues/} and expected to be available in Vizier in the future. 

\begin{deluxetable*}{cccccccccc}
\tablecaption{Summary of eFEDS observations. \label{table:spiders:efeds:sdss4}} 
\tablehead{\colhead{Plate}   & \colhead{MJD}  & \colhead{RA}  & \colhead{DEC}  & \colhead{Quality}  & \multicolumn{4}{c}{SN2} & \colhead{Tile ID}  \\ \colhead{Number}& \colhead{}& \colhead{(deg)}& \colhead{(deg)}& \colhead{}& \colhead{G1}& \colhead{I1}& \colhead{G2}& \colhead{I2} & \colhead{}}
\startdata
12525 & 58932 & 130.27 & -0.57 & B  &  7.6 & 17.8 &  9.2 & 26.0 & 17693 \\ 
12527 & 58930 & 130.02 & 1.65  & F  &  9.1 & 27.6 & 10.8 & 35.0 & 17694 \\ 
12531 & 58932 & 131.96 & 1.65  & B  &  3.2 &  9.1 &  4.0 & 11.5 & 17696 \\ 
12533 & 58931 & 134.15 & -0.57 & B  &  0.4 &  0.9 &  0.6 &  1.4 & 17697 \\ 
12544 & 58928 & 137.79 & 1.65  & F  &  9.3 & 21.9 &  9.8 & 29.7 & 17702 \\ 
12545 & 58930 & 140.02 & -0.55 & F  &  8.3 & 24.6 &  9.4 & 30.9 & 17703 \\ 
12547 & 58928 & 139.75 & 1.65  & G  & 28.8 & 70.3 & 27.1 & 81.4 & 17704 \\
\tablenotetext{}{Plate: plate number. MJD: Modified Julian Date of the observations. RA: Right Ascension of the center of the plate (degrees). Dec: Declination of the center of the plate (degrees). Quality: G: good; F: fair; B: bad. SN2: signal to noise reached in each spectrograph's arm.} 
\enddata
\end{deluxetable*}

\subsubsection{eFEDs in SDSS-V} \label{sec:efeds_s5}
The spectroscopic observations of the {\it eROSITA} sources in the eFEDS PV field, reduced at the end of the SDSS-IV program because of the March 2020 closure at APO, has been resumed within the early (plate-mode operations) phase of SDSS-V (see \S \ref{sec:future}). 
A much more complete description of the eFEDS program (SDSS-IV and SDSS-V) will be provided in A. Merloni et al., (in prep.). Among others, the more homogeneous coverage of the combined program will enable X-ray AGN clustering studies and galaxy-galaxy lensing measurements (J. Comparat et al. in prep) on an unprecedentedly large sample.

Eventually, with the commissioning of the robotic fibre positioners on both northern and southern SDSS-V sites, SPIDERS will deliver on its original promises of massive, systematic spectroscopic observations of the sources detected in the {\it eROSITA} all-sky survey as part of the SDSS-V `Black Hole Mapper' (BHM) program \citep{2017arXiv171103234K}. 
The experience and the scientific results obtained by the SPIDERS team within SDSS-IV, that we have briefly described above, represent a major milestone for the planning, execution and exploitation of the BHM survey program.

%% file: rm.tex
The Sloan Digital Sky Survey Reverberation Mapping (SDSS-RM) project is a dedicated multi-object optical reverberation mapping program \citep{Shen_etal_2015a} that has monitored a single 7 ${\rm deg}^2$ field (R.A. J2000=213.704, decl. J2000=+53.083) since 2014 during both SDSS-III and SDSS-IV, using the BOSS spectrographs  \citep[][]{Smee2013} at APO. 
The first-season SDSS-RM spectroscopic data were taken during January-July 2014 in SDSS-III \citep[][]{Eisenstein2011} and consist of a total of 32 epochs with an average cadence of $\sim 4$ days; each epoch had a typical exposure time of 2\,hr. The SDSS-RM program continued in SDSS-IV \citep[][]{2017AJ....154...28B}, with $\sim 12$ epochs per year (2 per month) with a nominal exposure time of 1\,hr each during 2015-2017, and $\sim 6$ epochs per year (monthly cadence) during 2018-2020. As of July 2020, SDSS-RM has obtained a total of 90 spectroscopic epochs over a spectroscopic baseline of 7 years (2014-2020). All SDSS spectroscopic data (pipeline reduced and calibrated) from SDSS-RM are included in the data releases of SDSS-III and SDSS-IV. The full technical details of SDSS-RM are provided in \citet{Shen_etal_2015a}. 

In addition to optical BOSS spectroscopy, accompanying photometric data in the $g$ and $i$ bands are acquired with the 3.6-m Canada-France-Hawaii Telescope (CFHT) and the Steward Observatory 2.3-m Bok telescope \citep{Kinemuchi_etal_2020}. The final photometric baseline spans 11 years (2010-2020) when including the 2010-2013 photometric light curves from the Pan-STARRS 1 \citep[][]{Kaiser_etal_2010} Medium Deep survey that covers the entire SDSS-RM field. The SDSS-RM sample includes 849 broad-line quasars with $i_{\rm PSF}<21.7$ and $0.1<z<4.5$ without any constraints on quasar properties. The detailed sample properties are described in \citet{Shen_etal_2019b}. 

The primary science goal of SDSS-RM is to measure reverberation mapping lags of different broad emission lines covered by optical spectroscopy across the full range of quasar luminosity and redshift probed by the sample. SDSS-RM has successfully measured short ($<6\, {\rm months}$) lags for the low-ionization broad lines \citep[e.g., \halpha, \hbeta, and \ion{Mg}{2};][]{Shen_etal_2016a,Grier2017} based on the 2014 data. \citet{Grier19} reported results on \ion{C}{4} lags using the first four years (2014-2017) of imaging and spectroscopy from SDSS-RM \citep[and for \ion{Mg}{2} lags in][]{Homayouni_etal_2020}, where the lags are typically longer than one season in the observed frame. Lag measurements based on more extended light curves are reported in \citet{Shen_etal_2019c}, and the analyzes of the final SDSS-RM dataset are underway. 

In addition to the main science goal of RM measurements, SDSS-RM also enables a diverse range of quasar science. Some notable examples are: measurements of continuum lags to constrain quasar accretion disk sizes \citep{Homayouni_etal_2019}; constraints on the black hole mass - host galaxy correlations at $z>0.3$ \citep{Shen_etal_2015b,Matsuoka_etal_2015,Yue_etal_2018,Li_etal_2021a}; spectral and variability properties of quasar emission and absorption lines \citep{Shen_etal_2016b,Sun_etal_2018,Hemler_etal_2019,Wang_etal_2019,Wang_etal_2020}; and extreme variability in quasars \citep{Dexter_etal_2019}. 

The SDSS-RM field continues to be monitored with optical imaging and spectroscopy as part of the BHM program in SDSS-V \citep[][and \S \ref{sec:future} below]{2017arXiv171103234K}. With more extended light curves, this program will be able to measure broad-line RM lags for the most luminous quasars at high redshift.

%% file: future.tex
\label{sec:future}
This paper documents the final data release from the SDSS-IV collaboration, and the seventeenth data release from SDSS programs as a whole (DR17). With this paper we make all SDSS-IV data public, concluding the program described in \citet{2017AJ....154...28B}. In the rest of this section, we give an update of SDSS-V which is now actively observing. 

\subsection{SDSS-V}

As SDSS-IV was wrapping up observations (in late 2020 at APO and early 2021 at LCO), the next generation of SDSS began --- SDSS-V\footnote{\url{www.sdss5.org}} \citep{2017arXiv171103234K}, with its first data taken using the existing plug plate system at APO in late 2020. Building on the legacy of earlier generations, SDSS-V is a multi-epoch spectroscopic survey to observe nearly six million sources using the existing BOSS/APOGEE spectrographs and 2.5~m telescopes, as well as very large swathes of the interstellar medium (ISM) in the Milky Way and Local Group using new optical spectrographs and small telescopes.  SDSS-V operates at both APO and LCO and will provide the first ``panoptic'' spectroscopic view of the entire sky, spanning a wide variety of target types, observing cadences, and science goals.

The scientific program is overseen by three ``Mappers'': 
\begin{enumerate}
\item The {\it Milky Way Mapper} (MWM) is targeting millions of stars and stellar remnants with both the APOGEE and BOSS spectrographs, probing stellar populations from the immediate solar neighborhood across the MW, to the far side of the Galactic disk and in the MW's satellite companions.  The MWM's primary goals are to explore (1) the formation and evolution of the MW, (2) the interior physics and evolutionary pathways of stars across all $T_{\rm eff}$ regimes, from combined asteroseismology and spectroscopy \citep[e.g.][]{Aerts2021}, and (3) the architecture of multi-star and planetary systems. 
\item The {\it Black Hole Mapper} (BHM) is targeting nearly half a million accreting SMBHs and other X-ray sources, including newly discovered systems from the {\it SRG/eROSITA} mission, with the optical BOSS spectrographs. The BHM seeks to characterize the X-ray sky, improve our understanding of accretion physics, and trace the evolution and impact of supermassive black holes across cosmic time.  
\item The \textit{Local Volume Mapper} is using a wide-field optical IFU, with new optical spectrographs (R $\sim$ 4000, $\lambda = 3600-9800$\AA) fed by a 16cm telescope, to map $\sim$2800 deg$^2$ of sky.  With their tens of millions of spectra sampling the ISM and embedded stellar populations in the MW and satellite galaxies, the LVM's maps will reveal the physics of star formation and the ISM, the complex interplay between stars and the ISM, and the connections between interstellar processes on parsec-sized to galaxy-wide scales.
\end{enumerate}

SDSS-V expands upon the operational infrastructure and data legacy of earlier SDSS iterations with several key developments. Among these are the installation of robotic fiber positioners in the focal planes of both 2.5~m telescopes at APO and LCO, which replace the now-retired SDSS plug plate system. In comparison to earlier SDSS surveys, these focal plane systems (FPS) enable improved observing efficiency, larger target densities, and more complex observing cadences for the MWM and BHM programs. In addition, the LVM is constructing additional small telescopes at LCO (with longer-term plans to expand to APO) that are linked to new optical spectrographs based on the DESI design \citep{Martini2018}.  SDSS-V is continuing the decades-long SDSS legacy of open data policies and efficient, well-documented public data access, boosted with the development of improved data distribution systems to serve its expansive time-domain, multi-object and integral-field data set to the world. The eighteenth data release of the SDSS (DR18), which will include the first SDSS-V data, mostly targeting and other information, is currently anticipated for 2022. 

\revision{SDSS-V will also build on the rich educational legacy established by previous SDSS generations. This will include continuing to distribute as many SDSS plug plates to schools and science centers as possible, and broadening the reach of the SDSS Voyages suite of educational activities. SDSS’s large online datasets are an ideal resource for supporting hands-on activities using real archival data, at a wide range of ability levels, and have the potential to act as a gateway for numerous data science and analysis topics.}

After twenty-one years of Sloan Digital Sky Surveys, the data released by SDSS-IV in DR17 has made significant contributions to our understanding of the MW, galaxy evolution, and the Universe as a whole, and it will continue to enable new research for years to come. The SDSS-IV project is now complete, but the new technology and exciting new surveys coming as part of SDSS-V mean that the future remains bright for the SDSS legacy.